%% file: article.tex
\RequirePackage{rotating}
\documentclass[manuscript,screen,nonacm]{acmart}

\usepackage{multirow}
\usepackage{rotating}
\usepackage{pifont}
\usepackage[utf8]{inputenc}
\usepackage[T1]{fontenc}
\usepackage{xcolor}
\usepackage{tikz}
\usetikzlibrary{positioning, fit, calc, shapes.geometric, arrows.meta, shadows}

\pgfdeclarelayer{background}
\pgfsetlayers{background,main}

\usepackage[utf8]{inputenc}
\usepackage[T1]{fontenc}
\usepackage{xcolor}
\usepackage{tikz}

\AtBeginDocument{%
  }

\setcopyright{acmlicensed}
\copyrightyear{2025}
\acmYear{2025}
\acmDOI{XXXXXXX.XXXXXXX}

\acmISBN{978-1-4503-XXXX-X/18/06}

\begin{document}

\title{Differential Privacy in Machine Learning: A Survey from Symbolic AI to LLMs}

\author{Francisco Aguilera-Martínez}
\email{faguileramartinez@acm.org}
\author{Fernando Berzal}
\email{berzal@acm.org}
\affiliation{%
  \institution{Department of Computer Science and Artificial Intelligence, University of Granada}
  \city{Granada}
  \country{Spain}
}

\renewcommand{\shortauthors}{Aguilera-Martínez and Berzal}

\begin{abstract}
Machine learning models should not reveal particular information that is not otherwise accessible. 
Differential privacy \cite{dwork2006dp} provides a formal framework to mitigate privacy risks by ensuring that the inclusion or exclusion of any single data point does not significantly alter the output of an algorithm, thus limiting the exposure of private information. 
This survey reviews the foundational definitions of differential privacy and traces their evolution through key theoretical and applied contributions. 
It then provides an in-depth examination of how DP has been integrated into machine learning models, analyzing existing proposals and methods to preserve privacy when training ML models. 
Finally, it describes how DP-based ML techniques can be evaluated in practice.
By offering a comprehensive overview of differential privacy in machine learning, this work aims to contribute to the ongoing development of secure and responsible AI systems.
\end{abstract}



\received{June 2025}
\received[revised]{March 2026}

\maketitle

\input{introduction}
\input{dp}
\input{dp-in-ml}
\input{dp-in-practice}
\input{conclusion}
\bibliographystyle{ACM-Reference-Format}
\bibliography{references}

\end{document}

%% file: introduction.tex
\section{Introduction}

The proliferation of large-scale data collection and analysis has brought unprecedented capabilities but also significant privacy risks. 
The tension between extracting meaningful insights from data and safeguarding individual privacy has become increasingly pronounced. 

Traditional privacy preserving techniques, such as anonymization, often rely on removing personally-identifiable information to protect sensitive data \cite{narayanan2007breakanonymitynetflixprize}. However, these methods have proven insufficient against sophisticated re-identification attacks, which exploit auxiliary information to link anonymized records back to individuals \cite{sp2008deanonymization}. 

The ``Fundamental Law of Information Recovery'' \cite{dwork2014tcs} reveals a trade-off between data utility and privacy: answering a sufficient number of seemingly innocuous queries about a database inevitably reveals sensitive information about it \cite{pods2003revealing}. As a consequence, any useful computation on a dataset necessarily results in some privacy loss. 

Heuristic approaches to privacy have been repeatedly shown to fail, so robust mathematical frameworks are desirable to ensure privacy without compromising utility. This need for formal guarantees has catalyzed the development of differential privacy (DP), a mathematically rigorous framework that quantifies and bounds the privacy risk posed by data analysis \cite{dwork2006dp}, offering a safeguard against inference attacks \cite{tods1980securedb} \cite{sp2017mia}.

The central tenet of differential privacy is indistinguishability. A randomized algorithm (or mechanism) is considered differentially private if its output behavior remains statistically similar regardless of whether any single individual data is included in or excluded from the input dataset. The intuition is that, when an observer cannot reliably determine whether a particular person's data were used in the computation, then the computation reveals little specific information about that person. This provides a strong guarantee: the privacy risk incurred by participating in a differentially-private analysis is minimal, roughly equivalent to the risk if one's data had not been included at all. This concept provides a form of plausible deniability for participants.

Differential privacy is a property of the algorithm or mechanism performing the analysis, not a property of the dataset itself. To achieve this indistinguishability, DP mechanisms must be randomized. Deterministic algorithms cannot satisfy the definition as their output is fixed for a given input. This inherent randomization introduces carefully calibrated noise or uncertainty into the computation, masking the contribution of any single individual while preserving the statistical utility of the overall result. The guarantee provided by DP is relative and probabilistic, bounding the change in output probabilities rather than promising absolute secrecy, acknowledging the fundamental trade-off between utility and privacy.

Differential privacy \cite{dwork2006dp} defines privacy through the lens of a worst-case guarantee: the probability of any output from a data analysis algorithm changes by at most a factor of \( e^{\epsilon} \) when a single individual’s data is added or removed, with a small failure probability  \( \delta \). This \(( \epsilon , \delta )\)-DP framework provides a quantifiable measure of privacy loss, enabling precise trade-offs between privacy and utility. Its strength lies in its resilience to arbitrary auxiliary information, making it robust against attacks such as membership inference, where adversaries attempt to determine whether a particular individual’s data were used in training \cite{sp2017mia}. Beyond its theoretical elegance, DP has spurred a rich ecosystem of algorithmic innovations tailored to machine learning, addressing challenges like high-dimensional data, iterative computations, and distributed systems \cite{dwork2014tcs}. 

The concept of tracker \cite{tods1979tracker}, an adversary that could learn the confidential contents of a database by creating a series of targeted queries and remembering their results, led to the conclusion that the privacy properties of a database can only be preserved by considering each new query in light of (possibly all) previous queries. This led to the inclusion of privacy accountants in DP-preserving ML algorithms. Privacy accountants track and manage the cumulative privacy loss, so that the total privacy budget is not exceeded over multiple data queries.

Subsequent research advancements have refined DP’s theoretical foundations and practical utility.
Methods have been proposed to optimize the privacy-utility trade-off, particularly in settings requiring iterative data access, such as the algorithms used to train ML models. 

Common use cases of DP algorithms include mobile devices, healthcare, and financial data. 
The first deployment of DP was made by the U.S. Census Bureau, for displaying the commuting patterns of the population of the United States \cite{icde2008commuting}. The first widespread use of local differential privacy was made by RAPPOR \cite{ccs2014rappor} for collecting Google Chrome browser telemetry data.
Microsoft also collects telemetry data privately in Windows \cite{nips2017windows}.
Predictive text or voice recognition models trained across user devices use privacy-preserving federated learning (e.g. Google’s Gboard uses DP to improve keyboard suggestions without compromising user privacy \cite{arxiv2023gboard}). 
Hospitals collaboratively train models on patient data without sharing sensitive records. Banks train fraud detection models without exposing customer data.
DP adoption in industry applications, from Google’s \cite{ccs2014rappor} or Uber's \cite{vldb2018uber} private query analysis to Apple’s emoji usage modeling \cite{apple2017learning}, underscores its practical relevance \cite{dwork2014tcs}.

The integration of differential privacy into machine learning has attracted increasing attention from both the research and practitioner communities, yet its theoretical foundations, practical mechanisms, and system-level implications remain scattered across disciplines. This survey aims to provide a structured and comprehensive overview of differential privacy in machine learning, with the goal of bridging formal definitions, practical mechanisms, and system-level considerations. 

Our survey on the use on DP in ML is structured as follows. First, we introduce the DP theoretical foundations, including its original definition, the role of sensitivity, and the multiple variants that have been proposed over the years, covering their formal properties with an emphasis on their composability, their robustness to post-processing, and their potential degradation in the presence of correlated data. Next, we examine how DP has been applied in different kinds of ML models, from symbolic and probabilistic AI to deep learning and model agnostic techniques, such as federated learning and PATE \cite{iclr2017pate}. We further discuss recent developments in applying differential privacy to large language models (LLMs), highlighting the new privacy risks introduced by LLM training and deployment pipelines. Finally, we discuss how the effectiveness of DP-based machine learning techniques can be evaluated in practice, focusing on privacy accounting, empirical privacy attacks, and the resulting privacy–utility trade-off. 

This survey is intended for researchers and practitioners working on privacy-preserving AI systems who seek a principled reference for understanding, implementing, and evaluating differential privacy across the full spectrum of machine learning models, including contemporary LLMs. Throughout our exploration, we seek to contribute to the ongoing research on privacy-preserving technologies in AI systems.

\begin{figure*}[htbp]
    \centering
    \Description{Conceptual roadmap of the survey.} 
    \resizebox{\textwidth}{!}{
        \input{diagram-roadmap-dp}
    }
    \caption{The scope of this survey, from the formal foundations of differential privacy to its deployment and trade-offs, with an emphasis on its integration into ML applications and LLMs.}
    \label{fig:roadmap_final}
\end{figure*}

%% file: diagram-roadmap-dp.tex
\definecolor{borderblue}{RGB}{0, 80, 150}
\definecolor{arroworange}{RGB}{180, 90, 40}
\definecolor{boxbg}{RGB}{245, 245, 245}

\begin{tikzpicture}[
        font=\sffamily,
        block/.style={
            rectangle, draw=gray!30, fill=boxbg,
            thick, rounded corners=5pt,
            text width=5.5cm, minimum height=3.2cm,
            inner sep=10pt, align=left,
            drop shadow={shadow xshift=2pt, shadow yshift=-2pt, opacity=0.2}
        },
        flow/.style={
            -{Latex[length=3mm]},
            arroworange, line width=2.5pt,
            rounded corners=12pt
        },
        labeltext/.style={font=\sffamily\small\bfseries, color=arroworange}
    ]

    \node[block] (b1) {
        {\bfseries Foundations\\of Differential Privacy}\\[0.2cm]
        \small
        $\bullet$ Formal definition \\
        $\bullet$ Sensitivity\\
        $\bullet$ Extensions and variants\\
        $\bullet$ Formal properties of DP mechanisms \\
    };

    \node[block, right=1cm of b1] (b2) {
        {\bfseries Differential Privacy\\in Machine Learning}\\[0.2cm]
        \small
        $\bullet$ ML Families: Symbolic, Probabilistic... \\
        $\bullet$ Deep Learning \\
        \hspace{0.4cm} \textendash \ DP-SGD \\
        \hspace{0.4cm} \textendash \ Privacy accounting vs. Privacy for free\\
        
    };

    \node[block, below=0.5cm of b1] (b4) {
        {\bfseries Deployment Paradigms \\ \& Empirical Evaluation }\\[0.2cm]
        \small
        $\bullet$ Model-agnostic DP frameworks \\
        $\bullet$ Measuring privacy effectiveness \\
        $\bullet$ Measuring utility effectiveness \\
        $\bullet$ Privacy–utility trade-offs \\
    };

    \node[block, right=1cm of b4] (b3) {
        {\bfseries Differential Privacy\\in LLMs}\\[0.2cm]
        \small
        $\bullet$ Privacy risks in LLM pipelines \\
        $\bullet$ DP fine-tuning \& data-centric approaches\\
        $\bullet$ RAG, knowledge transfer, and distillation \\
        $\bullet$ Scaling laws and fundamental limitations \\
    };

    \draw[flow] (b1.north) -- ++(0,0.55) -- node[above, labeltext]{ADAPTATION} ($(b2.north)+(0,0.55)$) -- (b2.north);
    \draw[flow] (b2.east) -- ++(0.6,0) -- node[pos=0.3, right, labeltext, rotate=-90, yshift=0.2cm]{SCALING} ($(b3.east)+(0.6,0)$) -- (b3.east);
    \draw[flow] (b3.south) -- ++(0,-0.55) -- node[below, labeltext]{EVALUATION} ($(b4.south)+(0,-0.55)$) -- (b4.south);
    \draw[flow] (b4.west) -- ++(-0.6,0) -- node[pos=0.8, left, labeltext, rotate=90, yshift=0.2cm]{REFINEMENT} ($(b1.west)+(-0.6,0)$) -- (b1.west);

    \node[draw=borderblue, line width=1.5pt, rounded corners=20pt, 
          fit=(b1) (b2) (b3) (b4), inner sep=1.2cm] (container) {};

    \node[above=0.2cm of container.north, font=\small\bfseries\color{borderblue}] {FUNDAMENTALS};
    \node[below=0.2cm of container.south, font=\small\bfseries\color{borderblue}] {CHALLENGES};
    \node[left=0.3cm of container.west, font=\small\bfseries\color{borderblue}, rotate=90] {THEORY};
    \node[right=0.3cm of container.east, font=\small\bfseries\color{borderblue}, rotate=-90] {PRACTICE};

    \end{tikzpicture}

%% file: dp.tex
\section{Differential privacy}

Differential privacy (DP) provides a mathematical definition for the privacy loss associated with any data release drawn from a database.\footnote{DP was originally described in terms of statistical databases \cite{dwork2006dp}, sets of data that are collected under the pledge of confidentiality for the purpose of producing statistics that, by their production, do not compromise the privacy of those individuals who provided the data. Since a person's privacy cannot be compromised by a statistical release if their data are not in the database, in differential privacy, each individual is given roughly the same privacy that would result from having their data removed. In other words, the statistical functions run on the database should not be substantially affected by the removal, addition, or change of any individual in the data.} DP establishes a mathematically rigorous framework to ensure that the presence or absence of a single individual in a dataset does not significantly affect the outcome of a computation. DP protects the privacy of individual data examples by injecting carefully calibrated noise into statistical computations. Statistical information from the dataset can still be used (e.g. aggregate patterns) while provably limiting what can be inferred about any individual in the dataset.

In the context of our survey, this section provides the background for for the subsequent discussion on privacy-preserving machine learning systems. We first introduce the formal definition of differential privacy and the notion of sensitivity, which together determine the privacy guarantees of a DP mechanism. We then review key DP extensions and variants, highlighting their motivation and typical use cases. Finally, we summarize the formal properties of DP mechanisms that enable composition, post-processing robustness, and privacy accounting in iterative and large-scale machine learning pipelines.

\subsection{Formal definition}

Intuitively, a model $M$ exhibits differential privacy if its answers are indistinguishable when trained over datasets that differ in only one individual data example.

\subsubsection{\( \epsilon \)-differential privacy:}  Cynthia Dwork, Frank McSherry, Kobbi Nissim, and Adam D. Smith introduced $\epsilon$-differential privacy in 2006 \cite{dwork2006dp}. Differential privacy measures privacy risk by a parameter $\epsilon$ that bounds the log-likelihood ratio of the output of a (private) algorithm under two databases differing in a single individual’s data. 

Formally, a non-interactive mechanism $M$ (e.g. a machine learning model) satisfies $\epsilon$-differential privacy if, for all neighboring datasets $x, x'$ (i.e., datasets differing in at most one example) and for all measurable subsets of outputs $S$, the following non-strict inequality holds:
$$
    \left| \log \left( \frac{ P[M(x) \in S] }{P[M(x') \in S]} \right) \right| \leq \epsilon
$$
which is typically rewritten as
$$
    P[M(x) \in S] \leq e^{\epsilon} P[M(x') \in S]
$$

The parameter $\epsilon > 0$, known as privacy loss or leakage, controls the level of privacy. Smaller values of $\epsilon$ provide stronger privacy guarantees, ensuring that the distributions induced by $M(x)$ and $M(x')$ remain nearly indistinguishable.

When $\epsilon$ is small, $\log (1+\epsilon) \approx \epsilon$, so the initial definition of $\epsilon$-indistinguishability, or $\epsilon$-differential privacy, is roughly equivalent to
$$
    \frac{ P[M(x) \in S] }{P[M(x') \in S]} \approx 1 \pm \epsilon
$$

Under $\epsilon$-differential privacy, a change to one example in the training data only creates a small change in the model output probability distribution, as seen by a potential attacker. Since deterministic mechanisms cannot satisfy this definition except in trivial cases, DP inherently requires randomization in the output generation process.

\subsubsection{\( (\epsilon, \delta) \)-differential privacy:}
While \( \epsilon \)-differential privacy provides strong guarantees, there are scenarios where it may be too restrictive. In practice, a relaxation known as \( (\epsilon, \delta) \)-differential privacy \cite{dwork2006adp} allows for a small probability \( \delta \) of the privacy guarantee being violated, making it more suitable for real-world applications. 

Formally, a randomized mechanism \( M \) 
satisfies \( (\epsilon, \delta) \)-differential privacy if, for all neighboring datasets $x, x'$ 
and for all measurable output subsets $S$, the following property holds \cite{dwork2014tcs}:
$$
P[M(x) \in S] \leq e^{\epsilon} P[M(x') \in S] + \delta.
$$

When $\delta = 0$, $(\epsilon, \delta)$-differential privacy reduces to pure $\epsilon$-differential privacy.

In $(\epsilon, \delta)$-differential privacy, also known as approximate differential privacy, $\delta$ represents the probability that the privacy guarantee may be broken, which is typically required to be negligible, often less than the inverse of any polynomial in the database size $|x|$. 

Differential privacy can be interpreted as bounding the error rates of a hypothesis test \cite{kairouz2015pmlr}. Given a random output $y$ of the mechanism $M$, we choose the null hypothesis $H_0$, when $y$ came from a database without a particular individual example, and the alternative hypothesis $H_1$, when $y$ came from a database with the particular example. We have two error rates: the false positive rate (FPR) or probability of a false alarm (i.e. type I error, the adversary guesses $H_1$ when $H_0$ is true) and the false negative rate (FNR) or probability of a missed detection (i.e. type II error, the adversary guesses $H_0$ when $H_1$ is true). The differential privacy condition on a mechanism $M$ is equivalent to the following set of constraints
on the probability of false alarm and missed detection for a fixed $(\epsilon, \delta)$:
$$
P_{\text{FP}} + e^{\epsilon } P_{\text{FN}} \geq 1-\delta
$$
$$
e^{\epsilon } P_{\text{FP}} + P_{\text{FN}} \geq 1-\delta 
$$
In other words, it is impossible to get both small false positive rates and false negative rates from data obtained via a differentially private mechanism \cite{kairouz2015pmlr}. In fact, this trade-off is common in many ML scenarios. Reducing false positives often leads to more false negatives, and vice versa.  If the criteria for identifying a positive case are very strict (to reduce false positives), we may miss some actual positive cases (leading to false negatives). Conversely, if the criteria are very lenient (to reduce false negatives), we may incorrectly identify negative cases as positive (leading to false positives).

\subsection{Sensitivity}

How much does any individual contribute to the result of a database query or a training example to the model output?

A key concept in the implementation of differential privacy is sensitivity, which quantifies the maximum possible change in the output of a function when a single example in the dataset is modified. Specifically, the sensitivity of a function $f$ is defined as \cite{dwork2006dp}:
$$
\Delta f = \max_{x, x': d(x,x')=1} \left\| f(x) - f(x') \right\|_1
$$
where $x$ and $x'$ differ by a single example. Since this sensitivity is determined by all possible pairs $(x, x')$, it is also known as global sensitivity \cite{nissim2007stoc}.

It should be noted that the noise magnitude depends on the global sensitivity and the privacy parameter $\epsilon$, but not on the particular instance $x$. This noise might be higher than needed, since it does not reflect the function’s typical insensitivity to individual inputs. Hence a similar local sensitivity \cite{nissim2007stoc} can be defined for a particular instance $x$:
$$
\Delta_{x} f = \max_{x': d(x,x')=1} \left\| f(x) - f(x') \right\|_1
$$

Sensitivity is essential for determining how much noise needs to be added to the output to ensure privacy \cite{haeberlen2011usenix}. In many common settings, when the sensitivity of a query depends on the dataset size (e.g., empirical averages), querying fewer examples requires adding more noise to achieve the same privacy guarantee. However, for queries with fixed global sensitivity, the noise magnitude is independent of the dataset size.

Adding noise proportional to the global sensitivity of a function can obscure individual contributions while preserving aggregate statistics, as demonstrated by the SuLQ framework \cite{dwork2005pods}.

For many functions, the global sensitivity framework yields unacceptably high noise levels, whereas adding noise proportional to a smooth upper bound to the local sensitivity might still be safe (the smooth sensitivity framework \cite{nissim2007stoc}).

To achieve \( \epsilon \)-differential privacy, noise is typically introduced into the results of the differential privacy mechanism. In other words, differential privacy relies on adding carefully calibrated random noise to the function that we want to compute. Common distributions for noise generation include the continuous Laplace and Gaussian distributions for functions that output real-valued numbers, as well as the geometric distribution for discrete outputs. Other mechanisms, such as the exponential mechanism and posterior sampling, sample from a problem-dependent family of distributions instead.

\begin{itemize}


\item
The {\it Laplace mechanism} adds noise sampled from a Laplace distribution with scale $\Delta f / \epsilon$ and mean $0$. Given a query function \( f \), the private response mechanism is:
$$
\mathcal {M}_{\text{Laplace}}(x) = f(x) + \text{Laplace} \left( \mu = 0, \lambda = \frac{\Delta f}{\epsilon} \right)
$$
where $\text{Laplace} \left(  \mu = 0, \lambda = \frac{\Delta f}{\epsilon} \right)$ represents the Laplace noise that provides the necessary randomness to ensure that the privacy requirement is met \cite{dwork2006dp}. 

\item
The {\it Gaussian mechanism} adds noise from the Gaussian distribution (which requires the L2 norm in the definition of sensitivity) instead of the Laplace distribution. The variance of the Gaussian distribution is calibrated according to the sensitivity and privacy parameters to provide $(\epsilon ,\delta )$-differential privacy \cite{dwork2014tcs}:
$$
{\mathcal {M}}_{\text{Gauss}}(x) = f(x) + {\mathcal {N}}\left(\mu =0,\sigma ^{2}={\frac {2\ln(1.25/\delta )\cdot (\Delta f)^{2}}{\epsilon ^{2}}}\right)
$$

\item
A discrete variant of the Laplace mechanism, called the {\it geometric mechanism}, is universally utility-maximizing \cite{ghosh2009arxiv} \cite{ghosh2009stoc} \cite{ghosh2012siam}, optimizing expected loss across various priors and symmetric, monotone univariate loss functions through a data-independent post-processing step. For any prior (such as auxiliary information or beliefs about data distributions) and any symmetric and monotone univariate loss function, the expected loss of any differentially private mechanism can be matched or improved by running the geometric mechanism followed by a data-independent post-processing transformation. This result also holds for minimax (risk-averse) consumers \cite{gupte2010pods}. However, no such universal mechanism exists for multi-variate loss functions \cite{brenner2010focs} \cite{brenner2014siam}.
 
\item
The {\it exponential mechanism} \cite{mcsherry2007mechanism} comes with guarantees about the quality of the output, even for functions that are not robust to additive noise, and those whose output may not even permit perturbation. 
The mechanism $\mathcal {E}(x)$ chooses an output $y$ with probability proportional to $e^{\epsilon q(x,y)} \times \mu(y)$, where $x$ is the input, $q$ is a scoring function that reflects the appeal of the $(x,y)$ pair and $\mu(y)$ is a base measure. This mechanism skews a base measure to the largest degree possible while ensuring differential privacy, focusing probability on the outputs of highest value. The mechanism $\mathcal {E}(x)$ provides $(2\epsilon \Delta q)$-differential privacy, where $\Delta q$ measures the change in $q$ due to a single change in $x$.

\item
A {\it posterior sampling mechanism} \cite{dimitrakakis2014alt} is also differentially-private and provides finite sample bounds for distinguishability-based privacy under
a strong adversarial model.

\end{itemize}

When adding noise to ensure privacy may not be appropriate, such as when the function being analyzed is not continuous or involves discrete structures like strings or trees, posterior sampling models, such as the exponential mechanism, are required for privacy preservation. 

Challenges also arise when sensitivity is difficult to assess, although both Laplace and exponential mechanisms remain useful in these cases. In iterative algorithms, such as clustering methods (e.g. k-means and DBSCAN density-based clustering), privacy can still be maintained by applying some algorithmic techniques. These include releasing approximate counts for each cluster instead of exposing individual data points (using the geometric mechanism, the optimal and universal discrete extension of the Laplace mechanism), implementing differentially-private data structures (e.g. a differentially private region quadtree \cite{ho2011springl} for protecting GPS locations while keeping a reduced local sensitivity), or generating synthetic data, where the synthetic dataset mimics the statistical properties of the original dataset.

However, it should also be noted that a vulnerability is present in many implementations of differentially private mechanisms \cite{mironov2012ccs}. This vulnerability is based on irregularities of the floating-point implementations of the privacy-preserving Laplacian mechanism. Unlike its mathematical abstraction, the textbook sampling procedure results in a porous distribution over double-precision numbers that leads to a breach of differential privacy with just a few queries to the implementation of differentially-private mechanisms. 

Pan-privacy extends differential privacy  \cite{dwork2011cacm}: A pan-private algorithm is private ``inside and out,'' remaining differentially private even if its internal state becomes visible to an adversary \cite{dwork2010ics}. Not only the outputs, but also the distributions over the internal states of the algorithm should be similar for slightly different input data.

Beyond its privacy-preserving benefits, differential privacy also imposes an inherent limitation on the influence of any single individual data on the final output. This characteristic has profound implications beyond privacy, particularly in strategic settings where individuals may have incentives to manipulate outcomes. Differentially-private mechanisms naturally exhibit approximate strategy-proofness. In mechanism design, no player or coalition can gain by significantly altering the outcome in their favor. This insight extends to machine learning applications, where training models with differential privacy ensures that no single data point disproportionately affects the model parameters, thus improving the model robustness and generalization capability \cite{mcsherry2007mechanism}.

\subsection{Extensions and variants}

In the landscape of privacy-preserving data analysis, the foundational framework of differential privacy, initially introduced with a strict worst-case guarantee parameterized by \( \epsilon \), set the stage for subsequent refinements. 

Instead of introducing fundamentally new privacy notions, most differential privacy extensions and variants relax or reinterpret its original definition in order to improve their composition properties, support alternative threat models, or enable their practical deployment in complex machine learning pipelines. Following \cite{pejo2022guide}, these variants can be understood as choosing different trade-offs across several conceptual dimensions:

\begin{figure*}[ht!bp]
    \centering
    \Description{Taxonomy diagram of Differential Privacy extensions.} 
    \resizebox{\textwidth}{!}{
        \input{diagram_taxonomia_dp}
    }
    \caption{Conceptual taxonomy of differential privacy variants and extensions (see Table \ref{tab:variants}).}
    \label{fig:taxonomia_dp}
\end{figure*}

\begin{itemize}

\item
Quantification of privacy loss for averaging risk and obtaining better composition properties: \\ How is the privacy loss quantified across outputs? 

\item
Neighborhood definition for protecting specific values or multiple individuals: \\ Which properties are protected from the attacker?

\item
Variation of privacy loss for modeling users with different privacy requirements: \\ Can the privacy loss vary across inputs?

\item
Background knowledge for using mechanisms that add less or no noise to data: \\ How much prior knowledge does the attacker have?

\item
Formalism change for exploring other intuitive notions of privacy: 
\\ Which formalism is used to describe the attacker's knowledge gain?

\item
Relativization of knowledge gain for guaranteeing privacy for correlated data: \\ What is the knowledge gain relative to?

\item
Computational power  for combining cryptography techniques with differential privacy: 
\\ How much computational power can the attacker use?
\end{itemize}

From a practical perspective, DP variants can be broadly grouped into three categories:
(i) variants designed to obtain tighter composition and privacy accounting (e.g., CDP, RDP, GDP); (ii) variants targeting alternative threat models or data distributions (e.g., dependent DP, Blowfish privacy); and (iii) variants tailored to specific deployment settings, such as local data collection (e.g., LDP). 

Variants designed for tighter composition and privacy accounting, such as CDP \cite{dwork2016concentrateddifferentialprivacy}, RDP \cite{mironov2017rdp}, and GDP \cite{dong2022gdp}, are typically preferred in iterative training scenarios involving gradient-based optimization, where privacy loss accumulates across multiple training steps.

Variants targeting alternative threat models or data distributions, such as dependent DP \cite{liu2016ndss} or Blowfish privacy \cite{he2014blowfish}, are more suitable when classical adjacency assumptions do not capture real-world privacy risks, including correlated data or domain-specific privacy policies.

From a deployment perspective, selecting the appropriate DP variant depends on the interaction between threat models, data characteristics, and system constraints. 
Variants tailored to specific deployment settings, such as local differential privacy \cite{focs2008local}, are generally selected when the data curator is not trusted or when privacy must be enforced at the data collection stage.

\begin{table}[t]
    \centering
    \begin{tabular}{rlc}
      \hline
      Definition &  Description (idea) & Reference \\ 
      \hline
         $\epsilon$-DP & Original idea ($\epsilon$-indistinguishability) & \cite{warner1965randomized} \cite{dwork2006dp} \\
         $(\epsilon, \delta)$-DP & Approximate DP ($\delta$ failure probability) & \cite{dwork2006adp} \\
         $(\epsilon, \delta)$-PDP & Probabilistic DP (probabilistic relaxation) & \cite{cryptoeprint2018pdp} \\ 
         $\epsilon$-KLP & Kullback–Leibler privacy (mutual information) & \cite{ccs2016klp} \\ 
         $(\alpha, \epsilon)$-RDP & Rényi differential privacy (Rényi divergence) & \cite{mironov2017rdp} \\ 
         $(\mu, \tau)$-mCDP & Mean-concentrated DP (tail distribution) & \cite{dwork2016concentrateddifferentialprivacy} \\
         $(\xi, \rho)$-zCDP & Zero-concentrated DP (tail distribution) & \cite{bun2016cdp} \cite{bun2016cdptr} \\ 
         $(f, \epsilon)$-DDP & Divergence DP (divergence function) & \cite{barber2014divergence} \\ 
         $(P, \epsilon)$-OSDP & One-sided DP (sensitive records) & \cite{icde2020osdp} \\
         $(R,c,\epsilon)$-DDP & Dependent differential privacy (data correlation) & \cite{liu2016ndss} \\
         $\epsilon$-FLP & Free-lunch privacy (strongest variant, no utility) & \cite{kifer2011sigmod} \\ 
         $(\mathcal{R},\epsilon)$-GDP & Generic DP (generic neighboring relation) & \cite{kifer2011sigmod} \\ 
         $(G, \mathcal{I}_Q, \epsilon)$-BP & Blowfish privacy (generic DP inspired by the Pufferfish framework) & \cite{he2014blowfish} \\
         $f$-DP & $f$-differential privacy (hypothesis testing interpretation) & \cite{dong2022gdp} \\
         GDP & Gaussian differential privacy (hypothesis testing interpretation) & \cite{dong2022gdp} \\
         $(D, \epsilon)$-IDP & Individual DP (for a particular dataset), a.k.a. conditioned DP & \cite{tifs2017idp} \cite{jcp2017cdp} \\ 
         LDP & Local differential privacy (for single data examples) & \cite{focs2008local} \\
         \hline
          & & \\
    \end{tabular}
    \caption{Some variants and extensions of the original DP definition.}
    \label{tab:variants}
\end{table}

Table \ref{tab:variants} includes some representative differential privacy variants and extensions that have been proposed in the literature. In the remainder of this section, we focus on those variants that have been more influential in machine learning applications, in particular for privacy accounting and large-scale model training. We review concentrated DP, Rényi DP, and Gaussian DP, which enable tracking privacy loss in iterative model training algorithms, and conclude with local DP for distributed scenarios.

\subsubsection{Concentrated Differential Privacy (CDP)} 

Concentrated Differential Privacy (CDP) \cite{dwork2016concentrateddifferentialprivacy} is a refinement of the traditional (approximate) differential privacy, i.e. $(\epsilon,\delta)$-DP, designed to provide stronger privacy guarantees while improving accuracy in data analysis, which makes it particularly useful in machine learning.

In essence, CDP fine-tunes the privacy mechanism to provide better results without compromising security. Instead of applying a uniform level of randomness, it allows for a tighter, more efficient balance between privacy and accuracy. This balance ensures that the privacy loss remains small most of the time, making it easier to get meaningful insights while still protecting individual identities.

Its name derives from shifting from a uniform bound on privacy loss to a probabilistic concentration approach, which offers enhanced flexibility and tighter composition bounds for iterative data queries. CDP introduces a sophisticated relaxation of the classical model, shifting the focus from absolute bounds on privacy loss to a probabilistic characterization that leverages concentration properties, an approach that merits rigorous examination for its theoretical and practical implications.

Formally, CDP defines privacy in terms of a bound on the moment generating function of the privacy loss random variable, rather than using a strict probabilistic constraint as in traditional differential privacy. Given a mechanism $\mathcal{M}$ that operates on a dataset, CDP ensures that for all possible outputs $S$, 
$$
E[ e^{\lambda L}] \leq e^{\frac{\lambda^2 \sigma^2}{2}}
$$
where $L$ is the privacy loss random variable, $\sigma^2$ is a privacy parameter controlling variance (a.k.a. a variance proxy), and $\lambda$ is any real number.

The privacy loss random variable measures how much information an adversary can gain about an individual's data when a privacy-preserving mechanism is applied. Given a randomized mechanism $\mathcal{M}$ and any pair of neighboring datasets $x$ and $x'$ differing in a single record, the privacy loss random variable is defined as 
$$
L(\mathcal{M},D,D') = \log  \frac{P[\mathcal{M}(x) \in S]}{P[\mathcal{M}(x') \in S]}  
$$
for the possible outputs $S$ of $\mathcal{M}$, where $P[\mathcal{M}(D) \in S]$ and $P[\mathcal{M}(x') \in S]$ represent the probabilities of obtaining $S$ when the mechanism is applied to $x$ and $x'$, respectively.

Formally, a mechanism $\mathcal{M}$ satisfies $(\mu, \sigma^2)$-mCDP, mean-concentrated differential privacy \cite{dwork2016concentrateddifferentialprivacy}, if
$$
E[e^{\lambda (L - \mu)}] \leq e^{\frac{\lambda^2 \sigma^2}{2}} \text{ for all } \lambda \in \mathbb{R},
$$
thereby ensuring that the moment-generating function of the centered privacy loss is dominated by that of a Gaussian distribution. 

We say that CDP exhibits sub-Gaussian tail behavior \cite{kahane1960proprietes}, with mean bounded by $\mu$\ ($E[L] \leq \mu$) and variance proxy $\sigma^2$, a.k.a. $\tau$ ($Var[L] \leq \sigma^2$) \cite{buldygin2000metric} \cite{rivasplata2012subgaussian}, when we refer to the way its privacy loss is controlled (the randomness in how much an individual's data can influence the outcome). In CDP, the privacy loss random variable satisfies a concentration property that limits the likelihood of extreme privacy breaches (large deviations from expected privacy loss are rare).\footnote{A random variable has sub-Gaussian tail behavior if its probability of taking extreme values (very large deviations from the mean) drops exponentially, rather than following a heavy-tailed distribution.}


Zero-concentrated differential privacy (zCDP) refines CDP by defining the privacy loss using a differential privacy-like bound in terms of a quadratic constraint on the Rényi divergence between distributions \cite{renyi1961measures}\cite{tit2013renyi-kl}. This allows for stronger composition properties (when multiple privacy-preserving operations are applied, zCDP ensures that the accumulated privacy loss remains well-controlled) and improved accuracy (tighter bounds on noise addition, leading to better utility in machine learning and statistical analysis).

A mechanism $\mathcal{M}$ satisfies $\rho$-zCDP, zero-concentrated differential privacy \cite{bun2016cdp}, if, for all neighboring datasets $x$ and $x'$,
$ D_{\alpha} \bigl( \mathcal{M}(x) \parallel \mathcal{M}(x') \bigr) \leq \rho (\alpha-1) $
where $D_{\alpha}$ represents the Rényi divergence of order $\alpha$ (for any order $\alpha > 1$). Larger values of $\alpha$ correspond to tighter privacy bounds, tuning sensitivity to rare events, whereas $\rho$ controls the strength of the privacy guarantees (a privacy budget that controls how much information can leak when analyzing data, so that smaller $\rho$ leads to stronger privacy).

The Rényi divergence is a way to measure how different two probability distributions are. The Rényi divergence of positive order ($\alpha \neq 1$) of a probability distribution $P$ from another probability distribution $Q$ is
$$
D_{\alpha} \bigl( P \parallel Q \bigr) = \frac{1}{1-\alpha} \log \sum_{x} p(x)^\alpha q(x)^{1-\alpha}
$$
where, for $\alpha>1$, $p(x)^\alpha q(x)^{1-\alpha} = p(x)^\alpha / q(x)^{\alpha-1}$. This definition generalizes to continuous spaces by replacing the probabilities by densities and the sum by an integral. When the order $\alpha \to 1$, the Rényi divergence reduces to the Kullback–Leibler (KL) divergence \cite{ams1951kl}, commonly used in information theory: $D_1 \bigl( P \parallel Q \bigr) = \sum_x p(x) \log \frac{p(x)}{q(x)}$.  
When $\alpha > 1$, the Rényi divergence gives more weight to larger differences between the distributions, making it more sensitive to rare events. 
As $\alpha \to \infty$, the divergence focuses only on the largest discrepancy between the two distributions, which is related to min-entropy (i.e. a measurement of the worst-case uncertainty, $H_\infty(X) = -\log \max_x p(x)$, how much uncertainty remains if we assume the most likely event will happen). The max-divergence is the worst-case relative difference between distributions: $D_\infty \bigl( P \parallel Q \bigr) = \log \max_x \frac{p(x)}{q(x)}$.

In $\rho$-zCDP, 
$\mathbb{E}\left[ e^{(\alpha - 1) L} \right] \leq e^{\rho \alpha}$ and $\mathbb{E}\left[ e^{\lambda L} \right] \leq e^{\lambda^2 \rho / 2} $.
The privacy loss random variable, whose moments are now bounded,\footnote{For a random variable $X$, the moment-generating function is defined as $M_X(t) = E[e^{tX}]$, where $e^{tX}$ transforms the probability distribution into a function that encodes all its moments. The MGF provides a systematic way to control and compute moments. The first derivative at $t=0$ gives the first moment, expected value, or mean: $E[X]=M'_X(0)$. The second derivative helps compute variance: $Var[X] = M''_X(0) - M'_X(0)^2$. The n-th derivative gives the n-th moment: $E[X^n]=M_X^{(n)}(0)$. In zCDF, the privacy loss random variable satisfies $\mathbb{E}\left[ e^{(\alpha - 1) L} \right] \leq e^{\rho \alpha}$ for all $\alpha > 1$, ensuring that privacy leakage is bounded across all moments. This leads to predictable privacy guarantees and strong composition properties.} is sub-Gaussian with mean approximately $\rho$ and variance bounded by $2\rho$. Its definition implies $(\epsilon, \delta)$-DP with $\epsilon = \rho + 2 \sqrt{\rho \ln (1/\delta)}$ \cite{bun2016cdp}.

In general, in $(\xi, \rho)$-zCDP, $\mathbb{E}\left[ e^{\lambda L} \right] \leq e^{\lambda \xi + \lambda^2 \rho / 2} $, where $\xi$ represents the expected privacy loss (mean privacy leakage) and $\rho$ controls the spread (variance) of the privacy loss, preventing extreme deviations. Its definition implies $(\epsilon, \delta)$-DP with $\epsilon = \xi +\rho + 2 \sqrt{\rho \ln (1/\delta)}$ \cite{bun2016cdp}.

The formulation of \(\rho\)-zCDP, i.e. $(0, \rho)$-zCDP, leverages the Rényi divergence with adjustable sensitivity through the parameter $\alpha$ to establish a flexible and analytically robust privacy guarantee.  The linear constraint $\rho \alpha$ on the divergence ensures tight concentration of the privacy loss, setting zCDP apart from pure differential privacy, which depends on a worst-case exponential bound, and approximate differential privacy, which tolerates a failure probability. 


Four more variants of concentrated differential privacy exist. 
Approximate zCDP \cite{bun2016cdp} relaxes zCDP by only taking the Rényi divergence on events with high probability instead of on the full distribution, so that $\delta$-approximate $(\epsilon,0)$-zCDP is equivalent to $(\epsilon,\delta)$-DP. 
Bounded CDP \cite{bun2016cdp} relaxes zCDP by changing the inequality in the zCDP definition so that $ D_{\alpha} \bigl( \mathcal{M}(x) \parallel \mathcal{M}(x') \bigr) \leq \xi + \rho \alpha$ for $\alpha \in \{1,m\}$ (i.e. the inequality holds only for the first $m$ orders of the Rényi divergence). Finally, two definitions share the truncated CDP name, where the first \cite{colisson2016l3} relaxes $(0, \rho)$-zCDP the same way that bounded CDP and the second \cite{stoc2018truncatedcdp} requires the Rényi divergence to
be smaller than $\min \{ \xi, \alpha \rho \}$ for all $\alpha \geq 1$.

\subsubsection{Rényi Differential Privacy (RDP)}

RDP is another variant that also uses moment generating functions to analyze privacy loss. Like CDP, it provides stronger composition properties, making it easier to track cumulative privacy loss in iterative computations (e.g., machine learning). 

RDP is closely related to zCDP but differs in how it constrains the privacy loss, offering a more concentrated bound. A randomized mechanism $\mathcal{M}$ is said to satisfy Rényi Differential Privacy (RDP) of order $\alpha > 1$ with privacy parameter $\epsilon$ if, for all neighboring datasets $x, x'$ that differ in at most one element, the following holds  \cite{mironov2017rdp}:
$$
D_{\alpha} \left( \mathcal{M}(x) \parallel \mathcal{M}(x') \right) \leq \epsilon
$$
where $D_{\alpha}$ is the Rényi divergence \cite{renyi1961measures} of order $\alpha$.  
The constraint $\leq \rho (\alpha-1)$ in zCDP is replaced just by $\leq \epsilon$ in RDP.

As zCDP, RDP extends the classical notion of differential privacy (DP) by leveraging the Rényi divergence, a generalization of the Kullback-Leibler (KL) divergence \cite{ams1951kl} that ranges from the Kullback-Leibler (KL) divergence itself when $\alpha \to 1$ to the max-divergence at $\alpha \to \infty$. From an information-theoretic perspective, RDP quantifies the privacy leakage by bounding the divergence between perturbed probability measures, thereby providing a principled and mathematically rigorous foundation for privacy-preserving algorithms in modern machine learning and statistical inference \cite{mironov2017rdp}.

RDP offers superior composition properties, which allow for a tighter and more analytically tractable characterization of the cumulative privacy loss in iterative processes such as differentially-private stochastic gradient descent (DP-SGD). 

RDP provides a more flexible framework for privacy accounting, particularly in high-dimensional settings where privacy amplification via subsampling plays a fundamental role. 

Furthermore, RDP retains key invariance properties under post-processing and can be converted to $(\epsilon, \delta)$-DP guarantees, thereby establishing a bridge between privacy notions that are crucial in practical deployments. Given an $(\alpha, \epsilon_{RDP})$-RDP mechanism, we can derive an equivalent $(\epsilon_{DP}, \delta)$-DP mechanism using the following relationship \cite{mironov2017rdp}:
$$
\epsilon_{DP} = \epsilon_{RDP} + \frac{\log(1/\delta)}{\alpha-1}
$$
where $\alpha$ is the Rényi divergence order, $\epsilon_{RDP}$ is the RDP privacy parameter, $\delta$ is the failure probability parameter in DP (indicating rare privacy breaches), and $\epsilon_{DP}$ is the traditional DP privacy loss parameter. The stricter the privacy requirement (lower $\delta$), the greater the privacy loss we account for in DP (higher $\epsilon_{DP}$). Larger Rényi divergence orders allow for better privacy composition over multiple steps, tightening DP bounds.

When a mechanism is $(\alpha, \epsilon_{RDP})$-RDP, it is also $(\epsilon_{DP}, \delta)$-DP for any $\epsilon_{DP} > \epsilon_{RDP}$ and $\delta = e^{-(\alpha-1)(\epsilon_{DP}-\epsilon_{RDP})}$. This relationship is used, for instance, in the moments accountant (MA) \cite{abadi2016dpsgd}, a privacy analysis technique for ML algorithms. However, the approximation is loose and does not hold for all $\epsilon_{DP} \ge 0$.

Given an $(\alpha, \epsilon_{RDP})$-RDP mechanism, what are the smallest $\epsilon_{DP}$ and $\delta$ so that the mechanism is also $(\epsilon_{DP}, \delta)$-DP? \cite{jsait2021variants} relates approximate DP to RDP based on the joint range of two f -divergences that underlie the approximate DP and RDP. This enables us to derive the optimal approximate DP parameters of a mechanism that satisfies a given level of RDP. An $(\alpha, \epsilon_{RDP})$-RDP mechanism provides $(0, \delta)$-DP if $\epsilon_{RDP} < \log \frac{\alpha}{\alpha-1}$ and $\delta \in [\zeta_\alpha e^{(\alpha-1)\epsilon_{RDP}}, \frac{1}{\alpha}]$, where $\zeta_\alpha = \frac{1}{\alpha} \left( 1 - \frac{1}{\alpha}\right)^{\alpha -1}$.

Given an $(\alpha, \epsilon_{RDP})$-RDP mechanism, what is the characterization of its privacy region? A relationship between RDP and the hypothesis test interpretation of DP allows us to translate the RDP constraint into a trade-off between type I and type II error probabilities of a certain binary hypothesis test \cite{jsait2021variants}.

Finally, it should also be noted that the Gaussian mechanism with variance $\sigma^2$ is $(\alpha, \epsilon_{RDP})$-RDP for $\epsilon_{RDP} = \rho \alpha$ with $\rho = \frac{1}{2\sigma^2}$. Given that, for Gaussian noise, the Rényi divergence of order $\alpha$ is $D_\alpha = \alpha / 2\sigma^2$ and RDP requires the Renyi divergence to be bounded by $\epsilon_{RDP} = \rho \alpha$, $\rho = \frac{1}{2\sigma^2}$.

\subsubsection{Gaussian Differential Privacy (GDP)}

Instead of using the Rényi divergence or the standard DP framework, GDP specifically models privacy loss as a Gaussian distribution \cite{dong2022gdp}. This allows privacy guarantees to be expressed using a trade-off function, ensuring optimal trade-offs between privacy and utility. While GDP has strong theoretical properties, CDP provides a more flexible framework and is applicable to a broader range of algorithms.

Given a mechanism $\mathcal{M}$ and two neighboring datasets $x$ and $x'$, the trade-off function $T(\mathcal{M}(x), \mathcal{M}(x'))$ describes how well an adversary can distinguish between them. 

In GDP (and its $f$-DP generalization), the trade-off function fixes the type I error at any level $\alpha$, and considers the minimal achievable type II error. 
Mathematically, the trade-off function is $T(P,Q)(\alpha) = \inf_\phi \{ \beta_\phi : \alpha_\phi \leq \alpha \}$, where the infimum is taken over all (measurable) rejection rules that take as input the released results of the mechanism, with their type I and type II errors defined as $\alpha_\phi = E_P[\phi]$ and $\beta_\phi = 1-E_Q[\phi]$, where $P$ and $Q$ represent the probability distributions over the outputs of the neighboring datasets, $\mathcal{M}(x)$ and $\mathcal{M}(x')$.

A trade-off function is always convex, continuous, non-increasing, and $f(x) \leq 1-x$ for $x \in [0,1]$. 
Swapping $P$ and $Q$ results in a reflected trade-off function $T(Q,P)(\alpha) = 1 - T(P,Q)(1-\alpha)$, which ensures that the privacy analysis accounts for adversaries testing either dataset as the true distribution.

A randomized mechanism $\mathcal{M}$ satisfies $\mu$-GDP if, for all neighboring datasets $x$ and $x'$ differing by a single element, the trade-off function $T(\mathcal{M}(x), \mathcal{M}(x'))$ is bounded by the trade-off between two Gaussian distributions, $N(0,1)$ and $N(\mu,1)$:
$$
T(\mathcal{M}(x), \mathcal{M}(x')) \geq T(N(0,1), N(\mu,1))
$$
where the mean parameter $\mu$ represents the distance between the two Gaussian distributions, which acts as a measure of privacy loss. The Gaussian trade-off function $G_\mu(\alpha) := T(N(0,1), N(\mu,1)) = \Phi ( \Phi^{-1}(1-\alpha)-\mu)$ is decreasing in $\mu$, in the sense that $G_\mu \leq G_{\mu'}$ if $\mu \geq \mu'$.

Gaussian Differential Privacy (GDP) is articulated as a specialized instance of $f$-differential privacy ($f$-DP), a generalization of differential privacy that frames privacy as a hypothesis testing problem. Given a trade-off function $f$, a mechanism $\mathcal{M}$ is said to be $f$-differentially private  if $T(\mathcal{M}(x), \mathcal{M}(x')) \geq f$ for all neighboring datasets $x$ and $x'$.

GDP characterizes privacy in terms of the best achievable trade-off between Type I and Type II errors.
For the Gaussian mechanism, this trade-off is bounded by a Gaussian trade-off function with parameter $\mu = \Delta f/\sigma$, where $\Delta f$ is the maximum possible change in the function output when one data point is modified and $\sigma$ determines how much noise is added (adding Gaussian noise inherently satisfies $\mu$-GDP).


GDP aligns privacy loss with the normal distribution, making composition analysis much smoother in large-scale applications, the key advantage of GDP. Privacy loss accumulates like a sum of Gaussian variables, following the central limit theorem: when applying multiple privacy-preserving steps, the cumulative privacy loss follows a Gaussian distribution. 


The GDP guarantee that privacy loss follows a normal (Gaussian) distribution with mean $\mu$ can be converted into the traditional DP framework by relating the GDP privacy parameter $\mu$ to the $(\epsilon,\delta)$-DP parameters. GDP can be converted to $(\epsilon,\delta)$-DP using the following relationship:
$$
\delta(\epsilon) = \Phi \left( -\frac{\epsilon}{\mu} + \frac{\mu}{2} \right)
  - e^\epsilon \cdot \Phi \left( -\frac{\epsilon}{\mu} - \frac{\mu}{2} \right)
$$
where $\Phi$ is the standard normal CDF. This relationship gives a tight bound on $\delta$ for given $\epsilon$ and $\mu$. You can also numerically invert it to find $\epsilon$ for a given $\delta$. For any desired $\delta$, the corresponding $\epsilon$ is given by $\epsilon = \mu + \sqrt{2 \log (1/\delta)}$. 

For stronger privacy, lower $\delta$ means fewer violations allowed and leads to higher effective $\epsilon$. Higher $\mu$ means that the privacy loss is greater and leads to weaker DP guarantees (higher $\epsilon$). The privacy parameter in GDP plays a similar role to $\epsilon$ in DP, representing how much information an adversary can gain.

The GDP formulation marks a profound shift in differential privacy \cite{dwork2006dp} by reinterpreting privacy through the lens of trade-off functions, offering a more granular and exact alternative to traditional metrics like R\'enyi divergence \cite{tit2013renyi-kl} or the $(\epsilon, \delta)$-DP bounds, which often rely on loose approximations under composition. Parameterization by $\mu$ encapsulates privacy as the effective separation in a Gaussian distinguishability test, providing an intuitive yet mathematically precise measure that simplifies analysis across diverse applications. Unlike prior frameworks, GDP reliance on the Gaussian trade-off function yields closed-form expressions for composition, eliminating the need for numerical approximations or conservative bounds. Its asymptotic convergence property, which is rooted in the central limit theorem, positions it as a universal endpoint for privacy loss accumulation. 



GDP is particularly useful in iterative algorithms, where multiple computations are performed sequentially, such as in stochastic optimization, ML model training, and data analysis workflows. The key advantage of GDP in these scenarios is its ability to provide accurate privacy composition over many iterations, while maintaining strong privacy guarantees. For instance, in stochastic gradient descent \cite{robbins1951stochasticApproximation}, tight privacy accounting is critical. 


GDP is also useful in subsampling scenarios, where a mechanism operates on a randomly chosen subset of a dataset rather than the full dataset. Subsampling improves privacy by reducing adversaries' ability to infer individual participation. 
In traditional DP, subsampling amplifies privacy guarantees, meaning the effective $(\epsilon, \delta)$-DP parameters improve.
When a mechanism selects a random subset of data instead of the full dataset, the sensitivity of individual data points decreases. 
GDP naturally captures this privacy amplification effect through its Gaussian trade-off function.
Given a mechanism that satisfies $\mu$-GDP, subsampling reduces the effective privacy loss according to $\mu_{subsampled} = q \cdot \mu$, where $q$ is the subsampling rate. When only a fraction $q$ of the data is used, the privacy guarantees scale proportionally, leading to better privacy protection.
In DP-SGD \cite{abadi2016dpsgd}, which subsamples mini-batches of data at each step, GDP provides a direct way to measure privacy loss over multiple subsampling rounds, making it ideal for deep learning applications.

\subsubsection{Local differential privacy}


Differential privacy can be described as a constraint on the algorithms used to publish aggregate information about a statistical database, which limits the disclosure of private information.
An algorithm is differentially private if an observer checking its output cannot tell whether a particular individual's information was used in the computation \cite{dwork2017dp}. 
In local differential privacy \cite{duchi2018ldp}, true data is kept hidden even from the data scientist working on it, by applying differentially-private mechanisms to the data, before any statistics are computed. 

Differentially private algorithms work in two main modalities \cite{bassily2017local}. The curator model assumes a trusted centralized curator that collects all the information and analyzes it. In contrast, the local model enables privacy without a trusted curator,
making it suitable for distributed systems. Each piece of information is randomized by its provider to protect privacy even if all information provided to the analysis is revealed. A disadvantage of the local model is that it requires introducing noise at a significantly higher level than the curator model \cite{bassily2017local}.

In Centralized DP (CDP) 
privacy guarantees are applied at an aggregate level.
Local differential privacy (LDP) \cite{focs2008local} is designed for scenarios where each user perturbs her own data before sharing it, making it more robust against direct attacks.
A randomized algorithm $\mathcal {M}$ that takes user's private data as input is said to provide $\epsilon$-local differential privacy if, for all pairs of users' possible private data ($x$ and $x'$) and for all outputs $S$:
$$
\frac {P[{\mathcal {M}}(x)\in S]}{P[{\mathcal {M}}(x^{\prime })\in S]} \leq e^{\epsilon }
$$

The difference between the definition of local differential privacy and the definition of standard (global) differential privacy is that the local differentially private algorithm considers a single user's data, whereas the standard definition of differential privacy considers pairs of neighboring datasets. A randomized response mechanism (e.g., flipping a bit with probability dependent on $\epsilon$) satisfies $\epsilon$-DP in the local model, where users privatize their data before sharing.

The randomized response survey technique proposed by Stanley L. Warner in 1965 is frequently cited as an example of local differential privacy \cite{warner1965randomized}. Warner's innovation was the introduction of what could now be called the ``untrusted curator'' model, where the entity collecting the data may not be trustworthy. Before users' responses are sent to the curator, the answers are randomized in a controlled manner, guaranteeing differential privacy while still allowing valid population-wide statistical inferences. In 2003, \cite{pods2003local} gave a definition equivalent to local differential privacy. In 2008, \cite{focs2008local} first used the term `local private learning' and showed it to be equivalent to randomized responses.

In LDP, noise is added before data collection (at the user level) instead of after collection. In traditional DP approaches, trust is required in the data processor, whereas no trust is needed in the data processor when using LDP, since individual data is protected at its source. This makes LDP suitable for decentralized data collection (e.g. Google's Chrome privacy statistics or Apple's data anonymization) at the cost of lower accuracy, since noise is added early (in contrast to traditional DP approaches, where noise is added once raw data is processed).

\subsection{Formal properties of differential privacy mechanisms}

The relaxation in $(\epsilon, \delta)$-differential privacy provides additional flexibility for designing differentially-private mechanisms by enabling a trade-off between privacy and utility. In particular, mechanisms that satisfy \( (\epsilon, \delta) \)-differential privacy, including zCDP, RDP, f-DP, and GDP, still inherit key properties of pure differential privacy, including composability, robustness to post-processing, and graceful degradation in the presence of correlated data, which ensure that privacy guarantees are preserved under transformations and repeated interactions with private data \cite{dwork2014tcs}.

\subsubsection{Composability} 

Composability refers to the fact that the joint distribution of the outputs of (possibly adaptively chosen) differentially-private mechanisms satisfies differential privacy. 

In sequential composition, if we query an $\epsilon$-differential privacy mechanism $k$ times, and the randomization of the mechanism is independent for each query, then the result would be $k \epsilon$-differentially private. 
In general, with $k$ different $\epsilon_i$-differential privacy mechanisms, any function of them is $( \sum_{i=1}^{k} \epsilon _{i} )$-differentially private \cite{mcsherry2009sigmod}.
When the mechanisms are computed on disjoint subsets of the private database, in parallel composition, then the function would be $\max_{i} \epsilon _{i}$-differentially private instead \cite{mcsherry2009sigmod}.

A common reason for adopting $(\epsilon, \delta)$-differential privacy arises in the context of $k$-fold adaptive composition \cite{dwork2010boosting}, a differentially private analogue of the ``left or right'' notion of security for encryption schemes \cite{bellare1998crypto}. In k-fold composition, a process is repeated multiple times (k times), each time using a different slice of the data. The process is adaptive in the sense that each time we ask a question, we might change what we ask based on the answers we got before. An $(\epsilon, \delta)$-differentially private mechanism ensures that the composite mechanism satisfies $(\epsilon', \delta')$-differential privacy, where $\epsilon' = \sqrt{2k \ln(1/\delta')} + k\epsilon(e^\epsilon - 1)$ and $\delta' = k\delta + \delta_0$ for some $\delta_0 > 0$ \cite{dwork2010boosting}. This theoretical result highlights the feasibility of producing accurate query responses across extensive query sets while preserving individual privacy through adaptive boosting techniques.

\begin{table}[t]
    \centering
    \begin{tabular}{lcl}
         \hline
         DP mechanism & Composition bound & Growth rate \\ 
         \hline
        $\epsilon$-DP  & $k \cdot \epsilon$ & Linear \\ 
        $(\epsilon,\delta)$-DP 
            & $\sqrt{2k\log(1/\delta')} \cdot \epsilon + k \cdot \epsilon (e^\epsilon -1) $ & Sublinear $\epsilon$, linear $\delta$ \\
        $(\mu,\tau^2)$-mCDP 
            & $ \mu_{total} = k \cdot \mu \quad \tau^2_{total} = k \cdot \tau^2$ & Linear \\
        $\rho$-zCDP 
            & $ k \cdot \rho + 2 \sqrt{k \cdot \rho \cdot\ln (1/\delta)} $ & Linear \\
        $(\alpha,\epsilon)$-RDP 
            & $ k \cdot \epsilon$ & Linear \\
        $\mu$-GDP 
            & $ \mu_{total} = \sqrt{k} \cdot \mu $ & Sublinear \\
        $\epsilon$-LDP
            & $k \cdot \epsilon$ & Linear \\ 
        \hline
            & & \\
    \end{tabular}
    \caption{Sequential composition privacy bounds for different DP frameworks: How privacy degrades when multiple DP mechanisms are applied sequentially on the same dataset.}
    \label{tab:composition}
\end{table}

If a mechanism satisfies $(\epsilon,\delta)$-DP, then, after $k$ sequential queries, the total privacy parameters are bounded as follows: $\epsilon_{total} = \sqrt{k} \epsilon$ and $\delta_{total} = k \delta$. $\epsilon$ grows sublinearly ($\sqrt{k}$ scaling), meaning privacy loss accumulates more slowly than in pure DP, while $\delta$ grows linearly, meaning the probability of privacy failure increases proportionally with the number of queries.

The concentrated DP (CDP) sub-Gaussian bound enables stronger composition properties, meaning that when multiple privacy-preserving mechanisms are applied successively, the accumulated privacy loss grows in a well-controlled manner. Since a sub-Gaussian distribution has tight concentration around its mean, the tail probabilities (i.e., the chance of large privacy loss) drop off exponentially. Even if you apply many mechanisms, the expected privacy loss grows slowly and predictably. Since privacy risks do not escalate uncontrollably, CDP provides a more efficient and scalable approach compared to traditional DP, making it particularly useful in repeated queries.

The mean-concentrated DP (mCDP) formulation departs from the stringent $\epsilon$-DP guarantee of pure differential privacy by emphasizing the concentration of privacy loss rather than its worst-case extremum. Under \( k \)-fold composition, the privacy loss accumulates as \( O(\sqrt{k} \cdot \mu + k \cdot \sigma) \), a sublinear improvement over the linear \( k \cdot \epsilon \) growth of the classical model. This enables more efficient noise calibration, enhancing data utility in iterative or adaptive analyses while preserving robust privacy guarantees.

The zero-concentrated DP (zCDP) has a more elegant privacy composition. If a mechanism satisfies $\rho$-zCDP with privacy parameter $\rho$, then after k sequential queries, the total privacy parameter scales additively: $\rho_{total} = k \cdot \rho$. When converted into $\epsilon$-DP, $ \epsilon_{total} = k \cdot \rho + 2 \sqrt{k \cdot \rho \cdot\ln (1/\delta)} $. While the total $\epsilon$ does grow with $k$, the dominant privacy risk (the part that matters most in practice) grows much more slowly than in pure DP, where $\epsilon$ grows strictly linearly, whereas the second term in zCDP scales with $\sqrt{k}$. If you fix $\epsilon$, $\epsilon$ grows roughly with $\sqrt{k}$ and $k$ grows like $\epsilon^2$. Given its quadratic accumulation of privacy loss, the number of queries you can make scales quadratically with the privacy budget. In other words, zCDP allows many more queries before hitting the same privacy budget as in traditional DP. In traditional DP, you get linear scaling (double the queries, double the privacy loss). In zCDP, you get quadratic scaling (double the privacy budget, four times the number of queries). 

In $(\alpha, \epsilon)$-RDP, after $k$ sequential queries, the total privacy parameter also scales linearly $\epsilon_{total} = k \epsilon$. One of the key strengths of RDP is its tight and clean composition rule: if you compose $k$ $(\alpha, \epsilon)$-RDP mechanisms, the composition satisfies $(\alpha, k \cdot \epsilon)$-RDP. It is still linear in $\epsilon$ but, crucially, it is independent of $\delta$, unlike approximate $(\epsilon,\delta)$-DP.

In $\mu$-GDP, privacy loss accumulates sublinearly, $\mu_{total} = \sqrt{k} \mu$. The probability of large privacy loss decays like a Gaussian tail, providing a tight and symmetric privacy guarantee. GDP provides stronger composition guarantees (privacy loss accumulates more smoothly over multiple queries) since traditional DP relies on worst-case bounds, which can lead to more extreme privacy loss fluctuations (its composition bounds can be looser, making privacy harder to manage over multiple queries).

In summary, pure DP, local DP, zCDP, RDP, and mCDP all show linear growth in privacy loss. Approximate DP grows faster due to the additional exponential term in its advanced composition. f-DP composition is not discussed here because its composition depends on the specific f-divergence used and often requires numerical methods. In practice, Gaussian DP and zCDP grow sub-linearly in terms of $\epsilon$-equivalent privacy loss ($\sqrt{k}$ scaling), making them much more efficient for many compositions. This property makes them valuable in iterative processes, including training deep learning models.

Each framework provides different trade-offs between privacy strength and composition efficiency. If you're working with iterative algorithms, zCDP and GDP tend to offer better composition guarantees than traditional DP.\footnote{In more realistic scenarios, the concurrent composition properties of interactive mechanisms, whereby an adversary can arbitrarily interleave its queries to the different DP mechanisms, have also been studied \cite{tcc2021concurrent} \cite{neurips2022concurrent} \cite{stoc2023concurrent}. In fact, computing the optimal bound for composing k arbitrary DP algorithms is, in general, \#P-complete \cite{tcc2015composition}.}


\subsubsection{Robustness to post-processing} 

The robustness to post-processing is a fundamental property of differential privacy (DP) frameworks. It ensures that, once data has been privatized, any further computation or transformation cannot weaken the privacy guarantee. In fact, robustness to post-processing is a key  property for the modular use of differential privacy. While post-processing could affect fairness or utility \cite{zhu2022postprocessing}, it should never compromise the privacy guarantees.

This robustness to post-processing property is defined as follows: for any deterministic or randomized function $F$ defined over the image of the mechanism $\mathcal{M}$, if $\mathcal{M}$ satisfies $\epsilon$-differential privacy, so does $F(\mathcal{M})$ \cite{dwork2006dp}.

Pure DP is robust to post-processing because the definition of $\epsilon$-DP is preserved under any data-independent transformation. Approximate DP is robust to post-processing because the $(\epsilon,\delta)$ bounds remain valid even after arbitrary post-processing (including randomized post-processing, as long as it does not access the original data). zCDP and RDP are defined via a Rényi divergence, which is non-increasing under post-processing, so that any function applied to the output of a zCDP mechanism preserves its $\rho$ bound in zCDP and you can safely apply further analysis or transformations without increasing privacy loss in RCP. Like zCDP, mCDP relies on properties of the privacy loss random variable (its moments), which are preserved under post-processing. GDP is based on the privacy loss being sub-Gaussian, and this property is preserved under post-processing. In general, f-divergences are contractive under post-processing (i.e., they do not increase), so the f-DP guarantee remains intact no matter how the output is used. Finally, in LDP, each user perturbs their data before sharing them, so any transformation after that does not affect the privacy guarantee.

Composition permits modular construction and analysis of differentially private mechanisms and motivates the concept of privacy loss budget. When all elements of a complex mechanism that access sensitive data are separately differentially private, so will be their combination, followed by arbitrary post-processing. The DP guarantee is composable and repeating invocations of differentially private algorithms lead to a degradation of privacy whose rate depends on the adopted composition framework, as described above.

\subsubsection{Graceful degradation in the presence of correlated data}

Differential privacy guarantees can be weak in some real-world scenarios. A no-free-lunch theorem \cite{kifer2011sigmod} shows that it is not possible to provide privacy and utility without making assumptions about how the data are generated. The privacy of an individual is preserved when it is possible to limit the attacker's inferences about the participation of the individual in the data generating process. 

Group privacy extends privacy guarantees to protect small groups, not just individuals. Obviously, an $\epsilon$-DP mechanism satisfies $k \epsilon$-DP for groups of size $k$, i.e. datasets differing in $k$ records \cite{dwork2006dp} \cite[Section 2.3]{dwork2014tcs}.

The concept of graceful degradation in differential privacy refers to how well a privacy framework maintains its guarantees when the assumption of independent data entries is violated, i.e., when data is correlated.

Different DP frameworks handle such situations with disparate degrees of success:

\begin{itemize}

\item 
In pure DP, $\epsilon$-DP, the privacy guarantee holds regardless of correlations, because it is defined over neighboring datasets.
However, when individuals' data are highly correlated (e.g., family members), removing one individual may still reveal information about others. So the guarantee is still valid, but less meaningful in practice.

\item 
Approximate DP, $(\epsilon,\delta)$-DP, is similar to pure DP. The core guarantee still applies to neighboring datasets and the $\delta$ term allows for a small probability of failure. However, correlations can make the $\delta$ failure events more likely to leak meaningful information.

\item 
Mean-concentrated DP, mCDP, tracks the mean and variance of privacy loss, which can be skewed by correlations. Therefore, the guarantees degrade more gracefully than in pure DP, but still suffer under strong dependencies.

\item
Zero-concentrated DP, zCDP, assumes a concentration of privacy loss, which may not hold under strong correlations. The theoretical guarantees remain, but the interpretation of $\rho$ becomes less tight.

\item 
Rényi DP, RDP, like zCDP, relies on divergence measures that assume independence for tightness. The bound still holds, but actual leakage may be higher than expected.

\item
Gaussian DP, GDP, assumes the privacy loss is sub-Gaussian, which may not hold under correlation. In other words, the Gaussian tail bounds may underestimate the true risk. In general, the degradation of f-DP mechanisms depend on the specific divergence used, since some f-divergences are more sensitive to correlations than others.

\item
Local DP, LDP, is inherently robust to correlations.
Since each user perturbs their data before sharing, correlations in the raw data do not affect the privacy guarantees.

\end{itemize}

In practice, the use of differential privacy can lead to privacy breaches. Differential privacy does not always adequately limit inference about participation in the data generating process. Data correlation, i.e., the probabilistic dependence relationship among examples in a database, can seriously deteriorate the privacy properties of DP, as demonstrated by Bayesian attacks on differentially private mechanisms \cite{liu2016ndss}. 

DP frameworks might underestimate the privacy risk and, therefore, the amount of noise that must be added to achieve a desired bound on the adversary's ability to make sensitive inferences. When correlation (or dependency) among data examples would break the desired privacy guarantees, dependent differential privacy (DDP) \cite{liu2016ndss} explicitly incorporates probabilistic dependency constraints among tuples:
$$
    P[\mathcal{M}(D(L,R)) \in S] \leq e^{\epsilon} P[\mathcal{M}(D'(L,R)) \in S]
$$
where $D$ and $D'$ are neighboring databases, when the modification of an example in database $D(L,R)$ causes change in at most $L-1$ examples in $D'(L,R)$ due to the probabilistic dependence $R$ between examples. $L$ is the dependence size, i.e. when any example in $D$ is dependent on at most $L-1$ other examples, whereas the relationship $R$ might be due to the data generating process \cite{kifer2011sigmod}.


In order to handle correlation using differential privacy, \cite{chen2014correlated} just multiplied the sensitivity of the query output by the number of correlated records. This naive approach to group privacy protects databases differing in $n$ examples, by setting $\epsilon$ to $\epsilon/n$: instead of having each example $\epsilon$-differentially private protected, now every group of $n$ examples is $\epsilon$-differentially private protected (and each example is $\epsilon/n$-differentially private protected).
However, excessive noise must then be added to the output, severely degrading the utility of the model. DDP provides a finer dependent perturbation mechanism (DPM) to achieve the desired privacy guarantees \cite{liu2016ndss}.  

The Pufferfish framework \cite{kifer2012pods} \cite{song2017sigmod} generalizes DP by incorporating adversarial beliefs about existing data relationships using a data generation model.
This provides rigorous privacy guarantees against adversaries who may have access to any auxiliary background information and side information on the database. Both DDP \cite{liu2016ndss} and Blowfish \cite{he2014blowfish}, which considers correlations due to deterministic constraints, are subclasses of the Pufferfish framework.



In common DP frameworks, privacy guarantees typically depend on the subject's ability to control their data. However, in some situations, privacy must persist even when the subject no longer has control over whether their data is revealed. For instance, we might be interested in keeping something a secret, even when the subject of the secret no longer has the ability to keep it a secret (i.e. withholding their data does not maintain the secret). This requires stronger guarantees than those offered by DP frameworks, which do not protect from public leaks or external correlation attacks. Computational differential privacy \cite{crypto2009cdp} frameworks go beyond standard DP by considering adversaries with computational constraints, meaning privacy can be maintained even if adversaries gain partial knowledge. Information-theoretic privacy \cite{allerton2014itp}\cite{jsait2021itsp} frameworks provide privacy guarantees that hold independent of any adversary’s computational power: even if all the subject’s data is exposed, critical secrets remain secure under predefined conditions.\footnote{Zero-Knowledge Proofs (ZKPs) \cite{stoc1985zkp}\cite{sjc1989zkp} are cryptographic techniques that allow the verification of a statement without revealing the underlying data. While not strictly a DP framework, ZKPs could contribute to situations where someone needs to prove a claim without exposing private information. Privacy-preserving cryptographic methods, such as secure multiparty computation (MPC) \cite{yao1982psc}\cite{stoc1987smc} and homomorphic encryption \cite{rivest1979data}\cite{stoc2009fhe}, help in ensuring privacy even when data is processed externally, reducing reliance on a subject’s ability to control their own data.} 
The real challenge arises when a secret becomes exposed due to external factors, meaning a stronger notion of ``future-proof'' privacy is required. Some privacy-preserving frameworks such as ``indistinguishability obfuscation'' \cite{crypto2001io}\cite{stoc2021io} or ``traceable anonymization'' \cite{sciadv2024anonymization}\cite{ijerph2021healthdata}\cite{sac2019decouples} attempt to achieve persistent secrecy even under unexpected conditions.


%% file: diagram_taxonomia_dp.tex
 \begin{tikzpicture}[
            node distance=0.8cm and 0.15cm,
            font=\sffamily\small,
            base/.style={rectangle, draw=black!70, thick, align=center, rounded corners=2pt, fill=white},
            header/.style={base, fill=gray!20, font=\sffamily\bfseries, minimum width=18cm, inner sep=8pt},
            dim/.style={base, fill=gray!5, font=\sffamily\scriptsize\bfseries, text width=2.5cm, minimum height=1.2cm},
            family/.style={base, fill=gray!15, font=\sffamily\bfseries, minimum width=5cm, inner sep=6pt},
            variant/.style={base, align=left, text width=6cm, inner sep=5pt},
            highlight/.style={draw=blue!80, line width=1.5pt, fill=blue!5},
            arrow/.style={-{Latex[length=3mm]}, thick},
            bluearrow/.style={-{Latex[length=3mm]}, blue!80, line width=1.5pt},
            groupbox/.style={draw=gray!50, dashed, fill=gray!5, rounded corners=5pt, inner sep=10pt}
        ]

        \node[header] (top) {CONCEPTUAL DIMENSIONS IN DIFFERENTIAL PRIVACY};
        
        \node[dim, below=0.6cm of top] (d4) {BACKGROUND\\KNOWLEDGE\\ASSUMPTIONS};
        \node[dim, left=of d4] (d3) {VARIATION OF\\PRIVACY LOSS\\ACROSS INPUTS};
        \node[dim, left=of d3] (d2) {NEIGHBORHOOD\\DEFINITION};
        \node[dim, left=of d2] (d1) {PRIVACY LOSS\\QUANTIFICATION};
        \node[dim, right=of d4] (d5) {FORMALISM\\CHANGE};
        \node[dim, right=of d5] (d6) {RELATIVIZATION\\OF KNOWLEDGE\\GAIN};
        \node[dim, right=of d6] (d7) {COMPUTATIONAL\\POWER};

        \foreach \i in {1,...,7} \draw[arrow] (top.south -| d\i.north) -- (d\i.north);

        \node[header, below=1.2cm of d4] (mid) {DIFFERENTIAL PRIVACY VARIANTS AND EXTENSIONS};
        \draw[-{Latex[length=4mm]}, line width=6pt, draw=blue!50!black, shorten <=3pt, shorten >=3pt] (d4.south) -- (mid.north);

        \node[family, below=0.8cm of mid.south west, anchor=north west, xshift=0.5cm] (f1) {ACCOUNTING \& COMPOSITION};
        
        \node[variant, highlight, below=0.8cm of f1, text width=5.3cm, align=center, anchor=north] (v1a) {Concentrated DP (CDP)};
        \node[variant, highlight, below=0.1cm of v1a, text width=5.3cm, align=center, anchor=north] (v1b_z) {Zero-Concentrated DP (zCDP)};
        \node[variant, highlight, below=0.1cm of v1b_z, text width=5.3cm, align=center, anchor=north] (v1b) {Rényi Differential Privacy (RDP)};
        \node[variant, highlight, below=0.1cm of v1b, text width=5.3cm, align=center, anchor=north] (v1c) {Gaussian Differential Privacy (GDP)};
        \node[variant, below=0.1cm of v1c, text width=5.3cm, align=center, anchor=north] (v1d) {f-Differential Privacy (f-DP)};
        \node[variant, below=0.1cm of v1d, text width=5.3cm, align=center, anchor=north] (v1e) {Kullback-Leibler Privacy (KLP)};

        \node[family, right=0.8cm of f1] (f2) {ALTERNATIVE THREAT MODELS};
        \node[variant, below=0.8cm of f2, text width=3cm, align=center, anchor=north] (v2a) {Blowfish Privacy};
        \node[variant, below=0.1cm of v2a, text width=3cm, align=center, anchor=north] (v2b) {Dependent DP};
        \node[variant, below=0.1cm of v2b, text width=3cm, align=center, anchor=north] (v2c) {One-Sided DP};
        \node[variant, below=0.1cm of v2c, text width=3cm, align=center, anchor=north] (v2d) {Individual DP};
        \node[variant, below=0.1cm of v2d, text width=3cm, align=center, anchor=north] (v2e) {Generic DP};

        \node[family, right=0.8cm of f2] (f3) {DEPLOYMENT MODELS};
        \node[variant, highlight, below=0.8cm of f3, text width=3cm, align=center, anchor=north] (v3a) {Local DP (LDP)};

        \node[groupbox, fit=(v1a) (v1e), fill opacity=0.2, text opacity=1] (box1) {};
        \node[groupbox, fit=(v2a) (v2e), fill opacity=0.2, text opacity=1] (box2) {};

        \draw[arrow] (mid.south -| f1.north) -- (f1.north);
        \draw[arrow] (mid.south -| f2.north) -- (f2.north);
        \draw[arrow] (mid.south -| f3.north) -- (f3.north);

        \draw[arrow] (f1.south) -- (box1.north);
        \draw[arrow] (f2.south) -- (box2.north);
        \draw[arrow] (f3.south) -- (v3a.north);

        \node[base, below=2.5cm of v2d, font=\bfseries, inner sep=8pt] (leg) {KEY VARIANTS IN MACHINE LEARNING APPLICATIONS};
        
        \draw[arrow] (leg.west) -- ++(-5.4, 0) |- (v1a.west);
        \draw[arrow] (leg.west) -- ++(-5.4, 0) |- (v1b.west);
        \draw[arrow] (leg.west) -- ++(-5.4, 0) |- (v1b_z.west);
        \draw[arrow] (leg.west) -- ++(-5.4, 0) |- (v1c.west);
        \draw[arrow] (leg.east) -| (v3a.south);

\end{tikzpicture}

%% file: dp-in-ml.tex
\section{Differential privacy in Machine Learning models}\label{chapter:dpml}

Differential privacy (DP) lets us release statistical information about datasets while protecting the privacy of individual data examples. Therefore, its use in Machine Learning enables the use of aggregate patterns in the data while limiting the amount of information that is leaked about specific individuals \cite{dwork2008survey}. By injecting carefully calibrated noise into statistical computations, the utility of the statistic is preserved while provably limiting what can be inferred about any individual in the dataset.


The incorporation of differential privacy into Machine Learning models relies on two key properties:

\begin{itemize}

\item
The composition properties of DP mechanisms enable the concept of a privacy budget: a total allowable privacy loss that can be allocated and spent across various computations performed on a dataset. Careful management and accounting of this budget using appropriate composition rules are essential for practical DP applications.

\item 
A cornerstone property of differential privacy is its robustness to post-processing.
Once data has been released via a DP mechanism, analysts can perform any kind of subsequent analysis on the released data without weakening the original privacy guarantee. This makes DP highly modular and practical, as the privacy analysis only needs to focus on the initial mechanism interacting with the sensitive data.

\end{itemize}

Several approaches can be used to train a privacy-preserving ML model \cite{ji2014dpml}. Several general strategies have been developed, with specific adaptations required for different classes of ML algorithms:

\begin{itemize}

\item
{\it Output perturbation:} First, learn a model on clean data, i.e., train a ML model $M$ using standard, non-private methods on the sensitive dataset to obtain its parameters $\theta$. Then, use a DP mechanism (e.g. exponential or Laplacian) to generate a noisy model ($\theta_{DP} = \theta + noise$), i.e. noise is added directly to the trained model parameters before releasing $M_{\theta_{DP}}$ ($\hat{y} = M_{\theta_{DP}}(x)$). Alternatively, noise can be added to the predictions made by the non-private model ($\hat{y} = M_{\theta}(x)+noise$). For approaches that have many iterations or multiple steps, the DP mechanism can be applied to the output parameters of each iteration/step. While conceptually simple, calculating the sensitivity of the entire training process for complex models such as deep neural networks is often intractable, leading to overly large noise and poor utility. This method is more feasible for simpler models with easily bounded sensitivity.

\item
{\it Objective perturbation:} Instead of perturbing the output, this method perturbs the objective function (e.g., loss function plus regularization term) that the ML algorithm seeks to minimize. The algorithm then finds the parameters $\theta_{DP}$ that minimize this noisy objective: $\theta_{DP} = \arg \min_{\theta} (L(\theta)+noise)$. This approach is commonly used for convex optimization problems found in algorithms such as SVM and logistic regression, as the sensitivity of the objective function can often be bounded more easily than the sensitivity of the final parameters.

\item
{\it Training mechanism perturbation:} Within iterative optimization algorithms like Stochastic Gradient Descent (SGD), this strategy is known as gradient perturbation. In each iteration, the gradient of the loss function is computed, perturbed with noise, and then used to update the model parameters. This is the dominant approach for training deep neural networks with DP, primarily through the DP-SGD algorithm \cite{abadi2016dpsgd}. It allows privacy to be incorporated incrementally throughout training.

\item
{\it Sample and aggregate:} First, split the dataset into many small subsets. Next, combine the results from all subsets to estimate a model, adding noise in this aggregation step. Typically, models are trained on disjoint data subsets and the training results (either model predictions or model parameters) are privately aggregated. In PATE (Private Aggregation of Teacher Ensembles) \cite{iclr2017pate}, an ensemble of teacher models are trained on disjoint private data partitions to provide noisy, aggregated labels for a public dataset, which is then used to train a final student model.

\end{itemize}

In ML, input perturbation is uncommon, even though it can be used successfully in signal processing, because noise addition at the input may incur too much perturbation for learning to be possible \cite{sp2013dpml}. 

The choice of strategy depends significantly on the particular ML algorithm to be used. Below we describe how DP has been incorporated into different ML techniques.



\subsection{Differential privacy in Symbolic AI} 
\label{sec:ai:symbolic}

Symbolic AI, Classical AI, or Good Old-Fashioned AI (GOFAI), uses explicit, human-readable rules, logic, and symbols to represent knowledge and solve problems. It was dominant AI approach from the 1950s to the 1990s.

Decision tree learning is the quintessential example of symbolic AI in machine learning and data mining. The most common strategy for learning decision trees from data is a greedy algorithm that, starting from the root of the decision tree, recursively partitions the training examples, a process known as top-down induction of decision trees (TDIDT) and used by ML algorithms such as ID3, C4.5, and CART. Alternatively, decision trees can also be built in a completely random way (RDT, random decision trees \cite{icdm2003rdt}).

When taking differential privacy into account, tree-based models require specialized handling of splits and leaves:

\begin{itemize}

\item 
PPID3 (Privacy-Preserving ID3 \cite{tkdd2008ppid3}) shows how to construct decision trees on vertically-partitioned data with an arbitrary number of parties where only one party has the class attribute, albeit not in a DP context and without an experimental study on its effectiveness in terms of privacy, utility, or their trade-off. 

\item
PPRDT (Privacy-Preserving Random Decision Trees \cite{tdsc2014rdpp}) leverages the fact that randomness in tree structure can provide strong privacy (better than simple perturbation of input/output) and requires significantly less time than alternative cryptographic approaches.

\item
SuLQ-based ID3 \cite{kdd2010tdidt} is a privacy-preserving ID3 that uses noisy counts (i.e., adds Laplace noise to the observed frequencies) and is based on the
SuLQ framework (Sub-Linear Queries \cite{dwork2005pods}), a predecessor of differential privacy. However, as the count estimates required to evaluate the splitting criterion must be carried out for each attribute separately, the privacy budget must be split among all those separate queries. Consequently, the budget per query is small, resulting in a wasteful use of the privacy budget.

\item
DiffPID3 \cite{kdd2010tdidt} provides a better approach: rather than evaluating each attribute separately, we can evaluate the attributes simultaneously in one query, the outcome of which is the attribute to use for splitting. 
A quality function $q$ provided to the exponential mechanism scores each attribute according to the splitting criterion. 

\item
In DiffPC4.5 \cite{kdd2010tdidt}, each time a partition reaches a pre-determined depth, or the number of samples in that partition is about the same scale as the random noise, or the sample space corresponding to that partition is too small, the mechanism stops operating on that partition. It then assigns to that partition a noisy count of samples for each label. After the partitioning process is complete, these noisy counts are used to decide whether to remove those nodes without having to consider privacy.

\item 
LPDT \cite{neurips2023tdidtldp} builds locally-private decision trees for non-parametric regression. On both synthetic and real-world data, LPDT exhibits superior performance when compared with other state-of-the-art LDP regression methods using private histograms \cite{ejs2021phist} \cite{arxiv2022phist} and de-convolution kernels \cite{farokhi2020deconvoluting}.

\end{itemize}

Combining multiple models, such as bagging, boosting, stacking, or meta-learning, is commonly used to improve the accuracy of single models in ML. Applying DP requires managing privacy across the ensemble:

\begin{itemize}

\item
{\it Ensembles of random decision trees} \cite{icdm2003rdt}: \cite{icdm2009dprdt} and \cite{tdp2012dprdt} use the Laplace mechanism to add noise to leaf frequencies and satisfy $\epsilon$-DP (smaller values of $\epsilon$ imply that more noise is added). 
\cite{eswa2017dpdf} proposes a query that outputs the most frequent label in some subset of the data with high probability (using the exponential mechanism) and uses this query in each leaf node for all random trees in a forest. 

\item
{\it Random forests} \cite{breiman2001random} are a popular method for building decision tree ensembles, based on the generation of bootstrap samples of training data and then, for each bootstrap sample, selecting a random subset of features to be used for each decision tree in the ensemble. Unfortunately, although ensemble approaches usually improve performance, they also weaken data privacy. 
A common strategy is to train each tree using a DP decision tree algorithm and divide the total privacy budget $\epsilon$ among the $T$ trees using sequential composition (i.e., each tree gets $\epsilon/T$). If trees are trained on disjoint subsets of the data, parallel composition can be used, allowing each tree to potentially use the full budget $\epsilon$, leading to better utility.
The variance-preserving framework in \cite{icdm2015dprf} is designed so that, when an adversary attempts to estimate a specific attribute of any individual data point, the variance of the estimation error crucially depends on the number of trees in the ensemble and, therefore, bounds can be derived on the maximum number of trees that are allowed in the ensemble while maintaining privacy. When the estimated variance may not be lower than the prior variance of the attributes, random forests may offer ``privacy for free.''

\item
{\it Gradient-boosted decision trees}: Applying DP to GBDT is challenging due to the sequential nature of training, since each tree corrects the errors of the previous ones. This creates dependencies and makes sensitivity analysis and budget allocation complex. The privacy budget accumulates additively across trees in the sequence and, therefore, naive application of noise can lead to significant accuracy loss, since the gradients (or residuals) used to train subsequent trees depend on previous trees and the data, making sensitivity hard to bound tightly. DPBoost \cite{aaai2020gbdt} employs a two-level
boosting structure called ``ensemble of ensembles'' (EoE) for privacy budget allocation, combining parallel composition within each ensemble (i.e. training with disjoint training subsets) and sequential composition across ensembles. DPBoost adaptively controls the gradients of training data for each iteration and clips leaf nodes to tighten the sensitivity bounds.
OpBoost \cite{arxiv2022opboost}\cite{vldb2022opboost} uses LDP for vertical federated learning, where users with non-overlapping attributes of the same data samples jointly train a model without directly sharing the raw data. 
FederBoost \cite{tdsc2024federboost} observes that GBDT training primarily depends on the relative order of feature values for finding splits, not their exact values.
Other alternatives offer different trade-offs among privacy, utility, and efficiency, such as SecureBoost \cite{is2021secureboost}, which relies on cryptography, or SimFL \cite{aaai2020pgbdt}, based on locality-sensitive hashing.

\end{itemize}

Applying DP to decision trees presents unique challenges due to their recursive structure and reliance on data queries at multiple stages \cite{fletcher2016tdidt} \cite{cs2019tdidt}: 

\begin{itemize}

\item
{\it Node splitting}: Selecting the best attribute and threshold to split a node typically involves calculating impurity measures, like the Gini index or information gain, which are based on data frequencies within each node. Such computations can be privatized using the Laplace mechanism (applied to the counts used by the splitting criterion) or the exponential mechanism, which is often preferred and  uses the impurity measure as its utility function to probabilistically select a split that is close to optimal while preserving privacy. 
Handling continuous attributes requires care, as selecting an exact threshold from the data leaks privacy. Potential solutions include selecting a noisy threshold or using the exponential mechanism to select a range.

\item
{\it Leaf value determination}: Leaf nodes typically predict the majority class among the instances reaching the leaf. Again, both the Laplace and the exponential mechanism can be used: add Laplace noise to the counts of each class within the leaf (and output the class with the highest noisy count) or directly output a (potentially non-majority) class label, sacrificing count information but potentially improving privacy.

\item
{\it Termination and pruning}: Standard criteria need adaptation. Termination can be based on reaching a maximum depth (since deeper trees consume more privacy budget), a minimum noisy count threshold, or when the noise magnitude overwhelms the signal. Pruning techniques can be adapted by using noisy counts or specifically removing nodes that were likely generated due to noise.

\end{itemize}


The privacy budget allocation is paramount in DP decision trees, since the total privacy budget must be distributed across all decisions made during tree construction, including deciding splits at internal nodes and determining the final leaf values. Additionally, when used in decision tree ensembles, the budget must be properly allocated among the individual decision trees within the ensemble.

\subsection{Differential privacy in Probabilistic AI} 
\label{sec:ai:probabilistic}

Probabilistic AI resorts to models that explicitly represent uncertainty through probability distributions. The probabilistic AI approach, rooted in Statistics, uses Bayes' Theorem to update the likelihood of a hypothesis as more data becomes available. It is used both for discriminative models, such as conditional random fields (CRFs) or maximum-entropy Markov models (MEMMs), and generative models, such as Bayesian networks, hidden Markov models (HMMs) or Gaussian mixture models (GMMs). 


Estimators play a central role in probabilistic AI. For instance, they are used to infer the parameters of probability distributions or models that best explain observed data. Essentially, estimators are functions that provide estimates of unknown parameters based on observed data, i.e., approximations of quantities of interest based upon the evidence in a given dataset. 

In probabilistic AI, estimators are used to learn model parameters (e.g., mean and variance in a Gaussian distribution), make predictions by estimating the most likely outcomes, quantify the uncertainty in predictions or model parameters, or support inference in Bayesian models or probabilistic graphical models.

\begin{itemize}

\item
A point estimator provides a single ``best guess'' value for an unknown parameter. It does not quantify uncertainty but gives a specific value based on the data (e.g., the sample mean $\hat{\mu} = \frac{1}{n} \sum_{i=1}^n x_i$ is a point estimator of the population mean).
The propose-test-release (PTR) \cite{stoc2009dpstats} mechanism can be used to obtain $(\epsilon,\delta)$-DP robust statistical estimators (robust in the sense that they are resilient against outliers and small errors in data measurement).
Convergence rates of differentially private approximations to statistical estimators \cite{icml2012dpstats} have also been studied within the smooth sensitivity framework (i.e., using local sensitivity \cite{nissim2007stoc}).

\item
M-estimators (short for maximum likelihood-type estimators) generalize point estimators by defining the estimate as the solution to an optimization problem: $\hat{\theta} = \arg \min_\theta \sum_{i=1}^n \rho(x_i,\theta)$, where $\rho(x_i,\theta)$ is a loss function that measures the discrepancy between the data and the model. Maximum likelihood estimators (MLEs) are a special case of M-estimators, where $\rho(x_i,\theta) = -\log p(x_i|\theta)$. M-estimators are useful for robust statistics, where you want to reduce the influence of outliers (e.g., using Huber loss instead of squared error).
For a large class of parametric probability models, one can construct an $\epsilon$-differentially private estimator whose distribution converges to that of the maximum likelihood estimator using the sample and aggregate strategy \cite{arxiv2008dpestimators}: split $D$ into $k$ disjoint random subsets (a.k.a. blocks), estimate parameters on each subset $\hat{\theta}_i$, and use a differentially-private mean of $\hat{\theta}_i$ to approximate $\theta$ (i.e., compute the average estimate $\bar{z}$, draw a random observation $R$ from a double-exponential distribution $Y \sim Lap(\frac{\Lambda}{2\epsilon})$, where $\Lambda$ is the diameter of the parameter space $\Theta$, and output $\theta^* = \bar{z}+R$). The estimator can also be made differentially private by replacing the empirical average with a differentially private aggregate using a widened Windsorized mean \cite{stoc2011dpestimators}. A third mechanism for obtaining $\epsilon$-differentially private M-estimators is based on the use of perturbed histograms \cite{nips2011dpmestimators}, i.e., adding double-exponential noise to each histogram count in order to obtain a noisy density function.

\end{itemize}


Estimators are used by many ML algorithms to learn from data, from Naive Bayes classifiers and Bayesian networks to k-means and GMMs, yet they serve different purposes due to the nature of the algorithms (probabilistic in Naive Bayes and Bayesian networks, geometric in k-means and GMMs):

\begin{itemize}

\item
{\it Naive Bayes} is a probabilistic classifier based on Bayes' theorem with the assumption of feature independence. Estimators are used to compute the probabilities needed for classification (i.e., class prior probabilities and conditional feature likelihoods). These estimates are plugged into Bayes' Theorem to compute posterior probabilities. 
An $\epsilon$-differentially private Naive Bayes model (DP-NB) \cite{wi2013dpnaivebayes} is quite simple: just estimate the sensitivity for each attribute appropriately and add Laplacian noise to the model parameters. 

A Naive Bayes classifier can also be learned jointly by different data owners in a distributed environment, without the help of a trusted curator (i.e., without breaking the privacy of each data provider), with the help of additive homomorphic encryption (AHE-NB) \cite{is2018dpnaivebayes}, just by adding noise in the aggregated ciphertexts, which is functionally equivalent to performing the mechanism that implements differential privacy on the corresponding plaintexts.
Also without a trusted data curator, individuals can send their perturbed class prior probabilities so that the data aggregator can estimate all the probabilities needed by
the Naive Bayes classifier satisfying local differential privacy (LDP-NB) \cite{arxiv2019ldpnaivebayes}.

\item
PrivBayes \cite{tods2017privbayes} first spends half the privacy budget to learn a Bayesian network structure that captures the dependencies in the data, and then uses the remaining privacy budget to measure the statistics necessary to learn the Bayesian network parameters. It has been used as a differentially private mechanism that generates synthetic data. Private-PGM \cite{icml2019privatepgm} is a general-purpose post-processing tool to infer a data distribution given noisy measurements that uses a compact representation of the data distribution to avoid exponential complexity. Private-PGM can be used to improve the original PrivBayes \cite{tods2017privbayes} and has also been used to win the 2018 NIST differential privacy synthetic data competition \cite{jpc2021mst}.

\item
{\it k-means } is an unsupervised clustering algorithm that partitions data into k clusters by minimizing within-cluster variance. The centroid of each cluster is given by a point estimator of the mean of the data points assigned to it.
A differentially-private k-means algorithm \cite{nissim2007stoc} can be implemented using the sample and aggregate strategy: randomly split the training set into many subsets, run the non-private k-means algorithm on each subset, choose the right aggregation function, and then use the smooth sensitivity framework (i.e., local sensitivity) to publish the final noisy output. This final step preserves privacy while the underlying k-means algorithm is essentially unchanged. 
Alternative privacy-preserving k-means variations have also been proposed, beyond the DP framework, by not allowing access to the true data vectors at any stage of the clustering algorithm \cite{cikm2019ppkmeans}.

Like other ML techniques, the k-means algorithm has also been adapted to guarantee local differential privacy in a distributed environment \cite{cs2020ldpkmeans}. In each iteration, the service provider (i.e., the untrusted curator) aggregates the received perturbed feature vectors of users based on the clusters that they belong to in the previous iteration, computes the $k$ centroids, and then it publishes the centroids of the current iteration to users. Also in each iteration, after being informed of the computed centroids, each user computes the distances to the centroids based on its exact data and finds the nearest corresponding cluster; then, each user sends the cluster assignment to the service provider using binary strings with noise that preserve user’s private information in terms of local differential privacy.

\item
{\it Gaussian Mixture Models (GMMs)} can be seen as a probabilistic generalization of the k-means clustering algorithm. In k-means, each cluster is represented by a centroid (a point in space) and all points are assigned to the nearest centroid using Euclidean distance. In GMMs, each cluster is represented by a Gaussian distribution (defined by a mean vector and a covariance matrix). This allows clusters to have elliptical shapes with different sizes and orientations.
Gaussian mixture models have inspired several differentially-private algorithms with applications in ML:

\begin{itemize}

\item
PGME \cite{neurips2019gmm} is a differentially private algorithm for recovering the parameters of an unknown Gaussian mixture, provided that the components are sufficiently well separated, that uses private PCA [Principal Component Analysis] and private clustering steps.

\item
DP-LMGMM \cite{tdsc2012gmm} learns multiclass Gaussian mixture model-based classifiers that preserve differential privacy using a large margin loss function with a perturbed regularization term. 

\item
MR \cite{ndss2025gmm} adapts the expectation-maximization (EM) algorithm for locally differentially private (LDP) settings. Using a reduction framework inspired by Gaussian mixture models (GMM) and focusing on density estimation under LDP constraints, mixture reduction (MR) provides an approach that is suitable for decentralized or federated ML.
\end{itemize}

\end{itemize}

In some situations, Bayesian learning can be inherently differentially private \cite{pmlr2015sgmc}. For instance, getting one sample from the posterior is a special case of the exponential mechanism and this sample as an estimator is near optimal for parametric learning (e.g., in logistic regression). Algorithmic procedures of stochastic gradient Langevin Dynamics (and variants) that attempt to sample from the posterior also guarantee differential privacy as a byproduct. There might be other cases where existing randomness could be exploited for privacy \cite{pmlr2015sgmc}.

\subsection{Differential privacy in Statistical AI} 
\label{sec:ai:statistic}

Apart from probabilistic AI models, other approaches to AI also have their roots in statistical techniques. Some non-probabilistic statistical AI models generate predictions or classifications based on direct data mappings (deterministically) rather than modeling probability distributions. They resort to statistical methods to model, infer, and predict patterns in data, without being symbolic GOFAI techniques. They bridge traditional statistics and modern machine learning. Key techniques include traditional regression models and kernel-based methods, such as Support Vector Machines (SVMs).\footnote{Pedro Domingos \cite{domingos2018master} identifies five tribes of machine learning, each representing a distinct school of thought: symbolists view learning as inverse deduction (Section \ref{sec:ai:symbolic}), connectionists aim to reverse-engineer the brain using artificial neural networks and deep learning (Section \ref{sec:ai:connectionist}), evolutionaries simulate the process of evolution (for either symbolic or connectionist models), Bayesians focus on probabilistic modeling (Section \ref{sec:ai:probabilistic}), and, finally, analogizers learn by recognizing similarities between new and old situations (Section \ref{sec:ai:statistic}).}


\subsubsection{Traditional regression models} 

Traditional regression models include linear regression for predicting a continuous outcome and logistic regression for predicting a binary outcome. In both cases, we can use objective perturbation in order to make them differentially-private, i.e., just add noise to the target function. For instance, adding a linear regularization term to the loss function can ensure privacy \cite{jmlr2011dperm} \cite{vldb2012dpregression}.

\begin{itemize}

\item 
Linear regression models are of the form $\hat{y} = \beta_0 + \beta_1 x_1 + ... + \beta_nx_n$. They are used to predict continuous variables in forecasting, trend analysis, or risk modeling. They are a linear model, adjusted to minimize the Mean Squared Error (MSE), so they assume linearity, as well as independence, homoscedasticity, and the normality of errors. 

Objective perturbation \cite{jmlr2011dperm} \cite{vldb2012dpregression} and the sample and aggregate framework \cite{stoc2009dpstats} can be used to design DP linear regression models. In linear regression, L2 regularization limits the magnitude of weights, reducing the L2-sensitivity of the output. Gaussian noise can then be added to achieve $(\epsilon,\delta)$-DP \cite{jmlr2011dperm}.

\item
Logistic regression models are of the form $P(y=1|x) = \hat{y} = 1 / {\left(1+e^{-(\beta_0 + \beta_1 x_1 + ... + \beta_nx_n)}\right)}$. They are used to predict a binary outcome in medical diagnosis, fraud detection, customer churn prediction, or spam detection. Unlike linear regression models, which predict a value directly, logistic regression models predict the probability of one class in binary classification problems. They use the sigmoid function to squash the output between 0 and 1 and minimize the log loss function (a.k.a. cross-entropy).

Logistic regression models can be made differentially-private by objective perturbation methods, including both the sensitivity and objective regularized $(\epsilon,\delta)$-DP logistic regression \cite{nips2008dplogistic} \cite{jmlr2011dperm}, as well as the functional mechanism (FM) $\epsilon$-DP regression \cite{vldb2012dpregression}.
Regularization (e.g., L2 regularization) can be used to bound the sensitivity of the model output (i.e., weights or predictions) and then noise can be added, proportional to this reduced sensitivity, to satisfy DP. Empirical results on logistic regression demonstrate improved utility \cite{jmlr2011dperm}.

\end{itemize}

\subsubsection{Support Vector Machines (SVMs)}

SVMs aim to find the hyperplane that best separates data points belonging to different classes. 
Linear SVMs work when training examples are linearly separable, whereas kernel SVMs are used when data are not linearly separable:

\begin{itemize}

\item 
Linear SVMs find the hyperplane that maximizes the margin between the two classes in binary classification problems by solving a convex optimization problem with constraints. Both output perturbation mechanisms \cite{jmlr2011dperm}\cite{arxiv2009dpsvm} and objective perturbation mechanisms \cite{jmlr2011dperm}  have been used for linear SVMs.

\item
Kernel SVMs apply a kernel function to map data into a higher-dimensional space where a linear separator can be found. This allows SVMs to create non-linear decision boundaries in the original space.
Several ideas have been explored for private kernel SVMs:

\begin{itemize}

\item
For all translation-invariant kernels,\footnote{A kernel function of the form $k(x,y) = g(x-y)$, for some function $g$, is called translation-invariant.} such as the RBF, Laplacian, and Cauchy kernels, you can approximate kernel functions in the original sample space with a linear kernel in another space, in order to avoid publishing training data \cite{arxiv2009dpsvm}. 

\item
For kernel functions that are not translation-invariant, such as the polynomial kernel or the sigmoid kernel, you can interpret a model as a function and use another function to approximate it, the basis of the Test Data-Independent Learner (TDIL) \cite{icml2013dpkernels}.

\end{itemize}

\end{itemize}

DP is typically applied to SVMs using objective perturbation \cite{jmlr2011dperm}\cite{icml2013dpkernels}, leveraging the convexity of the standard SVM objective function (hinge loss plus L2 regularization).
Noise, often Gaussian or Laplace, depending on the specific formulation and desired privacy guarantees, is added to the SVM objective function. The optimization algorithm then finds the parameters (hyperplane weights and bias) that minimize this perturbed objective. The amount of noise depends on the sensitivity of the objective function, which can often be bounded. 
Perturbing the regularized objective function directly with random noise is simple to implement and leverages regularization smoothing.
L2 regularization (strong convexity) reduces sensitivity in empirical risk minimization, enabling DP with less noise and improving utility \cite{jmlr2011dperm}.

While less common, output perturbation (adding noise to the final hyperplane parameters) \cite{nips2008dplogistic}\cite{jmlr2011dperm}\cite{arxiv2009dpsvm} or gradient perturbation (when using SGD to solve the SVM optimization) \cite{nips2017dperm} are also potential approaches. In both cases, care must be taken not to exacerbate bias and unfairness for different groups of individuals \cite{neurips2021dperm}.

For high-dimensional data, applying dimensionality reduction techniques before SVM training can improve utility and reduce computational overhead. 
Differentially-private dimensionality reduction techniques allow the private release of a low-dimensional approximation to a set of data records, e.g., by adding Gaussian noise to the covariance matrix in order to achieve a DP-compliant Principal Component Analysis algorithm \cite{stoc2014ag}. 
When combining dimensionality reduction with SVMs, end-to-end privacy can be ensured either by adding noise in the PCA phase (i.e., to the covariance matrix), as in DPPCA-SVM \cite{scn2021dppcasvn}, or by adding noise in the SVM phase (i.e., during SVM training), as in PCA-DPSVM \cite{scn2021dppcasvn}.
Privacy can be adopted in both the PCA-based dimensionality reduction phase and the SVM training phase, as shown by Differential Privacy-compliant Federated Machine Learning with Dimensionality Reduction, FedDPDR-DPML \cite{cmc2024dpsvm}, which enables multiple participants to collaboratively learn a global model based on weighted model averaging and knowledge aggregation, which is later distributed to each participant to improve local utility.

\subsection{Differential privacy in Deep Learning} 
\label{sec:ai:connectionist}

Deep learning models, with their vast number of parameters and non-convex optimization landscapes, pose significant challenges for differential privacy.
The use of differential privacy in deep learning models typically relies on gradient perturbation, using a differentially-private version of the stochastic gradient descent algorithm \cite{abadi2016dpsgd}.

\subsubsection{Differentially-Private Stochastic Gradient Descent (DP-SGD)} 

Differentially Private Stochastic Gradient Descent (DP-SGD) \cite{abadi2016dpsgd} adapts SGD to satisfy $(\epsilon, \delta)$-differential privacy by clipping per-example gradients to a fixed norm and adding Gaussian noise. This method builds on stochastic optimization methods \cite{robbins1951stochasticApproximation} \cite{kiefer1952StochasticStimation} \cite{rosenblatt1958perceptron} to ensure differential privacy in deep learning. 

Gradient perturbation via DP-SGD has become the standard technique for DP deep learning. The standard SGD is modified so that DP is incorporated in each training iteration using clipped noisy gradients. First, as in the standard SGD, we sample a random mini-batch of data and compute per-example loss gradients for all training examples in the mini-batch. Then, the first modification to the standard algorithm is introduced: gradient clipping (i.e., clipping the L2 norm of each per-example gradient). Clipped per-example gradients are averaged and a second modification completes the DP-SGD algorithm: Gaussian noise. Instead of updating the model parameters using the gradients as in SGD, we add Gaussian noise to the average gradients and update the model parameters using the noisy version of the gradients.

Both gradient clipping and gradient noise were incorporated into DP-SGD to achieve differential privacy: 

\begin{itemize}
    
\item 
{\em Gradient noise:} Gaussian noise is added to the gradients to obscure individual data contributions. 
That Gaussian noise is scaled by the gradient clipping threshold. 
The noise level introduced into the gradients allows the privacy loss to be distributed over multiple training steps, decreasing the per-step privacy cost and enabling more epochs to be run within the same cumulative privacy budget. 

The noise level has a significant impact on model performance, as larger noise levels reduce the risk of privacy leakage but can also degrade accuracy. DP-SGD, therefore, exhibits the privacy-utility trade-off. 

\item
{\em Gradient clipping} limits the gradient norm to a predefined threshold, bounding the sensitivity of each individual's contribution to the average gradient.
Gradient clipping has two important effects: on the one hand, clipping reduces the variance of gradients and helps maintain privacy by bounding the maximum contribution of any individual training example; on the other hand, it can distort the unbiasedness of the gradient estimate. 

The choice of the appropriate clipping bound is crucial, as a small bound can significantly affect the direction of the gradient (i.e., the gradients may deviate significantly from their true values), while a larger bound requires adding more noise to the gradients. A reasonable heuristic is taking the median of the norms of the unclipped gradients throughout the training process. This helps in balancing privacy and model accuracy, as excessive clipping may lead to a loss of useful information, while insufficient clipping could compromise privacy.

\end{itemize}


In deep learning, regularization is common practice.
In DP-SGD, regularization is implicitly used to stabilize training, reducing sensitivity. 
DP-SGD also exploits subsampling privacy amplification. Whereas regularization reduces sensitivity, subsampling and noise in SGD (both natural and added) are used provide DP guarantees. 


The original DP-SGD algorithm was sequential. Since SGD algorithms can be parallelized and executed asynchronously, distributed DP-SGD algorithms have been proposed, such as DSSGD, the Distributed Selective SGD \cite{ccs2015dpdeeplearning}. In selective SGD, the learner chooses a fraction of parameters to be updated at each iteration. This selection can be completely random, but selecting the parameters with larger gradients is smarter (i.e., those whose current values are farther away from their local optima). 
A sparse vector technique (SVT) \cite{focs2010svt} can be used to prevent indirect privacy leaks caused by participants sharing locally-updated parameters: randomly selecting a small subset of gradients whose values are above a threshold, and then sharing perturbed values of the selected gradients. 
DSSGD assumes two or more participants training independently and concurrently. After each round of local training, participants asynchronously share with each other the gradients they computed for some of the model parameters. Participants thus benefit from each other's training data, without actually seeing them, and produce more accurate models than they would have been able to learn in isolation, limited to their own training data.

In settings where only the labels are sensitive but the inputs are public, specialized techniques can be used, potentially achieving better utility than full DP-SGD \cite{neurips2021labeldp}.


Some challenges related to the use of DP-SGD include the following \cite{kdd2023dpfyml} \cite{das2024advances}:

\begin{itemize}

\item 
{\em Hyperparameter sensitivity}:
DP-SGD performance is highly sensitive to hyperparameters such as the clipping threshold (for gradient clipping), the noise multiplier (for gradient noise), the learning rate, and the batch size, even the activation function. Bounded activation functions (e.g., tanh) are sometimes preferred over unbounded ones (e.g., ReLU) to help control gradient norms. Tuning these hyperparameters is critical but complex, and the tuning process itself should be accounted for in the privacy budget when using private data \cite{das2024advances}. It should also be noted that architectural choices might affect privacy. For instance, batch normalization can break DP guarantees by creating dependencies across examples in a batch. Alternatives such as group normalization or layer normalization, or specific DP-compatible variants, are often required \cite{acmcs2025dpcdl}.

\item
{\em Computational cost:} Computing per-example gradients is significantly more computationally and memory intensive than standard batch gradients. Techniques like efficient vectorization or "ghost clipping" aim to mitigate this \cite{das2024advances}.

\item
{\em Fairness: } DP-SGD can amplify existing biases. Gradient clipping disproportionately affects examples from underrepresented groups or those with more complex features, which tend to have larger gradients, reducing their influence on the model and leading to lower accuracy for these groups \cite{das2024advances}.

\item
{\em Utility degradation:} DP-SGD often leads to a noticeable drop in model accuracy compared to non-private training, especially for smaller privacy budgets (low $\epsilon$). While DP-SGD is a robust method for ensuring privacy, it often leads to significant performance degradation, due to the added noise. Some variants, such as NoisySGD and NoisyAdam \cite{hdsr2020dlgdp}, can achieve higher accuracy under the same privacy budget.

\end{itemize}

An effective strategy for improving utility is to pre-train a large model on public data (non-privately) and then fine-tune it on the sensitive target task using DP-SGD. Larger pre-trained models often show better robustness to the noise introduced during private fine-tuning \cite{yu2022differentiallyprivatefinetuninglanguage}. It should be noted, however, that the assumption of large web-scraped datasets being truly ``public'' and free of privacy concerns is debatable \cite{arxiv2024dppretraining}.

The application of DP transforms the deep learning process. It requires careful algorithm design, introduces new hyperparameters, impacts computational requirements, and raises considerations about fairness and the very definition of privacy.

\subsubsection{Privacy accounting}

The privacy accountant tightly tracks cumulative privacy loss over all training iterations to ensure differential privacy in deep learning algorithms. Privacy accountants leverage composition theorems and privacy amplification by subsampling:

\begin{itemize}

\item
The {\em Moments Accountant (MA)} was a key innovation introduced with DP-SGD \cite{abadi2016dpsgd}.
MA bounds the moments of the privacy loss random variable, hence its name. 
MA provides tight bounds on the cumulative privacy loss over multiple iterations (compositions) of the DP mechanism in DP-SGD using the strong composition theorem, tighter than basic composition theorems.
MA is efficient for Gaussian mechanisms and DP-SGD with subsampling, yet it might not generalize well to other mechanisms. Widely adopted in practice (e.g., TensorFlow Privacy), it requires the numerical computation of moments. 
Whereas composing directly from the tail bound (i.e., the privacy loss in the sense of differential privacy) can result in quite loose bounds, MA computes the log moments of the privacy loss random variable, which compose linearly. Then, MA converts the moments bound into the $(\epsilon,\delta)$-DP guarantee. Since moments compose linearly, the resulting privacy loss is proportional to the number of iterations.

It should be noted that, although the growth of privacy loss calculated with moments accountant is much lower than the growth of privacy loss calculated by the strong composition theorem, when the total privacy budget $\epsilon_{total}$ is very small, the bounds computed by the strong composition theorem are actually lower than the bounds computed by the moments accountant.

\item 
The {\em Analytical Moments Accountant (AMA)} \cite{wang2019analyticalmoments} generalizes the moments accounting technique, originally developed for the Gaussian mechanism \cite{arxiv2019rdpgaussian}, to any subsampled RDP mechanism (e.g., Gaussian, Laplace, or exponential). AMA tracks the Rényi Differential Privacy (RDP) of a mechanism by computing the moments (expected values) of the privacy loss random variable and allows for the efficient conversion of RDP to $(\epsilon,\delta)$-DP. Using the relationship between RDP and the log-moment generating function, log moments can be computed for any mechanism with known RDP guarantees. Once the log-moments are computed, numerical optimization can be used to find the tightest $\epsilon$ for a given $\delta$.

\item
The {\em CLT-based privacy accountant} \cite{hdsr2020dlgdp}, as applied to NoisySGD and NoisyAdam, is a method for tracking and bounding the cumulative privacy loss under Gaussian differential privacy (GDP). It is called the CLT approach because the composition theorem of f-DP is tight and, more importantly, the privacy central limit theorem (CLT) is asymptotically exact.  This privacy accountant leverages the CLT to approximate the privacy loss distribution as a Gaussian (normal) distribution, which simplifies the analysis of privacy guarantees under composition.
CLT composition models the privacy loss random variable (PLRV) for each query and uses the CLT to approximate the sum of these losses. According to the CLT, the sum of many independent random variables tends toward a normal (Gaussian) distribution. This allows us to approximate the distribution of the total privacy loss. Using the Gaussian approximation, we can compute a tight bound on the total $\epsilon$ for a given $\delta$ (or vice versa).

\item 
The {\em Amortized Privacy Accountant (APA)} \cite{iclr2023amortized}, another generalization of the Moments Accountant, amortizes the privacy cost over multiple steps or queries, especially under subsampling.
APA provides tighter and more adaptive privacy guarantees by analyzing the expected privacy loss over multiple iterations rather than the worst-case loss at each step. 
Instead of computing the privacy loss at each step, the amortized accountant 
estimates the total $\epsilon$ after multiple steps, given the noise multiplier and sampling rate.
By averaging the privacy cost over time, the amortized accountant can yield significantly tighter bounds on the total privacy loss, improving the utility of the trained model.

Amortized analysis is especially relevant when the actual noise added varies across steps, the sensitivity of the function being computed changes over time, or the data access pattern is not uniform or deterministic.
It is particularly useful in iterative deep learning algorithms like NoisySGD and NoisyAdam, where privacy loss accumulates over many steps.
Using Gaussian Differential Privacy (GDP), which naturally leads to amortized analysis, APA builds on the idea of privacy loss distributions and uses tools like supermartingales and hockey-stick divergences to track privacy loss more precisely.
The amortized accountant uses privacy amplification theorems \cite{beimel2010bounds} \cite{beimel2014bounds} \cite{focs2008local} and strong composition theorems \cite{dwork2010boosting}\cite{bassily2014dperm} to provide an amortized bound of privacy loss.
Amortized accountants may be less precise than exact methods like RDP or CLT composition, but they are much faster and still provide valid privacy guarantees.

\item
The {\em Edgeworth Accountant} \cite{wang2022edgeworth} uses Edgeworth expansions to approximate the distribution of the privacy loss random variable under the f-DP framework, which generalizes traditional $(\epsilon, \delta)$-DP.
The Edgeworth Accountant provides very tight non-asymptotic bounds on $(\epsilon,\delta)$-DP and works with any noise mechanism.
It is computationally efficient even with many compositions, albeit it is more mathematically complex than the Moments Accountant. Its efficiency makes it suitable for large-scale applications such as LLMs.

The Edgeworth accountant is an advanced method for composing differential privacy guarantees, offering even tighter approximations of cumulative privacy loss than CLT-based methods. It builds on the Edgeworth expansion, a refinement of CLT, to more accurately approximate the distribution of the privacy loss random variable (PLRV). The Edgeworth expansion improves upon the CLT by incorporating higher-order moments (like skewness and kurtosis) of the distribution. While CLT approximates the sum of random variables as a normal distribution, Edgeworth adds correction terms to better match the true distribution.


Since the Edgeworth accountant does not rely on closed-form expressions for specific mechanisms (like Gaussian or Laplace mechanisms), it is mechanism-agnostic. As long as you can compute or estimate the moments of the PLRV for a mechanism, you can use the Edgeworth accountant. This includes non-standard or custom mechanisms, even if they do not have a known RDP or analytical form.
This makes the Edgeworth Accountant very flexible, especially for research and advanced applications.\footnote{DP in LLMs: Large language models (LLMs) present additional challenges due to their size and complexity. The EW-Tune framework \cite{behnia2022ewtune} has been proposed to implement DP in LLMs with the help of the Edgeworth Accountant \cite{wang2022edgeworth}. The Edgeworth series expansion models the distribution of noisy gradients by incorporating higher-order terms in its approximation. This approach enables a more precise calibration of noise injection for the differentially-private fine-tuning of large language models (LLMs) \cite{yu2022differentiallyprivatefinetuninglanguage}, effectively balancing model utility and privacy guarantees. Where DP-SGD \cite{abadi2016dpsgd} would introduce significant noise into gradients, EW-Tune reduces variance while maintaining strict differential privacy guarantees.}

\item
The {\em RDP Accountant} \cite{neurips2024rdpaccountant} is a holistic Rényi Differential Privacy (RDP) accountant designed to provide tighter and more efficient privacy accounting for differentially private algorithms, especially DP-SGD with fixed-size subsampling (i.e. mini-batches). 
The RDP accountant is designed to be modular, composable, and efficient, making it suitable for complex training pipelines. 
The RDP accountant supports both with replacement (FSwR) and without replacement (FSwoR) subsampling strategies.

How does it work? First, each mechanism is analyzed in terms of its Rényi divergence at various orders $\alpha$. For instance, the Gaussian mechanism has a closed-form RDP expression. Then, RDP composes linearly: the total RDP at order $\alpha$ is the sum of the RDPs of individual mechanisms. Finally, after composition, the total RDP is converted back to a standard $(\epsilon, \delta)$-DP guarantee using a tight conversion formula.


As the Analytical Moments Accountant, the RDP accountant requires that, for each mechanism, you can compute its Rényi divergence. If a mechanism has no known or computable Rényi divergence, or is non-parametric or adaptive in a way that breaks the assumptions, then it cannot be directly used with the RDP accountant.

\end{itemize}

\begin{table}[t]
    \centering

\begin{tabular}{llll}  
  \hline  
  Privacy    & DP        & Composition  & DP       \\
  Accountant & Framework & Type         & Mechanism \\
  \hline
  Moments \cite{abadi2016dpsgd} & $(\epsilon,\delta)$-DP & log moments & Gaussian \\
  Analytical \cite{wang2019analyticalmoments} & RDP & Subsampled RDP & Gaussian, Laplace, exponential \\
  CLT \cite{hdsr2020dlgdp} & GDP & Central Limit Theorem & Gaussian \\
  Amortized \cite{iclr2023amortized} & GDP & Adaptive & Gaussian \\
  Edgeworth \cite{wang2022edgeworth} & $f$-DP & Edgeworth expansion & Any \\
  RDP \cite{mironov2017rdp} \cite{neurips2024rdpaccountant} & RDP & Subsampled RDP & Gaussian, Laplace, exponential \\
  \hline
  & & & \\
\end{tabular}

    \caption{Privacy accountants for differential privacy in deep learning.}
    \label{tab:accountants}
\end{table}


Subsampling plays a crucial role in improving the privacy guarantees of differentially-private algorithms, especially in the context of privacy accountants. In the context of training deep learning models using stochastic gradient descent (e.g. DP-SGD, NoisySGD, or NoisyAdam), subsampling refers to selecting a random subset of the dataset at each training step, rather than using the full dataset (i.e., using a mini-batch).


Subsampling reduces the effective $\epsilon$ by a factor proportional to the sampling rate $q$ \cite{kasiviswanathan2008note}\cite{kasiviswanathan2014semantics}\cite[Section 3.6]{dwork2014tcs}. This phenomenon is known as privacy amplification by subsampling: when a mechanism $\mathcal{M}$ is $(\epsilon,\delta)$-DP on a dataset $D$, running the mechanism $\mathcal{M}$ on a random subsample of $D$ with sampling rate $q$ improves privacy to approximately $(q\epsilon,q\delta)$-DP.
In other words, subsampling inherently boosts privacy, which is useful in deep learning and large-scale data analysis.

Subsampling introduces randomness in the data access pattern, which leads to privacy amplification. If a mechanism is differentially private when applied to the full dataset, then applying it to a random subset reduces the effective privacy loss. Intuitively, each individual has a lower chance of being included in any given mini-batch, so their data is less exposed.

Subsampling is a privacy amplifier and a key enabler of practical DP training in deep learning. Privacy accountants leverage subsampling to tighten privacy bounds and enable efficient composition. Providing tighter and more realistic privacy guarantees and reducing per-step privacy cost enable longer training under fixed privacy budgets. 

Some accountants (e.g., the Amortized Accountant) are specifically designed to exploit subsampled mechanisms and provide accurate tracking over adaptive compositions. In fact, there are techniques that resort to privacy amplification by iteration. They efficiently handle multiple adaptive queries with minimal privacy budget. For instance, the Sparse Vector Technique (SVT) \cite{dwork2010continual}\cite[Section 4]{dwork2014tcs}, which answers whether a sequence of queries exceeds a noisy threshold, satisfies $\epsilon$-DP while adaptively halting after a fixed number of ``above-threshold'' answers. 

There are different types of subsampling. In Poisson subsampling, used by the Moments Accountant, each data point is included independently with some probability, which makes it easier to analyze. In uniform subsampling, a fixed-size mini-batch is selected uniformly at random, which is more realistic in practice, yet harder to analyze. When sampling a dataset, you can also choose whether elements can be selected more than once (i.e., whether sampling is with or without replacement). That decision can affect the tightness of privacy bounds (e.g., see the RDP Accountant \cite{neurips2024rdpaccountant}). 


In some contexts, including LLM training \cite{charles2025uls}, additional sampling approaches might be useful to protect user-level privacy, in order to ensure that the contribution of individual users to the model is protected. Whereas example-level sampling (ELS) involves clipping gradients at the example level, user-level sampling
(ULS) operates at the user level, allowing for gradient aggregation over all the examples provided by a single user. The distinction between ELS and ULS is also important in federated learning. In deep learning models, ELS is common practice in DP-SGD, since it is easier to analyze using tools such as moments and RDP accountants. ULS, on the other hand, requires more sophisticated privacy accounting (e.g., f-DP, Edgeworth, or user-level RDP accountants). In LLM fine-tuning,  user-level DP accountants \cite{charles2025uls} can provide tight user-level privacy guarantees even when using ELS, bridging the gap between the two approaches (by analyzing the privacy amplification effect of ELS at the user level).
Under fixed computational budgets, ULS tends to provide stronger privacy guarantees and better model performance, particularly when stringent privacy protection is required or when larger computational resources are available.

In general, when training deep learning models, the privacy accountant keeps track of the privacy loss spent after each training step that updates the model parameter values. If the accumulated privacy loss exceeds the allowed limits (i.e., the privacy budget), then training is stopped and no further updates are performed on the model. This is just a privacy-driven form of early stopping, a common regularization technique used in deep learning.

\subsubsection{Privacy for free}

Privacy for free, in the context of machine learning, refers to the idea that certain machine learning techniques can inadvertently provide some degree of privacy protection without requiring explicit privacy-preserving mechanisms like differential privacy \cite{sp2017mia} \cite{usenix2019memorization}. For instance, the regularization effect of dropout might also obscure individual data points in a dataset, making it harder to infer sensitive information about them \cite{nitish2014Dropout} \cite{csf2018mia}. 

``Privacy for free'' suggests that certain model architectures, training methods, or data handling strategies can offer privacy benefits inherently, without the need for heavy cryptographic tools or added noise (as in differential privacy). You can design and train models in a way that inherently protects privacy without significantly sacrificing performance or requiring complex privacy-preserving techniques.


Let us now examine some strategies that have been explored to obtain privacy benefits as a byproduct of good model design or training practices:

\begin{itemize}

\item 
{\em Regularization:} Over-parameterized models (like large neural networks) can memorize training data but, with proper regularization, they generalize better and leak less private information \cite{csf2018mia}. This reduced memorization offers a form of implicit privacy. Regularization prevents overfitting \cite{usenix2019memorization}, limits model exposure to training data, reduces memorization, and, in turn, can reduce the risk of leaking sensitive data:

\begin{itemize}

\item
L1 and L2 regularization (a.k.a. weight decay) adds a penalty term to the loss function. L1 (Lasso) encourages sparsity \cite{robert2018RegressionShrinkage}, i.e. many weights become zero, whereas L2 (ridge)  \cite{hoerl2000RidgeRegression}, encourages small, evenly distributed weights. Smaller weights reduce the model capacity to memorize specific data points.

\item
Dropout randomly drops units (along with their connections) during training, which prevents the co-adaptation of neurons and forces the neural net to learn more robust features \cite{nitish2014Dropout}. Reducing reliance on specific neurons that might encode sensitive patterns provides an implicit form of differential privacy. In fact, dropout stochasticity mimics noise injection, potentially reducing sensitivity \cite{pmlr-v48-gal16}. Dropout is computationally cheap and widely used in deep learning.

\item
Early stopping monitors validation loss and halts training when performance stops improving. 
Since training a model for too long can lead to memorization of specific data points, early stopping based on validation loss can reduce privacy leakage \cite{Nakkiran2021Deep}. Therefore, early stopping is another common (and simple) technique with privacy benefits.

\end{itemize}

\item
{\em Data minimization} using only the data necessary for the task (e.g., with the help of feature selection algorithms and dimensionality reduction techniques) can reduce the risk of exposing sensitive information \cite{staab2025sokdataminimizationmachine}. This aligns with the principle of privacy by design \cite{cavoukian2009privacy}.

\item
{\em Data condensation \cite{pmlr2022condensation}}, originally designed to improve training efficiency, can be used for private data generation, thus providing privacy for free. Instead of training models on the raw data, that can be attacked by MIA and suffer potential data leakages, we can use synthetic data for model training.

\item
{\em Data augmentation} applies transformations (e.g., rotation, cropping, noise) to training data with the goal of increasing data diversity without collecting more data. From the perspective of privacy, data augmentation makes it harder for the model to memorize specific examples \cite{zhang2017understanding} \cite{devries2017datasetaugmentationfeaturespace}, especially in image and text domains.

\item
{\em Adversarial training} trains the model on adversarial examples, i.e., inputs with small, intentional perturbations. As data augmentation, it is intended to improve robustness and generalization (this time, against adversarial attacks). From the privacy angle, both techniques discourage memorizing specific details \cite{milad2018MlMembership}.

\item 
{\em Data blending} blurs the identity of individual training samples, making memorization harder. For instance, you can combine two inputs and their labels linearly (``mixup'') \cite{zhang2018mixup} or cut and paste patches between training images, while mixing labels according (``cutmix'') \cite{yun2019CutMix}.

\item
{\em Label smoothing} replaces hard labels (e.g., [0, 1]) with soft labels (e.g., [0.1, 0.9]) to prevent the model from becoming overconfident \cite{szegedy2016rethinking}. As a form of input perturbation, it reduces the model tendency to memorize exact label associations.

\item
{\em Model compression and pruning:} Since smaller models tend to memorize less, techniques like pruning or knowledge distillation can reduce the model capacity to retain specific training examples, offering privacy as a side effect \cite{usenix2019memorization} \cite{Vitaly2020RequireMemorization}.

\item
{\em Unintended privacy from generalization:} Models that generalize well tend to memorize less. Research has shown that generalization and privacy can be correlated: improving one can help the other, sometimes for free \cite{csf2018mia}.

\end{itemize}

A caveat is in order: Even though all the aforementioned techniques might help improve privacy, they do not guarantee it. These methods reduce the privacy risk but do not eliminate it. Unlike differential privacy, these strategies do not provide formal privacy guarantees in the form of quantifiable privacy bounds. Moreover, they are often context-dependent, what works in one domain (e.g., image classification) may not apply in another (e.g., NLP with LLMs).

Can neural networks really provide privacy for free?
Unintended memorization is a persistent, hard-to-avoid issue that can have serious consequences, e.g., extracting unique secrets such as credit card numbers \cite{usenix2019memorization}. Even well-generalized models can memorize and leak sensitive data. Average data points are rarely leaked, but outliers are frequently subject to memorization and, consequently, privacy leakage. However, removing the layer of outliers that are most vulnerable to a privacy attack exposes a new layer of previously safe points to the same attack (the onion effect of memorization \cite{neurips2022memorization}).
Defending against memorization without training with rigorous privacy guarantees is unlikely to be effective \cite{neurips2022memorization}. In short, privacy for free is possible in some cases, but not reliable without formal mechanisms.

Sometimes, a DP variant of an existing technique can be devised. For instance, a Bayesian interpretation of dropout, whose noise is added with regularization in mind,  can lead to DP guarantees under the zCDP framework \cite{arxiv2017dpdropout}. The original dropout provides a form of randomization akin to DP, with implicit privacy benefits but no formal DP guarantee, even though the privacy-utility trade-off can be studied empirically for particular applications.

It has widely been hypothesized that regularization is a privacy enhancer. A theoretical discussion of sensitivity and stability \cite{dwork2014tcs}, noting that regularization-like constraints (e.g., bounded weights) align with DP goals, provides some theoretical foundation for designing DP strategies (without an empirical study in this case). 

As we already discussed, objective perturbation in SVMs \cite{jmlr2011dperm} uses regularization to achieve DP. Regularization reduces sensitivity in empirical risk minimization (ERM), enabling DP with less noise and improving utility.

Some studies \cite{lomurno2023regularization} have compared DP techniques with traditional regularization methods, such as dropout and L2 regularization, a.k.a. weight decay. Classic regularization
provides similar levels of protection against membership inference and model inversion attacks. Empirical results suggest that regularization methods may offer a more effective trade-off between privacy and utility (i.e., model performance). Unlike DP-SGD, which incurs in significant accuracy loss, regularization techniques can provide privacy protection with a minimal impact on performance.

\subsubsection{Privacy-preserving Deep Learning algorithms}

Several surveys cover recent advances in the application of differential privacy ideas within the field of deep learning \cite{arxiv2020pdlsurvey} \cite{neurocomputing2020ppdl} \cite{neurocomputing2024dpdl} \cite{acmcs2025dpcdl} \cite{das2024advances}.


In supervised learning, differential privacy has been used to train convolutional neural networks (CNNs) using the zero-concentrated (zCDP) \cite{dsc2019dpcnn} and the local (LDP) \cite{iotj2019ldpdl} differential privacy frameworks. Even without a formal underpinning, noise has been added to the inputs of LSTM recurrent neural networks to improve their robustness \cite{arxiv2019dplstm}. 

In distributed settings, multiparty learning mechanisms can make sharing input datasets unnecessary \cite{ccs2015dpdeeplearning}, whereas federated learning techniques can learn a shared model by aggregating locally computed updates \cite{arxiv2016federateddeepnetworks} \cite{aistats2017federateddeepnetworks}.

DarKnight \cite{micro2021darknight} is a framework for large DNN training that relies on cooperative execution between trusted execution environments (TEE) and accelerators, where the TEEs provide privacy and integrity verification, while the accelerators perform the bulk of the linear algebraic computation: customized data encodings based on matrix masking create input obfuscation within a TEE and, then, the obfuscated data is offloaded to GPUs. 


In unsupervised learning, DP techniques have been incorporated into autoencoders, generative adversarial networks, and other deep generative models.


An autoencoder is an artificial neural network used to learn efficient representations of data, typically for the purpose of dimensionality reduction, denoising, or feature learning. An autoencoder consists of two main parts: the encoder compresses the input into a smaller representation (called the latent space or bottleneck), while the decoder reconstructs the original input from the compressed representation. The goal is for the output to be as close as possible to the input, forcing the network to learn the most important features. There are different kinds of autoencoders: denoising autoencoders are trained to remove noise from input data, sparse autoencoders encourage sparsity in the latent space, and variational autoencoders (VAE) learn a probabilistic latent space, which is useful in generative modeling. How can we ensure that autoencoders do not inadvertently leak information about any individual data point in the training set? Apart from DP-SGD, several specific techniques have been proposed:

\begin{itemize}

\item
The deep private autoencoder (dPA) \cite{aaai2016dpautoencoders} enforces $\epsilon$-differential privacy by perturbing the objective function of the traditional
deep autoencoder, rather than its results. Objective perturbation, i.e., adding noise to the loss function, might be preferable to output perturbation, i.e., adding noise directly to the output of the encoder or decoder, which is simpler but may degrade reconstruction quality more significantly.

\item
A differentially-private autoencoder-based generative model (DP-AuGM) \cite{arxiv2018dpgenerative} is designed to defend against model inversion,
membership inference, and GAN-based attacks. An autoencoder is trained with
private data using a differentially private training algorithm and then the encoder is published (the decoder is dropped). New data will be generated (encoded) by feeding the user's own data (i.e., public data) into the encoder and the generated data can then be used to train any ML model. An analogous differentially-private variational autoencoder-based generative model (DP-VaeGM) \cite{arxiv2018dpgenerative} is also robust against membership inference attacks, result that leads to the conjecture that the key to defend against model inversion and GAN-based attacks is not due to differential privacy but the perturbation of training data.

\item
The autoencoder-based differentially private text transformation (ADePT) \cite{arxiv2021dpautoencoder} is a utility-preserving differentially
private text transformation algorithm designed to offer robustness against attacks and produce transformations with high semantic quality that perform well on downstream NLP tasks.
In this case, the autoencoder first transforms a given text input into some
latent representation. Clipping and noise are applied on the latent sentence representations returned by the encoder. Finally, text generation is performed by the decoder from the noisy latent representation.

\item
PATE-AAE \cite{interspeech2021pateaae}, with an adversarial autoencoder (AAE) that replaces the generative adversarial network (GAN) in a private aggregation of teacher ensembles (PATE), has been proposed to ensure differential privacy in speech applications. Following the PATE scheme (i.e., an ensemble of noisy outputs label synthetic samples and guarantee $\epsilon$-DP), the PATE-AAE framework consists of an AAE-based generator and a PATE-based classifier.

\end{itemize}


Generative adversarial networks (GANs) simultaneously train two competing neural network models. The first model, the generator, learns how to generate realistic synthetic data by mapping its input, which is a randomly sampled noise vector, to samples drawn from the distribution of its training data. The second model, the discriminator, learns how to distinguish between samples drawn from the training dataset and samples generated by the generator model, i.e., between real and generated data. Both models are trained until the generator produces data indistinguishable from real data. From the privacy point of view, the main problem with GANs is that the density of the learned generative distribution could concentrate on the training data points, memorizing specific training samples. Differential privacy can be embedded within the GAN discriminator training process, typically via DP-SGD with moments accountant:

\begin{itemize}

\item
Differentially Private Discriminator (DP-GAN) \cite{arxiv2018dpgan} is applied to the discriminator, the model that accesses the real data. The generator is trained using the feedback from the discriminator, never touching the real data.

\item
DP-WGAN \cite{arxiv2018dpgan} \cite{dpsdc2019dpwgan} addresses the challenges of instability in GAN training using the Wasserstein distance function as the GAN training objective.

\item
PATE-GAN \cite{iclr2019pategan} trains multiple discriminators (teachers) on disjoint subsets of the data. Then, their noisy, aggregated outputs are used to train a student discriminator. The the student discriminator trains the GAN generator.
Empirical studies \cite{arxiv2021dpsgdpategan} have shown that PATE, like DP-SGD, has a disparate effect on the under/over-represented classes (DP-WGAN evens the classes while PATE-GAN increases the imbalance).
For PATE, unlike DP-SGD, the privacy-utility trade-off is not monotonically decreasing but is much smoother and inverted U-shaped, meaning that adding a small degree of privacy actually helps generalization.

\item
G-PATE \cite{neurips2021gpate} is another framework based on PATE that trains a student data generator with an ensemble of teacher discriminators.
Private aggregation among different discriminators ensures strong privacy guarantees. 

\item
DataLens \cite{ccs2021datalens} generates synthetic data in a differentially-private (DP) way given sensitive input data. Also combining PATE with GANs, multiple discriminators are trained as teacher models and teachers vote using gradient compression techniques to train a student generator.

\item
GANobfuscator \cite{tifs2019ganobfuscator} uses GANs to obfuscate sensitive data, essentially generating synthetic data that mimic the statistical properties of the original data while preserving privacy. Sensitive data are fed into a discriminator with a privacy-preserving layer. As in previous proposals, random perturbation is applied to the discriminator.

\item
DTGAN \cite{arxiv2021dtgan} is specifically designed to generate synthetic tabular data with strong privacy guarantees. Based on a conditional Wasserstein GAN architecture, DTGAN confirms that applying DP-SGD on the discriminator achieves better data utility and training stability under strict privacy budgets and demonstrates strong resistance to membership inference attacks, keeping success rates close to random guessing ($\approx 50\%$).

\end{itemize}

Empirical studies \cite{ccs2024dpgenerative} profile how DP generative models for tabular data distribute privacy budgets across rows and columns. Since deep generative models spend their budgets per iteration, their behavior is less predictable with varying
dataset dimensions, but they are also more flexible. Probabilistic graphical models (e.g., MST or PrivBayes) outperform GANs at simple tasks like capturing statistics or marginal similarity, yet GANs perform better at more complex tasks like classification (and they are also more scalable than their probabilistic counterparts).

\subsection{Differential privacy in Large Language Models}

Large language models (LLMs) are trained with millions of gradient updates, each potentially leaking a small amount of information, so privacy accounting is paramount in LLMs.
Fortunately, subsampling amplifies privacy, reduces per-step privacy cost, and enables tighter accounting in LLM training.
The combined effect of privacy accounting and subsampling allows training LLMs with strong privacy guarantees.

Apart from the issues discussed in the previous section, which LLMs share with other deep learning models, LLMs introduce new privacy challenges due to their scale and multi-stage training and deployment pipelines. Their scale reinforces memorization. Modern LLMs are trained using massive textual corpora and refined through supervised fine-tuning, several flavors of reinforcement learning, and direct preference optimization (DPO). When deployed, their behavior is improved with techniques such as retrieval-augmented generation (RAG). Each additional stage in the LLM pipeline introduces potential privacy risks and increases the risk of leaking sensitive data, including data used to train the LLM and data from external RAG data sources. Those stages create additional attack surfaces that can be exploited, from straightforward gradient perturbation techniques to tailored lifecycle-aware strategies.

This section reviews differential privacy across the different stages of the LLM lifecycle. We first analyze LLM-specific privacy risks. Then, we review data-centric approaches, differentially-private fine-tuning strategies, privacy-preserving knowledge transfer, and privacy considerations in RAG. Finally, we discuss scaling constraints and fundamental limitations that affect privacy guarantees in LLMs.

\subsubsection{LLM-specific privacy challenges}

While DP-SGD \cite{abadi2016dpsgd} provides theoretical privacy guarantees, its high computational overhead makes it impractical at LLM scale. Even worse, scaling laws for differentially-private language models show DP-SGD leads to a sharp degradation in utility as model size and dataset scale increase \cite{pmlr-v267-mckenna25a}. Consequently, the focus has shifted focus towards differentially-private fine-tuning \cite{yu2022differentiallyprivatefinetuninglanguage} and parameter-efficient fine-tuning (PEFT) methods \cite{higashiDpPET2025}, such as LoRA adapters \cite{hu2021loralowrankadaptationlarge}.

LLMs are typically trained on massive Web-scale text corpora that include sensitive personal information. Unstructured natural language texts can leak private information through semantic inference and contextual cues, even when explicit personally identifiable information (PII) such as names or addresses is removed. Textual anonymization techniques are often insufficient, leading to unintended memorization and information leakage during both training and inference \cite{shi-etal-2022-selective}.

LLMs are increasingly embedded in complex pipelines that include retrieval-augmented generation (RAG), in-context learning, and tool-augmented reasoning. These pipelines introduce novel privacy risks through sensitive prompts, retrieved documents, and generated outputs that traditional DP training and fine-tuning mechanisms fail to address. Recent work has explored privacy interventions at the retrieval and inference stages, including differentially-private document selection and token-level aggregation during generation \cite{grislainRAG2025}. 

Other studies reveal the fragility of classical differential privacy assumptions when applied to unstructured text, such as record-level independence and well-defined neighboring datasets. Word-level differentially-private text sanitization mechanisms, while offering formal guarantees, are susceptible to contextual vulnerabilities: randomization at the word level leaves residual semantic and contextual clues that LLMs can exploit to infer the original content, enabling effective reconstruction attacks and undermining empirical privacy protections \cite{meisenbacherDouble-edge2025}. User-level contributions, correlated textual records, and long range dependencies in natural language complicate accurate privacy accounting.

Collectively, the aforementioned issued create unique research challenges for differential privacy in LLMs, beyond the mere scaling of deep learning techniques. Privacy in this domain demands a holistic approach that encompasses data curation,
model fine-tuning, deployment-aware pipelines, and evaluation under realistic threat models, particularly when constrained by fixed computational budgets and the need for user-level guarantees \cite{charles2025uls}.

\begin{table}[t]
\centering
\begin{tabular}{lll}
\hline
    \textbf{Pipeline stage} & \textbf{Privacy risk} & \textbf{DP strategies} \\ 
    \hline
    Pretraining \& data curation & Memorization of sensitive text & Selective DPSGD \cite{shi-etal-2022-selective} \\
     & & Local DP sanitization \cite{yue-etal-2021-differential} \\
     & & Embedding anonymization \cite{yuTextualDpContext2024} \\
     & & Contextual vulnerability mitigation \cite{meisenbacherDouble-edge2025} \\
    Fine-tuning & Memorization of sensitive text & DP-enhanced PEFT \cite{yu2022differentiallyprivatefinetuninglanguage} \cite{higashiDpPET2025} \\
    & Membership inference attacks & Loss landscape flatness \cite{chenPpMia2025} \\
    & Reconstruction attacks & HardLLM \cite{dengHardeningLLM2025} \\
    & Linguistic steganography \cite{coffey2024differential} & \\
    Knowledge transfer & Leakage from teacher model & DP distillation \cite{gargTaskSpecificK2024} \\
    RAG & Exposure of sensitive data & DP-RAG \cite{grislainRAG2025} \\
    Inference \& interaction & Prompt-based leakage & Context-aware textual DP \\
    System-level constraints & Utility collapse at scale \cite{pmlr-v267-mckenna25a} & Compute-aware DP accounting \cite{charles2025uls} \\
    \hline
\end{tabular}
\caption{Privacy risks and differential privacy in modern LLM pipelines.}
\label{tab:llm-dp-pipeline}
\end{table}

\subsubsection{Data-centric approaches}
\label{sec:llm:data-curation}

Data-centric approaches to differential privacy emphasize protecting or sanitizing textual data prior to or during processing, thereby addressing the unique privacy challenges posed by unstructured text, such as sparse sensitive information and long range semantic dependencies. 

A key innovation in this area is the notion of selective differential privacy, which provides rigorous guarantees on sensitive data portions rather than applying uniform protection across all training data. Selective DPSGD \cite{shi-etal-2022-selective}, proposed for RNN-based language models, identifies and privatizes sparse private elements (e.g., personally identifiable information).

In text sanitization under a local DP framework \cite{yue-etal-2021-differential}, sensitivity and semantic similarity are jointly considered to generate human-readable sanitized texts. By replacing sensitive terms with semantically-close alternatives under calibrated noise, this method enables privacy preserving text analytics without requiring global DP assumptions, 

For context-aware reasoning in LLMs, prompts and retrieved contexts can be safeguarded before transmission to service providers. Differential embedding hashing \cite{yuTextualDpContext2024} anonymizes sensitive embeddings while preserving reasoning capabilities, accompanied by a privacy loss quantification scheme that evaluates the privacy utility trade-off in applications like RAG. 

However, empirical analyses reveal limitations in word-level DP sanitization mechanisms, where randomization leaves residual contextual clues exploitable by reconstruction attacks. This ``contextual vulnerability'' undermines empirical privacy despite formal guarantees: LLMs can infer original semantics from sanitized texts, though they can also be leveraged for post sanitization refinement to enhance both privacy and utility \cite{meisenbacherDouble-edge2025}. 

\subsubsection {Differentially-private LLM fine-tuning}

Pretrained LLMs can be adapted to downstream tasks with privacy guarantees using parameter efficient fine-tuning (PEFT) techniques \cite{yu2022differentiallyprivatefinetuninglanguage} in order to mitigate privacy risks such as membership inference attacks (MIAs) \cite{sp2017mia} and data memorization. PEFT methods, like low-rank adaptation (LoRA) \cite{hu2021loralowrankadaptationlarge}, can achieve strong privacy-utility trade-offs by adding calibrated noise during gradient updates without full model retraining. Additional mechanisms to prevent data leakage during fine-tuning, such as adaptive noise injection based on sensitivity analysis and real-time monitoring of privacy budgets, help maintain model performance while limiting information disclosure \cite{Xiao_Zhang_Chen_Ren_Zhang_Xu_2025}. Empirical evaluations confirm the effectiveness of DP-enhanced PEFT for LLMs, showing that LoRA combined with DP-SGD can preserve competitive accuracy under moderate privacy levels (e.g., $  \epsilon \leq 8  $), although utility drops sharply for stricter bounds \cite{higashiDpPET2025}.

Approaches that promote loss landscape flatness during DP fine-tuning \cite{chenPpMia2025} address specific vulnerabilities like MIAs: sharpness-aware minimization encourages robust minima that resist adversarial inference on training data membership. LLMs can also be fine-tuned against linguistic steganography \cite{coffey2024differential}.
Comprehensive LLM hardening strategies \cite{dengHardeningLLM2025}, from differentially private data selection 
to trustworthy model quantization, can reduce deployment footprints while preserving DP guarantees and defending against reconstruction attacks.

\subsubsection{Privacy-preserving knowledge transfer and distillation}

Knowledge transfer techniques such as distillation provide an effective means to compress LLMs. Incorporating DP into model distillation enables the creation of compact, efficiently-deployable models that mitigate leakage risks from sensitive training data: a pre-trained teacher model can transfer domain-specific knowledge to compact student models under privacy constraints \cite{gargTaskSpecificK2024}. 
Private fine-tuning using DP-SGD followed by private task-specific distillation, also with DP-SGD, ensures differential privacy.
Experimental evaluations on the GLUE benchmark 
\cite{wang2019gluemultitaskbenchmarkanalysis} 
show that accuracy results are comparable to non-private distilled models. 

\subsubsection{Differential privacy in retrieval-augmented generation (RAG)}

When using RAG, external data sources and user prompts introduce additional exposure vectors for sensitive information. 
In privacy-aware retrieval and generation techniques \cite{grislainRAG2025}, during the retrieval stage, an exponential mechanism selects documents from a knowledge base with calibrated noise to ensure indistinguishability between neighboring databases, while maintaining relevance through semantic embeddings. In the generation phase, differentially private in context learning (DP-ICL) aggregates retrieved contexts at the token level to bound privacy loss, preventing inadvertent disclosure of confidential data.

In user prompts, sensitive embeddings can be anonymized using hashing algorithms \cite{yuTextualDpContext2024}. 
Additionally, word-level sanitization can mitigate contextual vulnerabilities \cite{meisenbacherDouble-edge2025}.
In short, the techniques used to protect training data (Section \ref{sec:llm:data-curation}) can be adopted to protect the external data sources used by RAG, as well as the user's own prompts.

\subsubsection{Scaling laws, computational constraints, and fundamental limitations}


Empirical and theoretical scaling laws for differentially-private language models reveal a pronounced degradation in utility as parameters and data volumes scale. ($  \epsilon  $, $  \delta  $)-DP calibrated-noise can overwhelm the signal in gradient updates, leading to a regime where model perplexity scales suboptimally, often plateauing or diverging from non-private counterparts. For instance, under realistic privacy budgets ($  \epsilon \approx 1-10  $), utility collapses beyond 1B parameters,
with the additional computation required to match non-private performance sometimes exceeding factors of 100x \cite{pmlr-v267-mckenna25a}. These results underscore the tension between privacy amplification via subsampling and the diminishing returns in large-scale pretraining, where privacy costs dominate learning dynamics.

User-level DP, where privacy guarantees apply per user rather than per example (i.e., user-level sampling and per-user gradient clipping instead of example-level sampling and per-example gradient clipping), has also been studied in LLMs \cite{charles2025uls}. In both synthetic and realistic fine-tuning tasks, ULS outperforms ELS in just about all settings. Moreover, the gap between them grows as the compute budget grows. Therefore, ULS is preferable with large compute budget settings.


Further research is needed on novel accounting methods and model-specific optimizations to surmount scaling barriers and extend DP applicability to ultra large LLM systems.

\subsection{Model-agnostic differential privacy} 


Different approaches can be applied to make any machine learning technique differentially private, including federated learning and PATE (Private Aggregation of Teacher Ensembles). 

\subsubsection{Differentially-Private Federated Learning (DP-FL)}

Federated learning is a machine learning technique that enables multiple devices or servers (often called clients) to collaboratively train a shared model without sharing their raw data. Instead of sending data to a central server, each client trains the model locally on its own data and only shares model updates (such as gradients or weights) with a central server. The server then aggregates these updates to improve the global model. Since raw data never leave the local device, federated learning is privacy-preserving by design.

DP implementations can follow two different strategies: central DP \cite{dwork2006dp}, when users trust the data curator, and local DP \cite{focs2008local} \cite{pods2003local}, which ensures privacy when clients do not trust the data curator. Both approaches can be used for DP federated learning. In local DP-FL, clients add noise to their model updates (or falsify their answers with some probability) before sending them to the server/curator. In central DP-FL, the server adds noise during the aggregation of updates.

Federated Learning (FL) enables collaborative model training across decentralized clients without centralizing sensitive datasets \cite{aistats2017federateddeepnetworks} by aggregating locally-computed updates to optimize a global objective (i.e., privately training ML models). Differentially-Private Federated Learning (DP-FL) \cite{arxiv2018dpfl} incorporates the principles of DP-SGD into the private training of deep learning models. 

Not sharing raw data is not enough for strong privacy protection. A survey of DP-FL ideas \cite{access2022dpflsurvey} organizes existing proposals into three loose categories. Some proposals try to protect users' privacy during model training by using mechanisms such as secure multiparty computing (SMC) \cite{arxiv2016smc} \cite{sp2017smc} or different cryptographic techniques (e.g., using homomorphic encryption \cite{nn2020homomorphic} or functional \cite{tnse2021ldpfunctional} encryption). A second set of proposals act on the network architecture of the FL system (e.g., introducing proxies, a latent layer, or P2P communication) or constrain communication (e.g. sharing partial information \cite{ccs2015dpdeeplearning}). A third category focuses on controlling resource consumption (e.g., transmission rate allocation and communication overhead) in different contexts (over multiple access channels and on the Internet of things, IoT, respectively).

Since FL requires multiple rounds of synchronization, some privacy accounting is also needed. For instance, several techniques have been proposed for adjusting the variance of the Gaussian noise to guarantee that the privacy loss  does not exceed the desired privacy budget $\epsilon$ after a given number of synchronization rounds using LDP \cite{icassp2021flldp} \cite{tmc2022udp}.

DP-FL \cite{arxiv2018dpfl} enables a set of clients, each with a private dataset, to collaboratively train a shared global model while ensuring client-level $(\epsilon, \delta)$-differential privacy using a distributed version of DP-SGD \cite{abadi2016dpsgd}. In each communication round, a random subset of clients is sampled to receive the current global model. Then, each client optimizes the model using its local data. Each local update of the model parameters is clipped to bound sensitivity, and the global model is updated by adding calibrated Gaussian noise to the averaged clipped updates.

The cost of privacy can be seen as the difference between the fitness of a privacy-preserving machine-learning model and the fitness of trained machine-learning model in the absence of privacy concerns. In DP-FL, this cost has an upper bound that is inversely proportional to the combined size of the training datasets squared and the sum of the privacy budgets squared \cite{tifs2021asynchronous}. 


Combining differential privacy with secure multiparty computation enables the reduction of noise injection as the number of parties increases without sacrificing privacy \cite{aisec2019ppfl}. This combination protects against inference threats over both the messages exchanged during model training and the final trained model. It can be used to train a variety of ML models, including decision trees, CNNs, and SVMs.


Frameworks like FedDPDR-DPML \cite{cmc2024dpsvm} combine DP-PCA and objective-perturbed DP-SVM training across multiple participants holding distributed data. Models are trained locally with DP, and a central server aggregates them to produce a global model (e.g., using weighted averaging based on local data size). 


As in central DP, regularization can act as a proxy for local differential privacy.
In federated learning, regularization stabilizes local models before DP aggregation. We can apply regularization techniques to user-level models before privatizing outputs (via randomized responses or added noise). Regularization ensures models generalize across users, reducing overfitting to individual data and thus enhancing privacy. 
In fact, since the first proposals on training deep networks from decentralized data it was conjectured that, in addition to lowering communication costs, model averaging produces a regularization benefit similar to that achieved by dropout \cite{aistats2017federateddeepnetworks}.


The distinction between example-level sampling (ELS) and user-level sampling (ULS) is especially important in federated learning:

\begin{itemize}

\item
Example-level sampling (ELS) randomly selects individual examples, i.e. each individual data point is sampled independently with some probability, regardless of which user it came from.
It is the common approach in centralized training, e.g. DP-SGD \cite{abadi2016dpsgd}, where mini-batches are formed by randomly selecting examples from the entire dataset, and per-example gradient clipping is applied to limit the influence of any single example.
ELS, therefore, protects individual examples.
It is simpler to implement and often yields better utility when users contribute a few examples.
However, ELS often leads to weaker privacy guarantees in federated settings, where users may contribute many examples. Additionally, ELS is vulnerable to correlated data leakage from the same user (it does not protect user-level privacy if users contribute multiple examples).

\item
User-level sampling (ULS) randomly selects entire users and all their data \cite{levyLearningUlp2021}. Entire users (and all their data) are sampled with some probability. When a specific user is selected, all their data is used in that round.
ULS is common in federated learning \cite{mcmahanlearning2018}, specially when data is naturally partitioned by user (e.g., mobile devices, health and financial data, or even LLMs).
ULS protects all users and all their data. 
It provides stronger privacy guarantees than ELS and it is more realistic for real-world privacy concerns. It is better suited for real-world applications where users contribute multiple, possibly correlated, examples.
However, ULS requires more sophisticated privacy accounting (e.g., f-DP, Edgeworth Accountant, or user-level RDP).

\end{itemize}

Whereas ELS is simpler and fits centralized training, ULS is more suitable for real-world privacy expectations, especially in federated settings. The choice affects both the privacy guarantees and the accounting method used. 

Empirical experiments on fine-tuning LLMs \cite{charles2025uls} found that ELS can outperform ULS in terms of utility when users contribute few examples or when computation budgets are tight (i.e., limited computational resources). On the other hand, ULS is superior when strong privacy guarantees are needed, users contribute many or diverse examples, and the computation budget allows for more sophisticated training.



\subsubsection{Private Aggregation of Teacher Ensembles (PATE)}

The PATE framework is a method for training ML models with strong differential privacy guarantees. 
PATE trains multiple models (teacher models) on disjoint subsets of the training data. Teachers models are aggregated to form an ensemble whose predictions are guaranteed not to leak those of individual models. The noisy aggregation of the teacher model outputs is then used to train a student model. This student model is learnt without direct access to the raw data nor the parameters of individual teacher models. The resulting student model can be shared and, due to the aggregator’s guarantee, is theoretically guaranteed not to leak information (from a DP point of view). 


The federated learning methods of the previous section protect the privacy of data privacy for multiple users in distributed ML, but the encryption algorithm and security protocols needed increase their computational overhead. PATE also provides data protection with some overhead, since it needs to train an additional student model. Training is more complex but can offer stronger privacy guarantees.

\begin{itemize}


\item
The original PATE \cite{iclr2017pate} uses an ensemble of teacher models trained on disjoint subsets of sensitive data. These teachers vote on labels for unlabeled public data, and a student model is trained on this labeled public data. Laplacian noise is added to the teacher votes to ensure $(\epsilon, \delta)$-differential privacy.

The noisy aggregated labels are used to train a student model, effectively shielding individual data contributions while preserving utility. A key detail is the use of a confidence-based thresholding mechanism, so that predictions with sufficient agreement among teachers are released, reducing noise variance and enhancing the quality of the transferred knowledge. PATE offers good performance when public unlabeled data is available.



The canonical PATE computes output labels by aggregating the predictions of a (potentially distributed) collection of teacher models via a voting mechanism. However, the addition of noise attains DP guarantees, makes PATE predictions stochastic, and also enables new forms of leakage of sensitive information \cite{neurips2022pate}.
For a given input, an adversary can exploit the stochastic PATE predictions to extract high-fidelity histograms of the votes submitted by the underlying teachers. From these histograms, the adversary can learn sensitive data, even when DP guarantees are not violated. Counterintuitively, this attack becomes easier as we add more noise to provide a stronger differential privacy. 
When learning from sensitive data, care must be taken to ensure that training algorithms address privacy norms and expectations.

An empirical evaluation of PATE robustness against adversaries using Monte Carlo sampling \cite{msc2024patemontecarlo} shows that teacher votes can be inferred. Experimental results confirm that adversaries are more successful in recovering voting information when the vote-aggregation mechanism introduces noise with a larger variance, revealing a trade-off between achieving confidentiality and differential privacy in collaborative
ML. Unfortunately, in practice, PATE provides virtually no privacy guarantees when an adversary is allowed unlimited access to the trained model.


\item
Individualized/personalized PATE (iPATE) \cite{arxiv2022ipate} adapts the privacy mechanism to provide personalized privacy budgets. Standard PATE assumes uniform privacy needs across all data points. However, in real-world scenarios, some individuals may require stronger privacy than others (e.g., when dealing with healthcare data, some patients may consent to broader data use, while others require strict privacy).

In iPATE, each data point (or individual) is assigned a customized privacy budget. The noise added during aggregation is adjusted based on the individual's privacy requirements. This adjustment allows for more utility (better model performance) while still respecting individual privacy constraints.
iPATE \cite{arxiv2022ipate} 
variants include:

\begin{itemize}

\item
{\em Upsampling PATE} replicates data points with higher privacy budgets across multiple teacher models. Unlike the original PATE, whose data partitions were disjoint, upsampling allows data with higher privacy budgets to be learnt by several teachers.
Upsampling increases the influence of less privacy-sensitive data while preserving privacy for more sensitive data. Upsampling either increases the number of teachers or allocates more data per teacher, with the former generally providing better utility by reducing vote count variance.

\item
{\em Vanishing PATE} reduces the influence of data points with stricter privacy requirements by limiting their participation in teacher training. Therefore, the privacy cost is minimized for sensitive individuals.

\item
{\em Weighting PATE} assigns weights to teacher votes based on the privacy budget of the data they were trained on. Weighting balances utility and privacy by giving more influence to less sensitive data. Instead of changing the composition of teacher model training datasets, as in the previous variants, the weighting mechanism is applied during the aggregation of teacher votes. As a result, teachers handling data with higher privacy budgets are given more influence, improving the accuracy of the final student model.

\end{itemize}

Given a set of teacher models, the aggregation mechanism in iPATE assigns a personalized privacy budget $\epsilon_i$ to each user, rather than enforcing a global $\epsilon$.
The aggregation mechanism adds personalized Laplacian or Gaussian noise to each teacher vote, calibrating that noise to an individualized sensitivity function that is adaptively determined from the user's privacy budget $\epsilon_i$.
By adapting noise levels individually, iPATE balances privacy and utility on a per-user basis.


\item  
Scalable PATE (sPATE) \cite{iclr2018spate} extends PATE to large scale tasks and uncurated, unbalanced training data with errors. 
The original PATE can be computationally expensive because many teacher models must be trained and their votes must be aggregated. Scalable PATE introduces some optimizations to make the framework more efficient and applicable to larger 
datasets. 

sPATE resorts to techniques such as parallel teacher training, model compression, and efficient vote aggregation. Training teacher models in parallel speeds up the process. Model compression reduces the size or number of teacher models. More efficient vote aggregation can be devised to reduce computation requirements. Specifically, sPATE introduces new noisy aggregation mechanisms for the ensemble of teacher models that require less noise to be added to provide privacy guarantees. The Confident Aggregator and the GNMax (Gaussian Noise Maximum) replace the LNMax (Laplace noise maximum) mechanism in the original PATE:


\begin{itemize}


\item
The Confident Aggregator \cite{iclr2018spate} only answers student queries when teacher votes are confident. As a benefit, privacy loss is reduced by avoiding noisy answers when teachers disagree.


When a query is made, we first check whether the consensus among teachers is sufficiently high. If the number of votes assigned to the most popular label among teachers (or the vote margin, i.e., the difference between the top two vote counts) exceeds a predefined threshold, we accept the query. If not, we reject it. The threshold itself is randomized in order to provide privacy during this checking process. 

Once a query has been accepted, the original noisy PATE aggregation mechanism is applied. Noise is added to each of the vote counts corresponding to each label and the label with more votes is returned. When the votes are not confident and the query is rejected, no label is returned.

The final result is a reduction of the number of queries that consume the privacy budget unnecessarily. 
Ambiguous or low-consensus queries are avoided to prevent wasting the privacy budget. In other words, the Confident Aggregator filters the queries that would actually deplete the privacy budget in the original PATE framework.
For comparable levels of student performance, the privacy budget afforded by the Confident Aggregator is smaller than in the original PATE noisy aggregation mechanism.


\item
The Gaussian Noisy Maximum Aggregator (GNMax) uses Gaussian noise instead of Laplace noise in the vote aggregation phase. After teachers vote, the vote tally is adjusted with Gaussian noise and the label with the highest noisy vote count is chosen.



The Gaussian distribution is more concentrated than the Laplace distribution used in the original PATE framework. Compared to Laplace noise, Gaussian noise often results in less distortion, especially when the number of classes is large. 
In addition, the chance of teacher consensus is increased by using more concentrated noise.
Under Rényi Differential Privacy (RDP), which is more suitable for composition over many queries, tighter privacy bounds can be provided for Gaussian noise. 


GNMax can be combined with smooth sensitivity analysis to adaptively adjust noise based on the sensitivity of each query. When using in conjunction with a confident aggregator, the Confident GNMax only releases labels when the noisy vote margin exceeds a threshold. There is also an Interactive GNMax that dynamically adjusts the noise or decision threshold based on query history. Yet another variant, called PV-PATE \cite{apweb2023pvpate}, uses an Improved Confident-GNMax Aggregator to further refine query filtering.

\end{itemize}

\end{itemize}



The PATE framework transfers the knowledge of an ensemble of teacher models to a student model, with intuitive privacy provided by training teachers on different subsets of data and strong privacy guaranteed by noisy aggregation of teachers' answers. 
The PATE framework has been combined with multiple techniques for different applications, including adversarial autoencoders for speech classification \cite{interspeech2021pateaae} and generative adversarial networks, in PATE-GAN \cite{iclr2019pategan}, G-PATE \cite{neurips2021gpate}, and DataLens \cite{ccs2021datalens}. 

Existing PATE variants, however, assume that teachers, who cast their votes, are honest. When they are corrupted by an adversary, attacks can become drastically more powerful as the attacker is able to control more participants \cite{msc2024patemontecarlo}. Defending against this kind of attacks might not be trivial. Even though PATE offers DP guarantees, its robustness is in question. Several vulnerabilities have been found (e.g., vote inference and corrupt teachers) and stronger noise mechanisms and vote aggregation strategies are needed.

%% file: dp-in-practice.tex
\section{Differential privacy in practice: Empirical evaluation of differential privacy methods}
\label{chapter:results}

\input{tableAlgorithms}

Measuring the effectiveness of the algorithmic techniques proposed to achieve differential privacy involves evaluating both their privacy guarantees and their utility (i.e., how well they preserve the usefulness of the data). Differential privacy is inherently a trade-off between these two aspects. Precise measurement requires a combination of theoretical analysis, empirical evaluation, and statistical tools. 

Table \ref{tab:algorithms} and its continuation, Table \ref{tab:algorithmsDL}, summarize many of the ML algorithms discussed in the previous Section, indicating the DL framework they are based on and the mechanism they employ to add differential privacy to the ML models they train (i.e., input, mechanism design, objective, and output perturbation, as well as the sample and aggregate strategy common in parallel and distributed ML). Those tables also report how privacy effectiveness, performance/utility, and the privacy-utility trade-off are evaluated through experiments in the publications where the different privacy-preserving algorithms were originally proposed.

\subsection{Measuring privacy effectiveness}


Most publications on DPML techniques tend to ignore the empirical evaluation of privacy, since it is theoretically guaranteed by the corresponding DP mechanism. They just report the calculated $(\epsilon,\delta)$ or other DP parametersachieved by the DP model, based on the DP mechanism and privacy accountant used  (e.g., RDP, zCDP).



However, implementing DP algorithms correctly requires careful attention to details, including numerical stability (e.g., avoiding floating-point vulnerabilities \cite{mironov2012ccs}), proper sensitivity analysis, and correct privacy accounting \cite{arxiv2022lessonslearned}. Subtle algorithmic errors or analytical mistakes, such as incorrect noise calibration or the misuse of composition rules (e.g., most variants of SVT are actually not private \cite{arxiv2016svt}), can silently undermine the claimed privacy guarantees \cite{neurips2022auditingdpml}. 

Even when the DP analysis and implementation are correct, a timing side-channel attack \cite{haeberlen2011usenix} may lead to the catastrophic compromise of a DP system.\footnote{DP guarantees are defined in the language of probability distributions. DP implementations use floating-point arithmetic precision. A single sample from a naïve implementation of the Laplace mechanism allows distinguishing between two adjacent datasets with probability more than 35\% \cite{mironov2012ccs}. Unlike operations on integers that are typically constant-time on modern CPUs, floating-point arithmetic exhibit input-dependent timing variability. Covert channels are notoriously
difficult to avoid and are devastating for DP: ``if a side channel allows the adversary to learn even a single bit of private information, the differential privacy guarantees are already broken!'' \cite{haeberlen2011usenix}. The snapping mechanism \cite{mironov2012ccs} enforces a discretized and bounded output domain and carefully rounds floating-point computations, ensuring that the implemented mechanism faithfully preserves the theoretical differential privacy guarantees.}

While DP provides theoretical upper bounds on privacy loss $(\epsilon,\delta)$, these bounds are worst-case estimates. Effectiveness in terms of privacy should be empirically measured by ensuring the algorithm adheres to the  $(\epsilon,\delta)$-DP definition and by analyzing how tightly this guarantee is achieved. Auditing DP implementations can help detect bugs and also potential misuse \cite{neurips2022auditingdpml}. 

The empirical validation of DPML techniques can be performed by simulation studies for testing DPML algorithms and adversarial testing for evaluating a particular ML model.


In simulation studies, you generate synthetic datasets $D$ and $D'$ differing in one record. Then, by running the ML training algorithm, you estimate the empirical ratio $P[M(D)\in S]/P[M(D')\in S]$. If it exceeds $e^\epsilon + \delta$, the mechanism is less effective than claimed. You can also vary dataset size, dimensionality, and noise levels to test robustness. Statistical tests (e.g., t-test, Kolmogorov-Smirnov) can be used to compare output distributions $M(D)$ and $M(D')$. Effective DPML algorithms should make these distributions statistically indistinguishable up to the $(\epsilon, \delta)$ bound. Such simulation studies are essential because, even when formal results might imply that learning while preserving differential privacy is possible, theoretical results often depend on technical assumptions, which do not always hold in practice \cite{kifer2011sigmod}.


For testing particular ML models, adversarial testing is a reasonable empirical validation strategy. You can simulate privacy attacks on the ML model to estimate the actual privacy loss and compare it to the theoretical $\epsilon$ it should provide. The experimental evaluation using simulated privacy attacks can provide a lower bound on the actual privacy leakage of a specific DP implementation \cite{jair2023dpfyml}, by measuring the actual success rate of adversarial attacks. 

\begin{itemize}

\item
Membership inference attacks (MIA) \cite{sp2017mia}, sometimes known as tracing attacks \cite{acomp2019dpdlsurvey}, pose a binary classification problem in which the attacker tries to predict whether a particular example was used to train the model from the model outputs (i.e., a black-box attack). They are the most common type of attack for the empirical privacy evaluation of DPML implementations. Attacks typically observe the model behavior on the target: when the model is highly confident or achieves a very low loss on the example, the example is more likely to have been in the training set (due to potential overfitting or memorization) \cite{jair2023dpfyml}. DP effectiveness is high if the attack success rate approaches the random guessing baseline (e.g., 50\% for binary membership).

\item
Attribute inference attacks (AIA) \cite{arxiv2015aia}, also called model inversion attacks \cite{ccs2015modelinversion}, pose a regression problem where the attacker attempts to predict the values of a sensitive target attribute, provided he has black-box access to the trained ML model. The attacker could even infer sensitive attributes that were not used directly as features during model training \cite{eurospw2021risks}.

\item
Data reconstruction/extraction attacks \cite{arxiv2025datareconstructionattacks} try to recover specific training data instances (or parts thereof) from the trained model. 
Privacy is effective if the reconstruction error remains high.

\end{itemize}

Other attacks try to infer specific information from the target model \cite{usenix2019attacks}: memorization attacks try to exploit the ability of high
capacity models to memorize sensitive patterns \cite{usenix2019memorization}, model stealing attacks try to recover the model parameters \cite{usenix2016stealing}, hyperparameter stealing attacks try to recover the hyperparameters used during model training, and property inference attacks try to infer whether the training data set has a specific property.

Common attack strategies include thresholding (loss < threshold or confidence > threshold) \cite{csf2018mia}, shadow models (auxiliary models trained on similar data learn a classifier that distinguishes members from non-members based on the target model output \cite{sp2017mia}), perturbation-based attacks (analyzing the change in loss when the input is slightly perturbed, i.e., the Merlin attack \cite{arxiv2020mia}), and combined attacks (combining loss and perturbation information, i.e., the Morgan attack \cite{arxiv2020mia}).


Attack success can be measured using standard classification metric (e.g., accuracy, precision, recall, AUC) or tailored metrics, such as membership advantage ($MA = TPR - FPR$). The precision, or positive predictive value ($PPV=TP/P$), is suitable for evaluating risk when the base rate of membership might be low (i.e., for unbalanced classes).
Instead of using averages across the whole dataset, attack success should focus on the most vulnerable examples, since privacy is often a per-instance property, e.g., TPR at low FPR for the most vulnerable sample \cite{ccs2024mia}.


Security evaluation curves \cite{ssc2018sec} \cite{pr2018sec} can also be used to evaluate the security of DPML systems. These curves characterize the system performance under different attack strengths and attackers with different levels of knowledge, thus providing a more comprehensive evaluation of the system performance.


Attempts to define purely deterministic privacy metrics (e.g., based on k-anonymity \cite{dalenius1986finding} \cite{samarati1998protecting} or non-stochastic information theory \cite{tifs2019deterministic}) generally fail to provide the robustness of DP, particularly against adversaries with auxiliary information. Some researchers have explored ``noiseless'' privacy by leveraging the inherent randomness in data \cite{arxiv2023empiricaldp}. However, DP strength lies precisely in its probabilistic nature and its ability to handle arbitrary side knowledge.


It should also be noted that the DP definition used by practitioners is not always the same. Neighboring datasets are allowed to change in different ways: add-or-remove one record, zero-out one record, or replace one record. The first two have comparable semantics for a fixed $\epsilon$, whereas the guarantee for replace-one is approximately twice as strong. Care should therefore be taken when comparing $\epsilon$ based on different criteria \cite{jair2023dpfyml}. 

According to the ``How to DP-fy ML'' guidelines \cite{jair2023dpfyml}, a large utility drop can be expected in large ML models under strong formal privacy guarantees ($\epsilon \leq 1$), so large that DP might be infeasible in practice.
However, DP offers a reasonable level of anonymization for many applications under reasonable privacy guarantees ($\epsilon \leq 10$).
DP guarantees are not enough for true data anonymization with a large $\epsilon$ ($\epsilon > 10$), even tough data memorization might be reduced  and, hence, offer some sort of privacy protection.
In any case, any finite $\epsilon$ is always better than no privacy at all. At least, privacy improvements can be quantified.

Additional measures should be considered before releasing or deploying ML models in practice \cite{jair2023dpfyml}, including removing privacy-sensitive training data, demonstrating robustness against known attacks, and privacy auditing. Empirical attacks, however, provide only a lower bound on privacy leakage. They demonstrate vulnerability against known attack strategies but do not guarantee protection against future, more sophisticated attacks. Their effectiveness also depends heavily on the adversary's assumptions and capabilities \cite{arxiv2024mia}.


Rather than selecting attacks in an ad hoc manner, structured privacy auditing protocols ensure coverage of realistic threat models, reproducibility, and comparability across deployments \cite{sp2017mia} \cite{usenix2019memorization}.
Auditing protocols for DPML systems typically include the following components:

\begin{itemize}

\item
{\it Threat model specification:} The evaluation should explicitly define the adversary’s goals, access level, and auxiliary knowledge \cite{sp2017mia} \cite{nasr2019comprehensive}. Clear threat model specification is central to empirical privacy analysis

\item
{\it Attack suite selection:} A representative set of attack variants should be executed. For example, membership inference evaluations should combine threshold-based, shadow model, and perturbation-based attacks to capture different leakage channels \cite{sp2017mia}. Attribute inference and reconstruction attacks should be included when sensitive attributes or generative models are involved \cite{ccs2015modelinversion} \cite{usenix2019memorization}.

\item 
{\it Canary and memorization testing:} Canary tests involve inserting unique or rare sequences into training data and probing whether these canaries can be extracted during inference. Canary-based evaluation has been used to demonstrate memorization in large models and to measure leakage \cite{carlini2021extracting}.

\item 
{\it Control of training confounders:} Experimental factors that influence privacy loss independently of the DP mechanism should be systematically controlled. For example, early stopping and overfitting have been shown to correlate strongly with privacy leakage \cite{song2020overlearningrevealssensitiveattributes}. Robust empirical evaluation must control hyperparameter tuning, dataset imbalance, and model capacity effects to avoid misleading conclusions \cite{jagielski2020auditing}.

\item 
{\it Evaluation metrics and reporting:} Attack performance should be measured using both global and worst-case metrics, including ROC-AUC, membership advantage, and precision-recall curves. Worst case vulnerability metrics help reflect the inherently worst-case nature of differential privacy \cite{csf2018mia}.

\item 
{\it Robustness analysis:} Auditing should be repeated across multiple privacy budgets, dataset sizes, and model architectures to assess the stability of empirical privacy under realistic deployment variations \cite{jagielski2020auditing}. 

\end{itemize}

Structured auditing protocols do not replace formal DP guarantees, but they provide an essential empirical validation layer that increases confidence in real-world deployments and helps identify implementation errors, modeling artifacts, or unexpected leakage channels \cite{usenix2019memorization} \cite{nasr2019comprehensive}.

Even under DP, classifiers can be used to infer private attributes accurately in realistic data, given that coarse properties of the population taken together can be combined to build a model that can be applied to individuals with high accuracy. Existing anonymization methods tend to ignore this issue. Why? Because attacks are inherently probabilistic \cite{kdd2011attack}. Although we can empirically determine that some attacks are correct in associating an individual with their true value a large fraction of the time, they do not indicate with certainty for which individuals this is a true inference and for which it is a false positive. The classifier might often be right, but we are never sure when.

In any case, it is not sufficient to naively apply differential privacy to data, and assume that this is sufficient to address all privacy concerns. You should always evaluate the privacy leakage for a given privacy budget $\epsilon$ before publishing a dataset or a ML model \cite{access2022dpflsurvey}.

\subsection{Measuring utility effectiveness}

In differential privacy, utility refers to how well the algorithm preserves the performance or usefulness of the data analysis. The metrics used to estimate utility depend on the specific task to be done, from statistical queries and machine learning to synthetic data generation.


The standard experimental methodology for evaluating DPML algorithms typically involves training a non-private version of the ML model on the target dataset to establish a baseline utility performance (baseline model) and training one or more versions of the DPML algorithm on the same dataset (DP models), varying the privacy parameters (e.g., $\epsilon$, $\delta$, noise multiplier, clipping threshold) and other learning algorithm hyperparameters to explore different points on the privacy-utility spectrum. Then, the utility of both the baseline and the DP models are evaluated on a held-out test dataset using standard ML metrics relevant to the task at hand.

Assessing the utility of a DPML model depends heavily on its intended use case.
For ML models, standard performance metrics are often used. In classification problems, common metrics include accuracy, precision, recall, F1-score (often macro-averaged for multi-class problems), the whole confusion matrix, the ROC curve, and the area under the curve (AUC) for both ROC and precision-recall curves. In regression problems, common metrics include Root Mean Squared Error (RMSE), Mean Absolute Error (MAE), or Mean Absolute Percentage Error (MAPE). For special-purpose applications (e.g., synthetic data generation using generative AI techniques), customized metrics are also used, such as normalized mutual information (NMI), the Inception score (IS) \cite{nips2017wgan}, Jensen-Shannon divergence (JS), total variation distance (TVD), Kullback–Leibler divergence (KL), Wasserstein distance (W), dimension-wise probability (DW), synthetic ranking agreement (SRA), average precision score (APR), or relative error (RE). 

Apart from using metrics suitable to the intended goal of the ML system, the difference in the objective function (e.g., cross-entropy loss) between private and non-private models could also be used. 
In some situations, such as synthetic data generation, downstream task performance is also evaluated (e.g., train models on synthetic data and compare their performance against models trained on real data). 

Sometimes, the sensitivity of those metrics is evaluated with respect to different ML algorithm hyperparameters, such as the number of dimensions in the input data, the number of training examples, the ensemble size, or even the number of training epochs. For instance, you can quantify how many additional samples are needed to achieve the same accuracy as a non-private model (i.e., sample complexity). Effective algorithms minimize this overhead \cite{abadi2016dpsgd}.

With respect to hyperparameter optimization, the recommended approach, when possible, is to perform all model architecture search and hyperparameter tuning on a proxy public dataset (with a distribution similar to the private data) and use only the private training dataset to train the final DP model \cite{abadi2016dpsgd}. 

When a single data set is used in multiple computations, the composition rule for privacy implies choosing a total privacy budget $\epsilon$ for all computations and restrict the privacy budget for each computation. When hyperpameter tuning must be performed on private data, any privacy guarantees reported should clearly state that all `data touches' are accounted for. This should include, at least, a guarantee that applies to the use of private data in training the final model, but ideally also a (weaker) guarantee that accounts for the use of private data in hyperparameter tuning \cite{jair2023dpfyml}.

No single utility metric is universally applicable \cite{nist2021utilitymetrics}. 
The choice of metrics should be guided by the specific goals of the task the ML model is intended for. A suite of metrics is often necessary to provide a comprehensive assessment.

Comparing performance metrics between the DP model and a non-private baseline quantifies the utility loss incurred for privacy. The cost of DP is a reduction in model utility. In other words, effectiveness is quantified by comparing the privatized output (the DP model) to the true non-private result (the non-private baseline model).
Effectiveness is high if the utility drop is small (a lower deviation indicates better utility). Effectiveness is excellent if DP models outperform baselines in utility for the same privacy budget $\epsilon$.

Finally, it has been observed that the cost of DP is not borne equally \cite{neurips2019accuracy}: DP model accuracy drops much more for underrepresented classes and subgroups. Critically, this gap is bigger in DP models than in non-DP models, i.e., if the original model is unfair, the unfairness becomes worse once DP is applied. 

\subsection{The privacy-utility trade-off}

Effectiveness is ultimately about optimizing the privacy-utility trade-off. In practice, this is quantified by measuring utility for different privacy budgets $\epsilon$:

\begin{itemize}

\item
{\em Privacy-utility curves}: $\text{utility}(\epsilon)$

Plot utility (e.g., accuracy, MSE) against $\epsilon$. Since lower $\epsilon$ means stronger privacy, the curve typically shows that, as privacy increases ($\epsilon$ decreases), utility decreases. Effective algorithms shift the curve upward (higher utility for the same $\epsilon$) or leftward (same utility for smaller $\epsilon$). The curve is usually monotonically decreasing: better privacy leads to worse utility  \cite{dwork2014tcs}. Small changes in $\epsilon$ can have large or small effects on utility depending on the particular DP mechanism and dataset, making it hard to predict outcomes without empirical testing. The resulting curve is often non-linear, with steep drops in utility at low $\epsilon$ values and diminishing returns at higher $\epsilon$. For instance, for the Laplace mechanism, utility decreases linearly with $1/\epsilon$, while the Gaussian mechanism with zCDP may offer sublinear degradation \cite{bun2016cdptr}. Since the privacy budget $\epsilon$ is not an intuitive concept, empirical curves help understand what a given $\epsilon$ means in practice.

\item
{\em Rate-distortion theory analogy}: $\epsilon(\text{utility})$

Treat privacy loss $\epsilon$ as ``rate'' and utility loss as ``distortion''. In rate-distortion theory, the goal is to compress data while minimizing the distortion (loss of fidelity) under a constraint on the bit rate (amount of information transmitted) \cite{coverRateDistortion2005}. The privacy-utility trade-off in DP is analogous to the rate-distortion trade-off in lossy compression \cite{du2012privacy}: stronger privacy (lower $\epsilon$) requires more noise and, therefore, produces higher distortion (lower utility); weaker privacy (higher $\epsilon$) requires less noise and leads to lower distortion (higher utility). Analogous to the rate-distortion function that gives the minimum rate needed to achieve a given distortion level, a privacy-utility function $\epsilon(\text{utility})$ gives the minimum privacy loss needed to achieve a certain utility level.
Effective algorithms achieve a lower distortion for a given rate, analogous to optimal compression. 

In rate-distortion theory, mutual information plays a central role in quantifying the trade-off between the amount of information retained and the distortion introduced during compression. Mutual information measures how much information the compressed version retains about the original data by quantifying the reduction in uncertainty about the original data given the compressed data. The mutual information  represents the minimum number of bits needed to encode the data so that the distortion does not exceed the desired level. Lower distortion (better fidelity) requires higher mutual information (more bits), and vice versa \cite{coverRateDistortion2005} \cite{ccs2016klp}.

The information bottleneck method, a generalization of rate-distortion theory, can also be applied to DP to formalize how much information about the input can be retained while satisfying privacy constraints. The information bottleneck (IB) framework \cite{tishby99information} formalizes how to extract the most relevant information from a variable while compressing it. Its goal is finding a compressed representation that retains as much information about the model output as possible and, at the same time, discards as much irrelevant information as possible about the input. The IB method aligns naturally with the DP goal of releasing information about a dataset while limiting what can be inferred about individuals.
In deep learning, IB-inspired objectives can be used to learn private embeddings that retain task-relevant information while discarding sensitive details \cite{makhdoumi2014Bottleneck} \cite{Tishby2015DeepLA}.

\item
{\em Pareto frontier:} $(\epsilon,\text{utility})$

Identify the set of $(\epsilon,\text{utility})$ pairs where no improvement in utility is possible without increasing $\epsilon$. Each point represents a DP mechanism with a specific privacy level $\epsilon$ and its corresponding utility. The Pareto-optimal points are the mechanisms for which no other mechanism is both more private and more useful. Points not on the frontier are dominated, i.e., there exists another mechanism that is better in both privacy and utility \cite{Avent2019AutomaticDO}.

The Pareto frontier helps identify the optimal trade-offs between privacy and utility, visualizing the cost of stronger privacy in terms of utility loss \cite{linCcc2024}. For mechanism selection, you just choose the particular DP mechanism on the Pareto frontier that best fits your privacy and utility requirements. For benchmarking, you compare new DP algorithm implementations against the Pareto frontier to assess their efficiency. Effective algorithms lie closer to the ideal frontier.

\end{itemize}


It is essential that experiments are performed to estimate the privacy-utility trade-off for each particular ML model and training dataset. 
The actual privacy loss might be overestimated. 
Conservative theoretical bounds may overestimate $\epsilon$, reducing perceived effectiveness.  
Formal privacy guarantees assume a worst-case adversary; real-world effectiveness may differ if adversaries are weaker. But privacy loss could also be underestimated. 
Apart from implementation subtleties, which might negatively affect privacy guarantees, effectiveness also varies with dataset size, dimensionality, or variance. This is particularly clear in small datasets, which might yield poor utility when strong privacy is required.

For a given privacy level, we need a larger sample size $n$ to achieve the same level of utility or approximation error. When data is scarce, we have to settle for the lowest $\epsilon$ for which the sacrifice in utility is acceptable. There is little consensus on how to choose that budget \cite{sp2013dpml}.  A KDD'2008 paper \cite{kdd2008composition} suggested appropriate $\delta$ should be less than $1/n^2$. 

Unlike metrics like accuracy or error rates, $\epsilon$ has no upper bound and no universally-accepted good or bad values. What counts as acceptable privacy varies by context. Selecting optimal values for $\epsilon$ and $\delta$ requires expertise, as overly restrictive parameters exacerbate utility loss, while lenient ones weaken privacy. Methods like individualized PATE require intricate tuning of upsampling and weighting parameters, with misconfigurations leading up to 10\% accuracy degradation. In distributed settings, DP-FL faces challenges from data heterogeneity, non-i.i.d. distributions amplify noise impact, reducing convergence rates by up to 20\% compared to centralized DP-SGD. 
Although client-level subsampling in federated learning provides privacy amplification, achieving a fixed global privacy budget $\epsilon$ in practice often requires injecting substantial noise due to per-client clipping, data heterogeneity, and repeated aggregation across rounds. As the number of participating clients increases, these effects can amplify the utility degradation induced by noise, even when the nominal privacy budget remains unchanged. With strict privacy constraints (i.e., low privacy budgets), dataset sampling and/or early stopping are likely to prove beneficial with respect to utility. For a specific ML algorithm, effectiveness might depend on its noise calibration, sensitivity handling, and composition properties, all of which must be tailored to the task at hand.

The aforementioned practical difficulties, rooted in the mathematical constraints of differential privacy and the practical demands of machine learning, highlight the need for experimental evaluation. 
The effectiveness of differential privacy techniques is the result of the interaction among theoretical privacy bounds, utility metrics, learning algorithm hyperparameters, and problem-specific details. Empirical validation and benchmarking are essential for the real-world application of DP in ML models. 

%% file: tableAlgorithms.tex
\newcommand{\xmark}{\ding{55}}

\begin{table}[t]
    \centering

\begin{tabular}{llllll}  
  \cline{2-6}
  & \multicolumn{2}{l}{DP} & \multicolumn{3}{l}{Empirical evaluation} \\
  \hline
  ML Algorithm & Framework & Perturbation & Privacy & Utility & Trade-off \\
  \hline
  
  & & & & & \\
  \multicolumn{6}{l}{\it Symbolic AI} \\
  PPID3 \cite{tkdd2008ppid3} \& PPRDT \cite{tdsc2014rdpp} & - & - 
    & - & - & - \\
  DiffPID3 \cite{kdd2010tdidt} & $\epsilon$ & mechanism 
    & - & $acc(|D|)$ & - \\
  DiffPC4.5 \cite{kdd2010tdidt} & $\epsilon$ & mechanism 
    & - & $acc(|D|)$ & $acc(\epsilon)$  \\
  Private-RDT \cite{icdm2009dprdt} \cite{tdp2012dprdt}  & $\epsilon$ & output 
    & - & -  & $acc(\epsilon)$ \\
  DPRF \cite{icdm2015dprf} & $\epsilon$ & mechanism
    & - & $AUC$ & $AUC(\epsilon)$ \\
  DPDF \cite{eswa2017dpdf} & $\epsilon$ & mechanism
    & $S$ & $acc(\Theta)$, $F_1$, $AUC$ & $acc(\epsilon)$ \\
  LPDT \cite{neurips2023tdidtldp} & LDP & aggregate
    & - & $MSE(\Theta)$ & $MSE(\epsilon)$ \\
  DPBoost \cite{aaai2020gbdt} & $\epsilon$ & mechanism
    & - & $acc(|E|)$ & $acc(\epsilon)$ \\
  OpBoost \cite{vldb2022opboost} & LDP & mechanism
    & - & - & $acc(\epsilon)$, $MSE(\epsilon)$ \\
  FederBoost \cite{tdsc2024federboost} & LDP & mechanism
    & - & $AUC(\Theta)$ & $AUC(\epsilon)$ \\

  & & & & & \\
  \multicolumn{6}{l}{\it Probabilistic AI} \\
  DP-NB \cite{wi2013dpnaivebayes} & $\epsilon$ & output
     & - & - & $acc(\epsilon)$ \\
  AHE-NB \cite{is2018dpnaivebayes} & $\epsilon$ & aggregate
     & - & - & - \\
  LDP-NB \cite{arxiv2019ldpnaivebayes} & LDP & aggregate
     & - & - & $acc(\epsilon)$ \\
  PrivBayes \cite{tods2017privbayes} & $\epsilon$ & aggregate
     & - & $acc(\Theta)$ & $SMI(\epsilon)$, $acc(\epsilon)$ \\    
  MST \cite{jpc2021mst} & RDP & mechanism 
     & - & - & - \\
  DPSyn \cite{jpc2021dpsyn} & $(\epsilon,\delta)$ & mechanism 
     & - & - & - \\
  PrivatePGM \cite{icml2019privatepgm} & $\epsilon$ & mechanism 
     & - & $acc$ & $acc(\epsilon)$ \\
  DPKMeans \cite{nissim2007stoc} & $\epsilon$ & aggregate
     & - & - & - \\
  PPKMeans \cite{cikm2019ppkmeans} & - & mechanism
     & - & $F_1(\Theta)$, $NMI(\Theta)$ & - \\
  LDPKMeans \cite{cs2020ldpkmeans} & LDP & aggregate
     & - & $NMI(\Theta)$, $RE(\Theta)$ & $NMI(\epsilon)$, $RE(\epsilon)$ \\
  PGME \cite{neurips2019gmm} & $(\epsilon,\delta)$ & mechanism
     & - & - & - \\
  DP-LMGMM \cite{tdsc2012gmm} & $\epsilon$ & mechanism
     & - & - & $acc(\epsilon)$ \\
  MR \cite{ndss2025gmm} & LDP & mechanism
     & - & - & $MAE(\epsilon)$ \\

  & & & & & \\
  \multicolumn{6}{l}{\it Statistical AI} \\
  PTR linear regression \cite{stoc2009dpstats} & $(\epsilon,\delta)$ & aggregate
     & - & - & - \\
  Objective LR \cite{nips2008dplogistic} \cite{jmlr2011dperm} & $(\epsilon,\delta)$ & objective
     & - & $acc(|D|)$  & $acc(\epsilon)$ \\
  FM linear regression \cite{vldb2012dpregression} & $\epsilon$ & objective
     & - & $MSE(|D|)$, $MSE(d)$ & $MSE(\epsilon)$ \\
  FM logistic regression \cite{vldb2012dpregression} & $\epsilon$ & objective
     & - & $acc(|D|)$, $acc(d)$ & $acc(\epsilon)$ \\
  PrivateSVM \cite{arxiv2009dpsvm} &  $(\epsilon,\delta)$ & output
     & - & - & - \\
  Objective SVM \cite{jmlr2011dperm} & $(\epsilon,\delta)$ & objective
     & - & $acc(|D|)$  & $acc(\epsilon)$ \\
  TDDL \cite{icml2013dpkernels} & $(\epsilon,\delta)$ & objective
     & - & - & $acc(\epsilon)$, $MSE(\epsilon)$ \\
  DPPCA-SVM \cite{scn2021dppcasvn} & $(\epsilon,0)$ & output
     & - & $acc(|D|)$ & $acc(\epsilon)$ \\
  PCA-DPSVM \cite{scn2021dppcasvn} & $(\epsilon,0)$ & output
     & - & - & $acc(\epsilon)$ \\
  DPDR-DPML \cite{cmc2024dpsvm} & $(\epsilon,\delta)$ & objective
     & - & $acc(|D|)$, $acc(d)$ & $acc(\epsilon)$ \\
          
  & & & & & \\
  \hline
  & & & & & \\
\end{tabular}

    \caption{DP in ML algorithms: Differential privacy in symbolic, probabilistic, and statistical AI. 
    Legend: $\epsilon$ (privacy budget), $S$ (sensitivity), $acc$ (accuracy, for classification problems), $MSE$ (mean-squared error, for regression problems), $MAE$ (mean absolute error), $F_1$ (F1 score), $AUC$ (area under the ROC curve), $NMI$ (normalized mutual information), $NMI$ (sum of mutual information), $RE$ (relative error), $d$ (number of dimensions), $|D|$ (training examples), $|E|$ (ensemble size), $\Theta$ (model hyperparameters).}

    \label{tab:algorithms}
\end{table}

\begin{table}[t]
    \centering

\begin{tabular}{llllll}  
  \cline{2-6}
  & \multicolumn{2}{l}{DP} & \multicolumn{3}{l}{Empirical evaluation} \\
  \hline
  ML Algorithm & Framework & Perturbation & Privacy & Utility & Trade-off \\
  \hline
  
  & & & & & \\
  \multicolumn{6}{l}{\it Deep Learning} \\
  DP-SGD \cite{abadi2016dpsgd} & $\epsilon$ & mechanism
     & - & $acc(\Theta)$ & $acc(\epsilon)$ \\
  NoisySGD \cite{hdsr2020dlgdp} & GDP & mechanism
     & $\delta(\epsilon)$, $\epsilon(\delta)$ & $acc(\Theta)$ & $acc(\epsilon)$ \\

  DPAGD-CNN \cite{dsc2019dpcnn} & zCDP & mechanism
     & - & - & $acc(\epsilon)$, $acc(\delta)$ \\
  LATENT \cite{iotj2019ldpdl} & LDP & mechanism
     & - & - & $acc(\epsilon)$ \\
  DP-LSTM \cite{arxiv2019dplstm} & - & input
     & - & $MSE$, $acc$ & - \\
  DSSGD \cite{ccs2015dpdeeplearning} & - & aggregate
     & - & $acc(\Theta)$ & - \\
  FedAvg \cite{aistats2017federateddeepnetworks} & - & aggregate
     & - & $acc(\Theta)$ & - \\

  dPA \cite{aaai2016dpautoencoders} & $\epsilon$ & objective
     & - & $acc(|D|)$ & $acc(\epsilon)$ \\
  DP-AuGM \cite{arxiv2018dpgenerative} & $(\epsilon,\delta)$ & mechanism 
     & - & $acc(\Theta)$ & $acc(\epsilon)$, $acc(\delta)$ \\
  DP-VaeGM \cite{arxiv2018dpgenerative} & $(\epsilon,\delta)$ & mechanism
     & - & $acc(\Theta)$ & $acc(\epsilon)$, $acc(\delta)$ \\
  ADePT \cite{arxiv2021dpautoencoder} & $(\epsilon,\delta)$ & mechanism 
     & - & - & $MIA(acc)$ \\
  PATE-AAE \cite{interspeech2021pateaae} & $(\epsilon,\delta)$ & aggregate 
     & - & - & $acc(\epsilon)$ \\
  DP-[W]GAN \cite{arxiv2018dpgan} & $(\epsilon,\delta)$ & mechanism
     & - & $W(\Theta)$ & $acc(\epsilon)$, $W(\epsilon)$, $DW(\epsilon)$ \\
  PATE-GAN \cite{iclr2019pategan} & $(\epsilon,\delta)$ & aggregate
     & - & $AUC(\Theta)$ & $AUC(\epsilon)$, $SRA(\epsilon)$  \\
  G-PATE \cite{neurips2021gpate} & RDP & aggregate
     & - & $AUC$, $acc(\Theta)$ & $acc(\epsilon)$, $IS(\epsilon)$ \\
  DataLens \cite{ccs2021datalens} & RDP & aggregate
     & $\epsilon(\Theta)$ & $acc(\Theta)$ & $acc(\epsilon)$, $IS(\epsilon)$  \\
  DTGAN \cite{arxiv2021dtgan} & RDP & mechanism 
     & $MIA$, $AIA$ & \multicolumn{2}{l}{$acc$, $F_1$, $AUC$, $APR$, $JS$, $W$} \\  
  GANobfuscator \cite{tifs2019ganobfuscator} & $(\epsilon,\delta)$ & mechanism
     & - & $acc(|D|)$ & $IS(\epsilon)$, $JS(\epsilon)$, $MIA(\epsilon)$ \\  

  & & & & & \\
  \hline
  & & & & & \\
\end{tabular}

    \caption{DP in ML algorithms (continued): Differential privacy in deep learning. Legend: $\epsilon$ (privacy budget), $S$ (sensitivity), $acc$ (accuracy, for classification problems), $MSE$ (mean-squared error, for regression problems), $MAE$ (mean absolute error), $F_1$ (F1 score), $AUC$ (area under the ROC curve), $APR$ (average precision score), $NMI$ (normalized mutual information), $IS$ (Inception score \cite{nips2017wgan} to measure the quality of generated data), $JS$ (Jensen-Shannon divergence), $W$ (Wasserstein distance), $DW$ (dimension-wise probability \& prediction), $SRA$ (synthetic ranking agreement), $RE$ (relative error), $MIA$ (membership inference attack success rate, as a quantifiable measure of privacy preservation), $AIA$ (attribute inference attack), $d$ (number of dimensions), $|D|$ (training examples), $|E|$ (ensemble size), $\Theta$ (model hyperparameters).}

    \label{tab:algorithmsDL}
\end{table}

%% file: conclusion.tex
\section{Conclusion}
\label{chapter:conclusion}


Differential Privacy (DP) is the leading theoretical framework for data privacy, offering mathematically rigorous and provable guarantees against a wide range of inference attacks. Its evolution from the foundational $\epsilon$-DP definition to more nuanced variants like $(\epsilon,\delta)$-DP, zCDP, RDP, GDP, and LDP reflects a continuous effort to enhance analytical tractability, particularly for composition, and capture the privacy properties of specific mechanisms. Core theoretical properties like sensitivity calibration, composition theorems, and post-processing immunity provide the tools for building complex privacy-preserving systems.

Differential Privacy provides an invaluable standard for reasoning about and quantifying privacy loss in an era of large-scale data analysis and machine learning. 
In machine learning, DP has been successfully adapted to a wide range of algorithms. While objective perturbation is common for convex models like SVMs, and specialized techniques exist for decision trees and ensembles, gradient perturbation via DP-SGD has become the de facto standard for deep learning. This involves careful gradient clipping, noise addition, and sophisticated privacy accounting over iterative training.

Evaluating DPML systems needs a multi-pronged approach. Theoretical $(\epsilon,\delta)$ bounds provide worst-case guarantees, standard ML metrics assess utility loss, and empirical privacy attacks (like MIA) offer practical lower bounds on information leakage. 

The significance of DP extends beyond technical achievements, as it addresses ethical and regulatory demands in data-sensitive domains. For instance, DP aligns with privacy regulations such as the European Union's General Data Protection Regulation (GDPR) \cite{gdpr2016}, which mandates strict controls on personal data processing \cite{bun2016cdp}.


ML security-related issues go beyond differential privacy. They can be broadly classified into five categories~\cite{access2020mls}: training set poisoning, backdoors in the training set, adversarial example attacks, model theft, and recovery of sensitive training data. Differential privacy directly targets the last category by providing formal protection against information leakage from trained models, including membership inference, attribute inference, and data reconstruction attacks. However, DP does not address integrity or availability threats such as data poisoning, backdoors, or adversarial examples, nor does it prevent model theft. In this sense, DP should be understood as a specialized privacy-preserving defense mechanism that complements, rather than replaces, other security techniques.

Deploying DP introduces additional challenges and trade-offs, including increased computational costs, potential utility degradation due to noise, and the complexity of tuning privacy parameters like \( \epsilon \) and \( \delta \). DP, however, is well-aligned with the general goals of machine learning. Memorizing a particular training example is a violation of privacy, but it is also a form of overfitting that harms ML models generalization capability. Hence, DP implies some form of stability in ML.

Once the models are trained, two additional concerns must be considered when models are deployed to make predictions \cite{network2018pnnl}: prediction requests might contain sensitive information (user privacy) and the knowledge embedded in the trained models should be protected (both training data and model hyperparameters). DP helps protecting training data, but additional measures should be taken to address the remaining privacy considerations.


A critical evaluation of DP’s benefits and liabilities is needed, particularly in the context of machine learning, where high-dimensional data and iterative model training amplify these issues. By evaluating DP's strengths and weaknesses, we conclude our comprehensive survey on the current state of DP in machine learning and its practical implications for data-sensitive applications.


On the positive side, DP use in ML offers some significant benefits in practice:

\begin{itemize}

\item
{\em Sensitive data analysis}: DP allows valuable statistical analysis and ML model training on sensitive datasets that might otherwise be inaccessible due to privacy regulations or concerns.

\item
{\em Provable privacy guarantees}: In sharp contrast with the heuristic nature of older anonymization techniques, DP provides a rigorous, mathematical definition of privacy, allowing for quantifiable upper bounds on information leakage concerning individuals' data.

\item
{\em Privacy accounting}: DP mechanisms are composable, allowing the construction of complex privacy-preserving workflows and algorithms from simpler building blocks. Composition theorems provide bounds on the cumulative privacy loss. Advanced privacy accounting enhances the tightness of $(\epsilon, \delta)$-guarantees.

\item
{\em Post-processing immunity}: Any computation performed on the output of a DP mechanism, without accessing the original private data, inherits the same DP guarantee. 

\item
{\em Robustness to auxiliary information}: DP formal guarantees hold irrespective of any side information an adversary might possess, now or in the future. This offers resilience that is not provided by other techniques.

\item
{\em Defense against inference attacks}: DP provides inherent protection against a range of common ML privacy attacks, including membership inference, attribute inference, and data reconstruction/extraction.

\item 
{\em Scalability}: DP methods extend privacy to large-scale and distributed systems. DP-SGD \cite{abadi2016dpsgd} can be used to train large neural networks, whereas EW-Tune \cite{behnia2022ewtune}
can be used for LLM fine-tuning. DP-FL \cite{aistats2017federateddeepnetworks} scales DP to federated settings.

\end{itemize}


Despite its many strengths, DP faces significant limitations and challenges in the context of ML:

\begin{itemize}

\item
{\em Privacy-utility trade-off}: The most fundamental DP challenge is the inherent trade-off between the strength of the privacy guarantee $(\epsilon,\delta)$ and the utility of the output (e.g., accuracy in classification, fidelity in data generation). Stronger privacy (smaller $\epsilon$) generally requires adding more noise, which degrades utility. Empirical studies indicate that the trade-off between model utility and privacy loss is complex. The fundamental trade-off between privacy and utility often leads to difficult choices, particularly for complex ML models where achieving strong privacy guarantees can substantially degrade performance. Achieving high utility with strong privacy guarantees remains difficult for many complex tasks.

\item
{\em Disparate impact and fairness}: A major concern is that DP can disproportionately affect ML model utility for underrepresented classes and subgroups, exacerbating existing biases. The potential for DP mechanisms to exacerbate unfairness and biases requires careful consideration and mitigation strategies. 

\item
{\em Interpretability of privacy guarantees}: The interpretation of the meaning of specific $(\epsilon,\delta)$ guarantees is far from intuitive. It can create communication barriers among project stakeholders, specially those without a theoretical background on DP.

\item 
{\em Sensitivity analysis}: Accurately bounding the global sensitivity of complex functions or ML training procedures can be difficult, potentially leading to overly conservative noise addition.
Many practical deployments use relatively large $\epsilon$ values, where theoretical worst-case guarantees are weak, and rely on empirical evidence of resistance to known attacks.

\item
{\em Implementation complexity and correctness}: The many subtleties involved in the proper implementation of DP might lead to unnoticed privacy problems and additional privacy and security risks. 
 
\item
{\em Hyperparameter tuning}: DPML models are often highly sensitive to many hyperparameters ($\epsilon$, $\delta$, clipping norm, noise scale, learning rate, batch size...). Finding optimal settings is complex, computationally expensive, and the tuning process itself can leak information if not done carefully or accounted for in the privacy budget.

\item
{\em Computational overhead:} DP mechanisms, e.g. DP-SGD for deep learning, can impose significant computational costs when  compared to non-private methods.
DP-SGD and EW-Tune can increase training time by 2x-3x, while EW-Tune’s Edgeworth Accountant requires an additional 20\% computational cost for iterative statistical computations. 
These computational costs limit DP applicability in resource-constrained situations.


\end{itemize}

The tension between DP theoretical foundations and practical issues with respect to  utility, complexity, and fairness is unlikely to disappear. While DP provides robust protection against certain privacy threats in theory, its application requires careful consideration of its many trade-offs and limitations.
Integrating DP with other privacy-enhancing technologies (PETs), such as cryptography, and considering it within broader ethical frameworks for responsible AI will be increasingly important.
Bridging the gap between the theoretical promises of DP and its practical, equitable, and widespread deployment requires continued innovation and interdisciplinary collaboration between computer scientists, statisticians, domain experts, ethicist, and policymakers. While not a panacea, DP is a critical component in the ongoing effort to harness the power of data responsibly.


The challenges outlined above suggest some avenues for further research, such as
improving the privacy-utility trade-off by designing more sophisticated DP algorithms,
addressing fairness and bias by understanding the mechanisms causing unfairness and designing interventions,
reducing the computational and memory footprint of DP algorithms,
and developing tailored DP algorithms for newer ML techniques like Large Language Models (LLMs) or Graph Neural Networks (GNNs).
From a theoretical point of view, 
improved privacy accounting and tighter composition theorems would enable a more efficient use of the privacy budget.
Formal verification tools and techniques are needed to increase trust in the implementation of complex DP algorithms.
Open questions remain regarding fundamental limits (lower bounds), the relationships between different DP variants, the optimal choice of privacy parameters, and the interplay between privacy and other constraints like computation and communication.
Future progress of DP in ML will likely involve context-specific solutions tailored to particular algorithms, data types, and application domains, moving beyond generic mechanisms.


Some final considerations regarding DP should be mentioned, as they are related to the opportunities and risks of DP adoption in diverse fields and applications. 
The trade-off between individual privacy and model utility is inherent to DP and it should be always be made explicit: higher privacy sacrifices model utility, higher utility sacrifices individual privacy. Appropriate best practices and guidelines on privacy-preserving technologies such as DP should be promoted. 
Even though DP quantifies privacy loss, as a theoretical upper bound, it might also encourage more data (and model) sharing. Due to the many subtleties associated to the use of DP, its use can create a false sense of security and that can actually increase actual privacy risks. For auditing and evaluating DP robustness, the use of privacy attacks (i.e., MIAs, AIAs, and reconstruction attacks) is recommended to assess the actual privacy offered by DP systems in practice and identify potential implementation flaws.